\documentclass[prd,showpacs,preprintnumbers,twocolumn,amsmath,nofootinbib,amssymb]{revtex4}
\usepackage{graphicx,color,dcolumn,booktabs,bm}
\usepackage{longtable,lscape}
\usepackage{txfonts}
\usepackage{overpic}
\usepackage{amssymb}
\usepackage{epstopdf}
\usepackage{indentfirst}
\usepackage{feynmf}   %{feynmp}
\usepackage{slashed}  %for Feynman symbols
\usepackage{cases}
\usepackage{color}
\usepackage{float}
\usepackage{multirow}
\usepackage{graphicx,color,dcolumn,booktabs,bm}
\usepackage{epsfig,dsfont,amssymb,amsmath,amsfonts,amsbsy,mathrsfs}
\usepackage{ulem}

\graphicspath{{Figures/}} %

\usepackage{hyperref}
\hypersetup{colorlinks,citecolor=blue,anchorcolor=red,menucolor=red, linkcolor=red,filecolor=red,runcolor=red,urlcolor=blue,frenchlinks=red}

\makeatletter
\@addtoreset{equation}{section}
\makeatother

\allowdisplaybreaks

\begin{document}

\title{Probing new types of $P_c$ states inspired by the interaction between an $S$-wave charmed baryon and an anticharmed meson in a $\bar T$ doublet state}

\author{Fu-Lai Wang$^{1,2}$}
\email{wangfl2016@lzu.edu.cn}
\author{Rui Chen$^{1,2,3,4}$}
\email{chen$_$rui@pku.edu.cn}
\author{Zhan-Wei Liu$^{1,2}$}
\email{liuzhanwei@lzu.edu.cn}
\author{Xiang Liu$^{1,2}$}
\email{xiangliu@lzu.edu.cn}
\affiliation{$^1$School of Physical Science and Technology, Lanzhou University, Lanzhou 730000, China
\\
$^2$Research Center for Hadron and CSR Physics, Lanzhou University and Institute of Modern Physics of CAS, Lanzhou 730000, China\\
$^3$School of Physics and State Key Laboratory of Nuclear Physics and Technology, Peking University, Beijing 100871, China\\
$^4$Center of High Energy Physics, Peking University, Beijing 100871, China
}

\begin{abstract}
Inspired by the observations of three $P_c$ states, we systematically investigate interactions between an $S$-wave charmed baryon $\mathcal{B}_{c}^{(*)}=\Lambda_c/\Sigma_c/\Sigma_c^{*}$ and an anticharmed meson $\bar T=\bar D_1/\bar D_2^*$ with the one-pion-exchange potential model and the one-boson-exchange potential model, and search for possible new types of $P_c$ states with the structures of $\mathcal{B}_{c}^{(*)}\bar T$. Both $S$-$D$ wave mixing and coupled channel effects are considered. Our results suggest that in some $\mathcal{B}_{c}^{(*)}\bar T$ systems there are ideal candidates of new types of $P_c$ states, i.e., the $\Sigma_c\bar{D}_1$ state with $I(J^P)=1/2(1/2^+)$, the $\Sigma_c\bar{D}_2^*$ state with $I(J^P)=1/2(3/2^+)$, the $\Sigma_c^*\bar{D}_1$ state with $I(J^P)=1/2(1/2^+)$, and the $\Sigma_c^*\bar{D}_2^*$ states with $I(J^P)=1/2(1/2^+, 3/2^+)$, and we suggest that these predicted new types of $P_c$ states can be detected in the process $\Lambda_b^0 \to \psi(2S) p \pi^{-}$.  Meanwhile, we also extend our study to the interactions between an $S$-wave charmed baryon and a charmed meson in a $T$ doublet, and we predict a series of double-charm molecular pentaquarks.
\end{abstract}

 %Potential models, Charmed mesons, Charmed baryons

\maketitle

\section{introduction}\label{sec1}

Exploration of exotic hadronic configurations, which are beyond conventional mesons and baryons in the unquenched quark model, is the academic frontier of hadron physics since these exotic hadronic states have a special position in hadron spectroscopy.  Studying exotic hadronic states can provide useful hints to deepen our understanding of the nonperturbative behavior of quantum chromodynamics (QCD). In the past 16 years, experimentalist have observed more and more charmoniumlike $XYZ$ states and $P_c$ states, which have stimulated the development of investigation of exotic hadronic states (see review papers \cite{Chen:2016qju,Liu:2013waa,Liu:2019zoy,Olsen:2017bmm,Guo:2017jvc,Hanhart:2017kzo} for more details).

In March of 2019, the LHCb Collaboration announced new progress on pentaquarks. By revisiting the process $\Lambda_b^0\to J/\psi p K$ with the combined data set collected in Run 1 plus Run 2 \cite{Aaij:2019vzc}, it was found that three resonance structures $P_c(4312)$, $P_c(4440)$, and $P_c(4457)$ exist in the $J/\psi p$ invariant mass spectrum. This observation shows that the $P_c(4450)$ structure reported by LHCb in 2015 contains two substructures $P_c(4440)$ and $P_c(4457)$ \cite{Aaij:2015tga}. This updated result of $P_c$ states provides strong evidence to support the molecular pentaquark \cite{Chen:2015loa,Chen:2015moa,Karliner:2015ina,Roca:2015dva,Mironov:2015ica,He:2015cea,Meissner:2015mza,Burns:2015dwa,Shimizu:2016rrd,Chen:2016heh,Eides:2015dtr,Huang:2015uda,Chen:2016otp,Yang:2015bmv,He:2016pfa,
Yamaguchi:2016ote,Richard:2017una,Yang:2011wz,Wu:2010jy,Wang:2011rga,Wu:2012md,Chen:2019asm,Chen:2019bip,Liu:2019tjn,He:2019ify,Xiao:2019aya,Guo:2019kdc,Huang:2019jlf,Xiao:2019mst,Shimizu:2019ptd,Meng:2019ilv,
Burns:2019iih,Pan:2019skd,Liu:2019zvb,Xu:2019zme,Valderrama:2019chc,Wang:2019hyc,Yamaguchi:2019seo,Cheng:2019obk,Du:2019pij}.

When facing such new and exciting results of exploring exotic hadronic states, we  propose the question, Where are we going? Let us review the charmoniumlike $XYZ$ from $B$ meson decays. As the first reported charmoniumlike state, $X(3872)$ has inspired extensive discussion of the $D\bar{D}^*$ molecule, which is related to an $S$-wave charmed meson interacting with an $S$-wave anticharmed meson. Of course, one must be curious about other hidden-charm charmoniumlike molecular states which are composed of an $S$-wave charmed meson and a $P$-wave anticharmed meson. Especially, the observation of $Z(4430)$ directly results in the discussion of whether it can be a $D^*\bar{D}_1$ or $D^*\bar{D}_1^\prime$ molecule \cite{Liu:2007bf,Liu:2008xz}. Later, our group carried out a systematical calculation of molecular systems composed of an $S$-wave charmed meson and a $P$-wave anticharmed meson with $J^P=0^+,1^+$ \cite{Shen:2010ky}, and predicted the possible heavy molecular states from the interaction between a pair of excited charm-strange mesons with $J^P=0^+$ or $1^+$ \cite{Hu:2010fg} and a pair of excited charmed mesons $J^P=1^+$ or $2^+$ \cite{Chen:2015add}. The observations of charmoniumlike structures in the $J/\psi\phi$ and $J/\psi\omega$ invariant mass spectra also made us combine these experimental data with former theoretical results to carry out further phenomenological work \cite{Liu:2010hf}. To date, more and more higher charmoniumlike $XYZ$ states from $B$ meson weak decays not only make the charmoniumlike $XYZ$ family become abundant, but also inspire interest of researchers in studying the interaction of more highly excited charmed mesons \cite{Chen:2016qju}.

\begin{figure}[!htbp]
\centering
\begin{tabular}{c}
\includegraphics[width=0.46\textwidth]{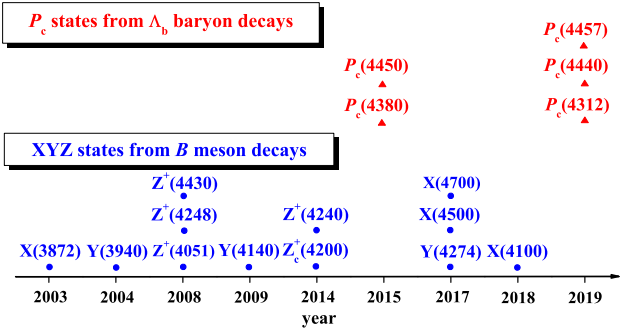}
\end{tabular}
\caption{(color online) The observed $P_c$ states from $\Lambda_b$ baryon decays \cite{Aaij:2019vzc,Aaij:2015tga} and charmoniumlike $XYZ$ states from $B$ meson decays \cite{Choi:2003ue,Abe:2004zs,Choi:2007wga,Mizuk:2008me,Aaltonen:2009tz,Aaij:2014jqa,
Chilikin:2014bkk,Aaij:2016iza,Aaij:2016nsc,Aaij:2018bla} with the corresponding first observation year.}\label{XYZPc}
\end{figure}

 When checking the production processes of $P_c$ states from $\Lambda_b$ baryon decays and $XYZ$ states from $B$ meson decays, there exists a similarity. At present, LHCb has only observed few $P_c$ states, which are closely related to the $S$-wave charmed meson and $S$-wave anticharmed baryon (see Fig. \ref{XYZPc}). According to the experiments behind discovered $XYZ$ states, we have enough reason to believe that more $P_c$ states with larger mass will be reported by LHCb in the near future. Theorists should pay more attention to this interesting issue and give more suggestions for future experimental searches for $P_c$ states. In fact, the present status of the theoretical study of $P_c$ is also similar to that of $XYZ$ states around 2008. Thus, it is a good starting point to investigate how an $S$-wave charmed baryon interacts with a more highly excited anticharmed meson, by which we can provide valuable information of new type of hidden-charm molecular pentaquarks.

In this work, we will investigate the interactions between an $S$-wave ground charmed baryon $\mathcal{B}_c^{(*)}=\Lambda_c/\Sigma_c/\Sigma_c^*$ and an anticharmed meson in a $\bar T$ doublet $(\bar{D}_1,\bar{D}_2^*)$, where the four $P$-wave charmed mesons have narrow widths \cite{Tanabashi:2018oca}. We also extend our work to study systems with a charmed baryon $\mathcal{B}_c^{(*)}$ and an excited charmed meson ${D}_1/{D}_2^*$. Here, we respectively adopt the one-pion-exchange (OPE) model and the one-boson-exchange (OBE) model, including the exchange contributions from $\pi$, $\sigma$, $\eta$, $\rho$, and $\omega$ in the latter model, which is similar to the study of nuclear force. In addition, the $S$-$D$ wave mixing effects and coupled channel effects are taken into account in our calculations. We hope that valuable information provided here can be helpful for further experimental searches for the predicted $P_c$ states with larger masses.

This paper is organized as follows. After this Introduction, we present the detailed framework in Sec.~\ref{sec2}. New types of $P_c$ states with the structures of $\Sigma_{c}^{(*)}\bar T$ and $\Lambda_{c}\bar T$  are given in Sec.~\ref{sec3} and Sec.~\ref{sec4}, respectively. In Sec. \ref{sec5}, we show the numerical results for doubly charmed molecular pentaquarks. The paper ends with a summary in Sec. \ref{sec6}.

\section{Framework}\label{sec2}

To answer whether or not an $S$-wave ground charmed baryon $\mathcal{B}_c^{(*)}=\Lambda_c/\Sigma_c/\Sigma_c^*$ and an anticharmed meson in $\bar T$ doublet $(\bar{D}_1,\bar{D}_2^*)$ can be bound together to form a molecular state, we first investigate the interactions between $\mathcal{B}_c^{(*)}$ and $\bar{D}_1/\bar{D}_2^*$. In this work, we adopt the OPE model and the OBE model, which were developed to study the interactions between hadrons with observed $XYZ$ charmoniumlike states and $P_c$ states \cite{Liu:2019zoy}, to quantitatively describe the strong interactions between $\mathcal{B}_c^{(*)}$ and $\bar{D}_1/\bar{D}_2^*$.

In general, there are three typical steps for extracting the effective potentials which depict the strong interactions between hadrons:

\begin{enumerate}
\item  By the effective Lagrangians approach, we can write out the scattering amplitudes $\mathcal{M}(h_1h_2\to h_3h_4)$ of the scattering process $h_1h_2\to h_3h_4$. Later, we list the effective Lagrangians adopted in the present work.

\item  Due to the Breit approximation, the effective potentials in momentum space can be related to the scattering amplitudes, i.e.,
\begin{eqnarray}\label{breit}
\mathcal{V}_E(\bm{q}) &=&-\frac{\mathcal{M}(h_1h_2\to h_3h_4)}{\sqrt{\prod_i2M_i\prod_f2M_f}},
\end{eqnarray}
where $M_{i(f)}$ denotes the mass of the initial (final) state.

\item By the Fourier transformation, the effective potentials $\mathcal{V}_E(\bf{q})$ can be transformed into $\mathcal{V}_E(\bm{r})$ in coordinate space, i.e.,
\begin{eqnarray}
\mathcal{V}_E(\bm{r}) =\int\frac{d^3\bm{q}}{(2\pi)^3}e^{i\bm{q}\cdot\bm{r}}\mathcal{V}_E(\bm{q})
\mathcal{F}^2(q^2,m_E^2),
\end{eqnarray}
where $m_E$ and $q$ are the mass and four-momentum of the exchanged particle, respectively. The monopole type form factor $\mathcal{F}(q^2,m_E^2) = (\Lambda^2-m_E^2)/(\Lambda^2-q^2)$ \cite{Tornqvist:1993ng,Tornqvist:1993vu} compensates for the effects from the off-shell exchanged mesons and the more complicated structure of vertices. The cutoff $\Lambda$ as a phenomenological parameter reflects the intrinsic size of hadrons or the typical hadronic scale. In addition, the introduced form factor also regularizes the effective potentials to cure the ultraviolet singularity.
\end{enumerate}
We can obtain the properties for some possible bound states by solving the Schr\"odinger equation with the effective potentials.

%\subsection{The effective Lagrangians}

To write out the relevant scattering amplitudes, one needs the effective Lagrangians to describe the interactions between $\mathcal{B}_c^{(*)}/\bar{D}_1/\bar{D}_2^*$ and the light scalar, pesudoscalar, or vector mesons \cite{Liu:2011xc,Ding:2008gr}, i.e.,
\begin{eqnarray}
\label{lag1}
\mathcal{L}_{\bar {T}}&=&g''_{\sigma}\langle \overline{T}^{(\bar Q)\mu}_a\sigma T^{\,({\bar Q})}_{a\mu}\rangle+ik\langle \overline{T}^{(\bar Q)\mu}_{b}{\cal A}\!\!\!\slash_{ba}\gamma_5{T}^{\,(\bar Q)}_{a\mu}\rangle\nonumber\\
    &&-i\beta^{\prime\prime}\langle \overline{T}^{(\bar Q)}_{b\lambda}v^{\mu}({\cal V}_{\mu}-\rho_{\mu})_{ba}T^{\,(\bar Q)\lambda}_{a}\rangle\nonumber\\
    &&+i\lambda^{\prime\prime}\langle \overline{T}^{(\bar Q)}_{b\lambda}\sigma^{\mu\nu}F_{\mu\nu}(\rho)_{ba}T^{\,(\bar Q)\lambda}_{a}\rangle,\\
\mathcal{L}_{\mathcal{B}_{\bar{3}}} &=& l_B\langle\bar{\mathcal{B}}_{\bar{3}}\sigma\mathcal{B}_{\bar{3}}\rangle
          +i\beta_B\langle\bar{\mathcal{B}}_{\bar{3}}v^{\mu}(\mathcal{V}_{\mu}
          -\rho_{\mu})\mathcal{B}_{\bar{3}}\rangle,\\
\mathcal{L}_{\mathcal{S}} &=& l_S\langle\bar{\mathcal{S}}_{\mu}\sigma\mathcal{S}^{\mu}\rangle -\frac{3}{2}g_1\varepsilon^{\mu\nu\lambda\kappa}v_{\kappa}\langle\bar{\mathcal{S}}_{\mu}A_{\nu}
\mathcal{S}_{\lambda}\rangle\nonumber\\
    &&+i\beta_{S}\langle\bar{\mathcal{S}}_{\mu}v_{\alpha}\left(\mathcal{V}^{\alpha}
    -\rho^{\alpha}\right) \mathcal{S}^{\mu}\rangle +\lambda_S\langle\bar{\mathcal{S}}_{\mu}F^{\mu\nu}(\rho)\mathcal{S}_{\nu}\rangle,\nonumber\\\\
\mathcal{L}_{\mathcal{B}_{\bar{3}}\mathcal{S}}&=&ig_4\langle\bar{\mathcal{S}^{\mu}}
\mathcal{A}_{\mu}\mathcal{B}_{\bar{3}}\rangle+i\lambda_I\varepsilon^{\mu\nu\lambda
\kappa}v_{\mu}\langle \bar{\mathcal{S}}_{\nu}F_{\lambda\kappa}\mathcal{B}_{\bar{3}}\rangle+H.c., \label{lag2}
\end{eqnarray}
which can be constructed in heavy quark symmetry and the chiral symmetry \cite{Liu:2011xc,Wise:1992hn,Casalbuoni:1992gi,Casalbuoni:1996pg,Yan:1992gz,Ding:2008gr}. Here, $v=(1,\bm{0})$ is the four velocity under the non-relativistic approximation. The multiplet field $T$ is composed of an axial-vector meson $\bar D_{1}=\left(\bar D_{1}^0, D_{1}^-, D_{s1}^-\right)^T$ and a tensor meson $\bar D_{2}^{*}=\left(\bar D_{2}^{*0}, D_{2}^{*-}, D_{s2}^{*-}\right)^T$ \cite{Ding:2008gr}:
\begin{eqnarray}
T^{(\bar Q)\mu}_a &=& \left[\bar D^{*\mu\nu}_{2a}\gamma_{\nu}-\sqrt{\frac{3}{2}}\bar D_{1a\nu}\gamma_5\left(g^{\mu\nu}-\frac{1}{3}(\gamma^{\mu}-v^{\mu})\gamma^{\nu}
\right)\right]\frac{1-\slash \!\!\!v}{2}.\nonumber
\end{eqnarray}
Its conjugate field satisfies $\overline{T}^{(\bar Q)\mu}_a=\gamma^0T^{(\bar Q)\mu\dag}_a\gamma^0$. Superfield $\mathcal{S}_{\mu}$ includes $\mathcal{B}_6$ with $J^P=1/2^+$ and $\mathcal{B}^*_6$ with $J^P=3/2^+$ in the $6_F$ flavor representation, i.e., $\mathcal{S}_{\mu} =-\sqrt{\frac{1}{3}}(\gamma_{\mu}+v_{\mu})\gamma^5\mathcal{B}_6+\mathcal{B}_{6\mu}^*$.
Axial current $\mathcal{A}_\mu$ and vector current ${\cal V}_{\mu}$ are defined as
${\mathcal A}_{\mu}=\left(\xi^{\dagger}\partial_{\mu}\xi-\xi\partial_{\mu}\xi^{\dagger}\right)/2$ and
${\mathcal V}_{\mu}=\left(\xi^{\dagger}\partial_{\mu}\xi+\xi\partial_{\mu}\xi^{\dagger}\right)/2$, respectively.
Here, $\xi=\exp(i\mathbb{P}/f_\pi)$ and $f_\pi=132~\rm{MeV}$ is for the pion decay constant. In the above formula, vector meson field $\rho_{\mu}$ and its strength tensor $F_{\mu\nu}(\rho)$ are $\rho_{\mu}=i{g_V}\mathbb{V}_{\mu}/{\sqrt{2}}$ and $F_{\mu\nu}(\rho)=\partial_{\mu}\rho_{\nu}-\partial_{\nu}\rho_{\mu}+[\rho_{\mu},\rho_{\nu}]$, respectively. Here, matrices $\mathcal{B}_{\bar{3}}$ and $\mathcal{B}_6^{(*)}$ are
\begin{eqnarray*}
\mathcal{B}_{\bar{3}} = \left(\begin{array}{ccc}
        0    &\Lambda_c^+      &\Xi_c^+\\
        -\Lambda_c^+       &0      &\Xi_c^0\\
        -\Xi_c^+      &-\Xi_c^0     &0
\end{array}\right),\quad
\mathcal{B}_6^{(*)} = \left(\begin{array}{ccc}
         \Sigma_c^{{(*)}++}                  &\frac{\Sigma_c^{{(*)}+}}{\sqrt{2}}     &\frac{\Xi_c^{(',*)+}}{\sqrt{2}}\\
         \frac{\Sigma_c^{{(*)}+}}{\sqrt{2}}      &\Sigma_c^{{(*)}0}    &\frac{\Xi_c^{(',*)0}}{\sqrt{2}}\\
         \frac{\Xi_c^{(',*)+}}{\sqrt{2}}    &\frac{\Xi_c^{(',*)0}}{\sqrt{2}}      &\Omega_c^{(*)0}
\end{array}\right),
\end{eqnarray*}
respectively. Additionally, matrices $\mathbb{P}$ and $\mathbb{V}_{\mu}$ have the conventional form which can be found in Ref. \cite{Chen:2019asm}.

We pick the interaction terms which contain only one light meson by expanding the effective Lagrangians (\ref{lag1})-(\ref{lag2}); such terms are collected in Appendix~\ref{app01}.

We need several input parameters to obtain the numerical results. Firstly, the coupling constants existing in Eqs. (\ref{lag1})-(\ref{lag2}) are $l_S=-2l_B=7.30$, $g_1={2\sqrt{2}}g_4/3=0.94$, $\beta_S g_{V}=-2\beta_B g_{V}=-12.00$, $\lambda_Sg_V=-2\sqrt{2}\lambda_Ig_V=-19.20 ~\rm {GeV^{-1}}$, and $k=-0.59$ \cite{Liu:2011xc}, which are widely used to discuss hadronic molecular states \cite{He:2015cea,Chen:2019asm,He:2019ify}. Here, the coupling constant $g_1$ can be extracted from the strong decay $\Sigma_c^{(*)} \to \Lambda_c \pi$ \cite{Tanabashi:2018oca}. The coupling constant $k$ for $D_1D_2^*\pi$ vertex is the same as the $DD^*\pi$ coupling constant within the quark model \cite{Falk:1992cx}, which is determined from the experimental width $\Gamma(D^* \to D \pi)$ \cite{Tanabashi:2018oca}. As is well known, information about the coupling constants $g_{\sigma}^{\prime\prime}$, $\beta^{\prime\prime}$, and $\lambda^{\prime\prime}$ is very scarce at present \cite{Ding:2008gr}. In this work, we estimate these coupling constants in the quark model \cite{Riska:2000gd}, and they can be extracted from the $D^{(*)}$-$D^*$ interactions \cite{Chen:2019asm}. Thus, we use $g_{\sigma}^{\prime\prime}=-0.76$,  $\beta^{\prime\prime} g_{V}=-5.25$, and $\lambda^{\prime\prime} g_{V}=3.44 ~\rm {GeV^{-1}}$ in the following  numerical analysis \cite{Liu:2011xc,Wang:2019aoc}, and the signs for these coupling constants are fixed by the quark model. In addition, the adopted mass values are $m_\sigma=600.00$ MeV, $m_\pi=137.27$ MeV, $m_\eta=547.85$ MeV, $m_\rho=775.49$ MeV, $m_\omega=782.65$ MeV, $m_N=938.27$ MeV, $m_{D_1}=2422.00$ MeV, $m_{D_2^*}=2463.05$ MeV, $m_{\Lambda_{c}}=2286.46$ MeV, $m_{\Sigma_{c}}=2453.54$ MeV, and $m_{\Sigma_{c}^*}=2518.07$ MeV \cite{Tanabashi:2018oca}. In our calculation, only the cutoff $\Lambda$ is a free parameter, and we attempt to find bound state solutions by varying the cutoff parameter in the range of $0.79$-$4.00~{\rm GeV}$.

In general, the loosely bound states are sensitive to the details of the effective potentials. Because of the uncertainties of the coupling constants $g_{\sigma}^{\prime\prime}$, $\beta^{\prime\prime}$, and $\lambda^{\prime\prime}$, we should be cautious when studying the existence possibility of the $\mathcal{B}_{c}^{(*)}\bar T$-type hadronic molecular states with the OBE potential model. On the other hand the $\pi$ exchange interaction, which is clear and widely accepted,  contributes to the long-range force in the formation of hadronic molecular states. However, the scalar and vector meson exchange interactions are used to provide the intermediate and short-range contributions, which are not determined very precisely. If the $\pi$ exchange interaction is dominant and strong enough to generate a loosely bound state, such a loosely bound state can be regarded as a relatively reliable candidate for a hadronic molecular state. Therefore, we study the $\mathcal{B}_{c}^{(*)}\bar T$ molecules within both OPE and OBE models. In our numerical analysis, we first give the numerical results with the OPE potential model. After that, we further take into account other pseudoscalar meson exchange interactions and the scalar and vector meson exchange interactions, and repeat the numerical analysis to check the effects. By comparing the numerical results, we can discuss which bound states are more reliable and less dependent on models.

As is known well, the deuteron is a loosely bound state composed of a proton and a neutron, and may be regarded as an ideal molecular state. Here, we briefly reproduce the binding energy $E$ and root-mean-square radius $r_{\rm RMS}$ for the deuteron via the OPE and OBE models; detailed information on the effective potentials is provided in Ref. \cite{Chen:2017jjn}. We summarize the numerical results in Table \ref{reproduced}. A reasonable cutoff in the study of the deuteron is estimated to be around 1.0 GeV, but varies among different scenarios \cite{Tornqvist:1993ng,Tornqvist:1993vu}. The tensor forces from the $S$-$D$ wave mixing effects are important for the study of the deuteron. Nevertheless, the $S$-wave contribution plays the leading role and the $D$-wave correction is subleading. There are about $95\%$ $S$-wave and $5\%$ $D$-wave $NN$ in the deuteron, which is consistent with the results from chiral effective field theory \cite{Epelbaum:1999dj,Bsaisou:2014zwa}. In addition, the deuteron is a loosely bound state, and its typical size should be much larger than the size of all the component hadrons. Knowledge of the deuteron helps us to study other molecular states.
\renewcommand\tabcolsep{0.40cm}
\renewcommand{\arraystretch}{1.50}
\begin{table}[!htbp]
\caption{Reproducing the binding energy $E$ and root-mean-square radius $r_{\rm RMS}$ for the deuteron with four different scenarios: (A) OPE without $S$-$D$ mixing effects, (B) OPE with $S$-$D$ mixing effects, (C) OBE without $S$-$D$ mixing effects, and (D) OBE with $S$-$D$ mixing effects.}\label{reproduced}
\begin{tabular}{c|ccc}\toprule[1.5pt]
Models&$\Lambda{\rm {(GeV)}}$&$E{\rm {(MeV)}} $&$r_{\rm RMS}{\rm {(fm)}}$\\\midrule[1.0pt]
OPE($S$-$D$\,$\times$)&/&/ &/ \\
OPE($S$-$D$\,\checkmark) &1.07&$-2.42$ &3.63\\
OBE($S$-$D$\,$\times$)&1.17&$-2.06$ &3.79\\
OBE($S$-$D$\,\checkmark)&0.87&$-2.56$ &3.58\\
\bottomrule[1.5pt]
\end{tabular}
\end{table}

\section{New types of $P_c$ states:$\Sigma_{c}^{(*)}\bar D_{1}(\bar D_{2}^{*})$ systems}\label{sec3}

In this section, we focus on the $\Sigma_{c}^{(*)}\bar D_{1}(\bar D_{2}^{*})$ systems in detail to illustrate how we search for bound states. First, we need to construct the flavor and spin-orbital wave functions for these discussed $\Sigma_{c}^{(*)}\bar{T}$ pentaquark systems. In Table \ref{flavor}, we summarize the flavor wave functions $|I,I_{3}\rangle$ for the discussed $\Sigma_c^{(*)}\bar{T}$ systems.
\renewcommand\tabcolsep{0.43cm}
\renewcommand{\arraystretch}{1.60}
\begin{table}[!htpb]
\centering
\caption{Flavor wave functions for the discussed $\Sigma_{c}^{(*)}\bar T$ systems. Here, $I$ and $I_3$ are their isospin and the third components, respectively. }\label{flavor}
{\begin{tabular}{c|lc}\toprule[1.0pt]
Systems  &$|I,I_3\rangle$    & Configurations \\\midrule[1.0pt]
\multirow{6}{*}{$\Sigma_{c}^{(*)}\bar T$}&$|\frac{3}{2},\frac{3}{2}\rangle$ & $\left|\Sigma_c^{(*)++}{T}^{-}\right\rangle$\\
                        &$|\frac{3}{2},\frac{1}{2}\rangle$ & $\sqrt{\frac{1}{3}}\left|\Sigma_c^{(*)++}{T}^{-}\right\rangle
                        +\sqrt{\frac{2}{3}}\left|\Sigma_c^{(*)+}\bar{T}^{0}\right\rangle$\\
                        &$|\frac{3}{2},-\frac{1}{2}\rangle$ & $\sqrt{\frac{2}{3}}\left|\Sigma_c^{(*)+}{T}^{-}\right\rangle +\sqrt{\frac{1}{3}}\left|\Sigma_c^{(*)0}\bar{T}^{0}\right\rangle$\\
                        &$|\frac{3}{2},-\frac{3}{2}\rangle$ & $\left|\Sigma_c^{(*)0}{T}^{-}\right\rangle$\\
                        &$|\frac{1}{2},\frac{1}{2}\rangle$ & $\sqrt{\frac{2}{3}}\left|\Sigma_c^{(*)++}{T}^{-}\right\rangle
                        -\sqrt{\frac{1}{3}}\left|\Sigma_c^{(*)+}\bar{T}^{0}\right\rangle$\\
                        &$|\frac{1}{2},-\frac{1}{2}\rangle$ & $\sqrt{\frac{1}{3}}\left|\Sigma_c^{(*)+}{T}^{-}\right\rangle
                        -\sqrt{\frac{2}{3}}\left|\Sigma_c^{(*)0}\bar{T}^{0}\right\rangle$\\
\bottomrule[1.0pt]
\end{tabular}}
\end{table}

For these discussed $\Sigma_c^{(*)}\bar{D}_1(\bar{D}_2^*)$ systems, their spin-orbital wave functions read as
\begin{eqnarray}
|\Sigma_{c}\bar{D}_{1}({}^{2S+1}L_{J})\rangle&=&\sum_{m,m',m_Sm_L}C^{S,m_S}_{\frac{1}{2}m,1m'}
C^{J,M}_{Sm_S,Lm_L}\chi_{\frac{1}{2}m}\epsilon^{m'}|Y_{L,m_L}\rangle,\nonumber\\
|\Sigma_{c}\bar{D}_{2}^{*}({}^{2S+1}L_{J})\rangle&=&\sum_{m,m'',m_Sm_L}
C^{S,m_S}_{\frac{1}{2}m,2m''}C^{J,M}_{Sm_S,Lm_L}\chi_{\frac{1}{2}m}\zeta^{m''}
|Y_{L,m_L}\rangle,\nonumber\\
|\Sigma_{c}^{*}\bar{D}_{1}({}^{2S+1}L_{J})\rangle&=&\sum_{m,m',m_Sm_L}
C^{S,m_S}_{\frac{3}{2}m,1m'}C^{J,M}_{Sm_S,Lm_L}\Phi_{\frac{3}{2}m}\epsilon^{m'}
|Y_{L,m_L}\rangle,\nonumber\\
|\Sigma_{c}^{*}\bar{D}_{2}^{*}({}^{2S+1}L_{J})\rangle&=&\sum_{m,m'',m_Sm_L}
C^{S,m_S}_{\frac{3}{2}m,2m''}C^{J,M}_{Sm_S,Lm_L}\Phi_{\frac{3}{2}m}\zeta^{m''}
|Y_{L,m_L}\rangle,\nonumber\\
\end{eqnarray}
where $C^{J,M}_{Sm_S,Lm_L}$, $C^{S,m_S}_{\frac{1}{2}m,1m'}$, $C^{S,m_S}_{\frac{1}{2}m,2m''}$,  $C^{S,m_S}_{\frac{3}{2}m,1m'}$, and $C^{S,m_S}_{\frac{3}{2}m,2m''}$ are the Clebsch-Gordan coefficients. $|Y_{L,m_L}\rangle$ is the spherical harmonics function. $\chi_{\frac{1}{2}m}$ and $\Phi_{\frac{3}{2}m}$ are defined as  the spin wave functions for fermions with $S=1/2$ and $3/2$, respectively. $\epsilon^{m'} (m'=0\,,\pm1)$ and $\zeta^{m''}(m''=0\,,\pm1\,,\pm2)$ \cite{Cheng:2010yd} are the polarization vector and tensor, respectively. The explicit expressions for these wave functions are
\begin{eqnarray}
\left.\begin{array}{ll}
\epsilon^{0}= \left(0,0,0,-1\right),\quad
&\zeta^{m''}=\sum_{m,n}C^{2,m''}_{1m;1n}\epsilon^{m}\epsilon^{n},\\
\epsilon^{\pm1}= \frac{1}{\sqrt{2}}\left(0,\pm1,i,0\right),\quad
&\Phi_{\frac{3}{2}m}=\sum_{m_1,m_2}C^{\frac{3}{2},m}_{\frac{1}{2}m_1;1m_2}
\chi_{\frac{1}{2},m_1}\epsilon^{m_2}.
\end{array}\right.
\end{eqnarray}

With the above preparation, we further write out the spin-orbit wave functions when the $S$-$D$ wave mixing effects are considered in our calculation, which becomes more complicated. The general expressions corresponding to several typical $J^P$ quantum numbers are
\begin{eqnarray}
|J^P={1}/{2}^+\rangle:&&\Sigma_{c}\bar D_{1}|{}^2\mathbb{S}_{\frac{1}{2}}/{}^4\mathbb{D}_{\frac{1}{2}}\rangle,\quad\Sigma_{c}^*\bar D_{1}|{}^2\mathbb{S}_{\frac{1}{2}}/{}^4\mathbb{D}_{\frac{1}{2}}/{}^6
\mathbb{D}_{\frac{1}{2}}\rangle,\nonumber\\
&&\Sigma_{c}^{*} \bar D_{2}^{*}|{}^2\mathbb{S}_{\frac{1}{2}}/{}^4\mathbb{D}_{\frac{1}{2}}/{}^6\mathbb{D}_{\frac{1}{2}}
\rangle,\nonumber\\
|J^P={3}/{2}^+\rangle:&&\Sigma_{c}\bar D_{1}|{}^4\mathbb{S}_{\frac{3}{2}}/{}^2\mathbb{D}_{\frac{3}{2}}/{}^4\mathbb{D}_{\frac{3}{2}}
\rangle,\nonumber\\
&&\Sigma_{c}\bar D_{2}^{*}|{}^4\mathbb{S}_{\frac{3}{2}}/{}^4\mathbb{D}_{\frac{3}{2}}/
{}^6\mathbb{D}_{\frac{3}{2}}\rangle,\nonumber\\
&&\Sigma_{c}^*\bar D_{1}|{}^4\mathbb{S}_{\frac{3}{2}}/{}^2\mathbb{D}_{\frac{3}{2}}/{}^4
\mathbb{D}_{\frac{3}{2}}/{}^6\mathbb{D}_{\frac{3}{2}}\rangle,\nonumber\\
&&\Sigma_{c}^{*} \bar D_{2}^{*}|{}^4\mathbb{S}_{\frac{3}{2}}/{}^2\mathbb{D}_{\frac{3}{2}}/{}^4
\mathbb{D}_{\frac{3}{2}}/{}^6\mathbb{D}_{\frac{3}{2}}/
{}^8\mathbb{D}_{\frac{3}{2}}\rangle,\nonumber\\
|J^P={5}/{2}^+\rangle:&&\Sigma_{c}\bar D_{2}^{*}|{}^6\mathbb{S}_{\frac{5}{2}}/{}^4\mathbb{D}_{\frac{5}{2}}/
{}^6\mathbb{D}_{\frac{5}{2}}\rangle,\nonumber\\
&&\Sigma_{c}^*\bar D_{1}|{}^6\mathbb{S}_{\frac{5}{2}}/{}^2\mathbb{D}_{\frac{5}{2}}/
{}^4\mathbb{D}_{\frac{5}{2}}/{}^6\mathbb{D}_{\frac{5}{2}}\rangle,\nonumber\\
&&\Sigma_{c}^{*} \bar D_{2}^{*}|{}^6\mathbb{S}_{\frac{5}{2}}/{}^2\mathbb{D}_{\frac{5}{2}}/
{}^4\mathbb{D}_{\frac{5}{2}}/{}^6\mathbb{D}_{\frac{5}{2}}/
{}^8\mathbb{D}_{\frac{5}{2}}\rangle,\nonumber\\
|J^P={7}/{2}^+\rangle:&&\Sigma_{c}^{*} \bar D_{2}^{*}|{}^8\mathbb{S}_{\frac{7}{2}}/{}^4\mathbb{D}_{\frac{7}{2}}/
{}^6\mathbb{D}_{\frac{7}{2}}/{}^8\mathbb{D}_{\frac{7}{2}}\rangle,
\end{eqnarray}
where the notation $|^{2S+1}L_J\rangle$ is used, in which $S$, $L$, and $J$ denote the spin, relative angular momentum, and total angular momentum of the corresponding systems, respectively.

Take the $\Sigma_{c}\bar D_{1}$ state with $I(J^{P})=1/2(1/2^{+})$ as an example to illustrate how we consider the $S$-$D$ wave mixing effects. The corresponding spatial wave function $|\psi\rangle$, the kinetic term $\mathcal{K}$, and the effective potential $\mathcal{V}$ can be expressed as
\begin{eqnarray}
|\psi\rangle&=&\left(\Sigma_{c}\bar D_{1}|{}^2\mathbb{S}_{\frac{1}{2}}\rangle,\,\Sigma_{c}\bar D_{1}|{}^4\mathbb{D}_{\frac{1}{2}}\rangle\right)^T,\\
\mathcal{K}&=&\mathrm{diag}\left(-\frac{\triangledown^2}{2\mu},\, -\frac{\triangledown^2}{2\mu}+\frac{3}{\mu r^2}\right),\\
\mathcal{V}&=&{\left(\begin{array}{cc}
\mathcal{V}^{\Sigma_{c}\bar D_{1}|{}^2\mathbb{S}_{\frac{1}{2}}\rangle\to\Sigma_{c}\bar D_{1}|{}^2\mathbb{S}_{\frac{1}{2}}\rangle}&\mathcal{V}^{\Sigma_{c}\bar D_{1}|{}^2\mathbb{S}_{\frac{1}{2}}\rangle\to\Sigma_{c}\bar D_{1}|{}^4\mathbb{D}_{\frac{1}{2}}\rangle}\\
\mathcal{V}^{\Sigma_{c}\bar D_{1}|{}^4\mathbb{D}_{\frac{1}{2}}\rangle\to\Sigma_{c}\bar D_{1}|{}^2\mathbb{S}_{\frac{1}{2}}\rangle}&\mathcal{V}^{\Sigma_{c}\bar D_{1}|{}^4\mathbb{D}_{\frac{1}{2}}\rangle\to\Sigma_{c}\bar D_{1}|{}^4\mathbb{D}_{\frac{1}{2}}\rangle}\\\end{array}\right)},\label{Vs}\nonumber\\
\end{eqnarray}
respectively. $\mu=m_{\Sigma_{c}}m_{D_1}/(m_{\Sigma_{c}}+m_{D_1})$ is the reduced mass of the $\Sigma_{c}\bar D_1$ system. The superscripts in Eq. (\ref{Vs}) represent the corresponding scattering processes, which are presented in Appendix \ref{app0201}.  By solving the coupled channel Schr\"odinger equation, i.e., $(\mathcal{K}+\mathcal{V})\,|\psi\rangle=E\,|\psi\rangle$ with $E$ the binding energy, we can obtain the properties of the bound state $\Sigma_{c}\bar D_{1}$ with $I(J^{P})=1/2(1/2^{+})$ as new types of $P_c$ state. Now we can present the numerical results for the $\Sigma_{c}^{(*)}\bar D_{1}(\bar D_{2}^{*})$ systems with the OPE and OBE models, respectively.

\subsection{$\Sigma_{c}^{(*)}\bar D_{1}(\bar D_{2}^{*})$ systems with the OPE model}

Generally speaking, the $\pi$ exchange interaction plays an irreplaceable role in forming loosely bound molecular states. In Table \ref{jg1p}, we present the binding energy $E$,  root-mean-square (RMS) radius $r_{RMS}$, and probabilities for different components for the bound states composed of $\Sigma_{c}\bar D_1$ and $\Sigma_{c} \bar  D_2^*$ with the OPE model. In order to check the specific roles of the $S$-$D$ wave mixing effects, we show the numerical results without and with the $S$-$D$ wave mixing effects.
\renewcommand\tabcolsep{0.08cm}
\renewcommand{\arraystretch}{1.50}
\begin{table}[!htbp]
\caption{Bound state properties (binding energy $E$,  root-mean-square radius $r_{RMS}$, and probabilities for different components) for the $\Sigma_{c}\bar D_{1}(\bar D_{2}^{*})$ systems with the OPE model. Cutoff $\Lambda$, binding energy $E$, and root-mean-square radius $r_{RMS}$ are in units of $ \rm{GeV}$, $\rm {MeV}$, and $\rm {fm}$, respectively. The largest probability of the quantum number configuration for a bound state is indicated by bold typeface. In Tables~\ref{jg1p}-\ref{jg2b} and Tables~\ref{jg1pu}-\ref{jg2bu}, we use same convention when numerical results are presented, the second column group shows results without considering the $S$-$D$ wave mixing effects while the last column group shows the relevant results with the $S$-$D$ wave mixing effects.}\label{jg1p}
\begin{tabular}{c|ccc|cccl}\toprule[1.5pt]
\multicolumn{8}{c}{$\Sigma_{c}\bar D_{1}$}\\\midrule[1.0pt]
$I(J^P)$&$\Lambda$ &$E$ &$r_{\rm RMS}$    &$\Lambda$ &$E$ &$r_{\rm RMS}$ &P(${}^2\mathbb{S}_{\frac{1}{2}}/{}^4\mathbb{D}_{\frac{1}{2}}$)\\
$\frac{1}{2}(\frac{1}{2}^+)$&1.04&$-0.23$ &4.79                &0.99&$-0.22$ &4.92 &\textbf{99.52}/0.48\\
                            &1.15&$-3.96$ &1.55                &1.10&$-3.65$&1.64&\textbf{99.05}/0.95\\
                            &1.26&$-12.69$ &0.93                &1.21&$-11.82$&1.00&\textbf{98.90}/1.10\\\hline
$I(J^P)$&$\Lambda$ &$E$&$r_{\rm RMS}$&$\Lambda$ &$E$&$r_{\rm RMS}$  &P(${}^4\mathbb{S}_{\frac{3}{2}}/{}^2\mathbb{D}_{\frac{3}{2}}/{}^4\mathbb{D}_{\frac{3}{2}}$)\\
$\frac{1}{2}(\frac{3}{2}^+)$&$\times$&$\times$ &$\times$                 &2.22&$-0.27$ &4.98&\textbf{97.06}/0.38/2.56 \\
                            &$\times$&$\times$ &$\times$                        &2.48&$-3.50$&1.87&\textbf{92.10}/0.95/6.95\\
                           &$\times$&$\times$ &$\times$                       &2.74&$-12.05$ &1.13&\textbf{88.28}/1.35/10.37\\
$\frac{3}{2}(\frac{3}{2}^+)$&3.51&$-0.19$ &4.96                &2.22&$-0.28$ &4.68&\textbf{98.33}/0.39/1.28\\
                            &3.67&$-3.88$ &1.46                &2.40&$-3.91$&1.59&\textbf{95.69}/1.00/3.31\\
                            &3.83&$-12.49$ &0.84                &2.57&$-11.97$ &0.97&\textbf{94.01}/1.39/4.60\\\midrule[1.0pt]
\multicolumn{8}{c}{$\Sigma_{c}\bar D_{2}^{*}$}\\\midrule[1.0pt]
$I(J^P)$&$\Lambda$ &$E$&$r_{\rm RMS}$ &$\Lambda$ &$E$&$r_{\rm RMS}$ &P(${}^4\mathbb{S}_{\frac{3}{2}}/{}^4\mathbb{D}_{\frac{3}{2}}/{}^6\mathbb{D}_{\frac{3}{2}}$)\\
$\frac{1}{2}(\frac{3}{2}^+)$&1.12&$-0.19$ &4.96                &1.02&$-0.23$ &4.91&\textbf{99.11}/0.12/0.77\\
                            &1.24&$-4.10$ &1.51                &1.14&$-3.80$&1.64 &\textbf{98.16}/0.25/1.59\\
                            &1.35&$-12.52$ &0.93                &1.26&$-12.44$ &0.99&\textbf{97.84}/0.31/1.86\\\hline
$I(J^P)$&$\Lambda$ &$E$&$r_{\rm RMS}$ &$\Lambda$ &$E$&$r_{\rm RMS}$ &P(${}^6\mathbb{S}_{\frac{5}{2}}/{}^4\mathbb{D}_{\frac{5}{2}}/{}^6\mathbb{D}_{\frac{5}{2}}$) \\
$\frac{1}{2}(\frac{5}{2}^+)$&$\times$&$\times$ &$\times$                &2.13&$-0.30$ &4.88&\textbf{96.59}/0.89/2.51\\
                            &$\times$&$\times$ &$\times$                &2.38&$-3.72$ &1.83&\textbf{90.75}/2.33/6.92\\
                           &$\times$&$\times$ &$\times$                &2.63&$-12.86$ &1.11&\textbf{86.13}/3.41/10.46\\
$\frac{3}{2}(\frac{5}{2}^+)$&2.94&$-0.27$ &4.60                &2.08&$-0.26$ &4.74&\textbf{98.59}/0.57/0.84\\
                            &3.09&$-3.95$ &1.45                &2.26&$-4.05$&1.55&\textbf{96.36}/1.47/2.16\\
                            &3.24&$-12.27$ &0.85                &2.43&$-12.57$ &0.94&\textbf{95.07}/1.99/2.95\\\hline
\bottomrule[1.5pt]
\end{tabular}
\end{table}

%The $\pi$ exchange effective potentials provide strongly attractive force in the range $r<1.0~\rm{fm}$ for the $\Sigma_{c}\bar D_1$ state with $I(J^{P})=1/2(1/2^{+})$ and the $\Sigma_{c} \bar  D_2^*$ state with $I(J^{P})=1/2(3/2^{+})$, and the strongly attractive interactions lead to reasonably loose bound states.
As shown in Table \ref{jg1p}, these states can be bound with several MeV binding energy when cutoffs are around 1.1 GeV, which are reasonable cutoff values. Their RMS radii are all around a few fm, which is consistent with the typical size of a hadronic molecular state. Thus, such loosely bound states can be good candidates of $P_c$ states with larger masses.

In contrast, the $\pi$ exchange interactions result in repulsive potentials at long range for the $\Sigma_{c}\bar D_1$ states with $I(J^{P})=3/2(1/2^{+}), 1/2(3/2^{+})$ and the $\Sigma_{c} \bar D_2^*$ states with $I(J^{P})=3/2(3/2^{+}), 1/2(5/2^{+})$, and therefore such states cannot be bound with the OPE model if we do not consider the $S$-$D$ wave mixing effects. For the $\Sigma_{c}\bar D_1$ state with $I(J^{P})=3/2(3/2^{+})$ and the $\Sigma_{c} \bar D_2^*$ state with $I(J^{P})=3/2(5/2^{+})$, the OPE potentials are slightly attractive at short range. Compared to the $\Sigma_{c}\bar D_1$ state with $I(J^{P})=3/2(3/2^{+})$, the $\Sigma_{c} \bar D_2^*$ state with $I(J^{P})=3/2(5/2^{+})$ is more easily bound because of stronger OPE attractions and hadrons with heavier masses, and we can obtain bound solutions for these two states when the cutoffs are taken to be around 3.6 and 3.0 GeV, respectively. In general, a loosely bound state with smaller cutoff corresponds to more attractive force. For the $\Sigma_{c}\bar D_1$ state with $I(J^{P})=1/2(3/2^{+})$ and the $\Sigma_{c} \bar D_2^*$ state with $I(J^{P})=1/2(5/2^{+})$, these weakly bound state solutions with the cutoff around 2.2 GeV disappear after removing the contributions of the $D$-wave channels, which implies that the $S$-$D$ wave mixing effects play an important role in generating these loosely bound states. Here, recall that the tensor force in the effective potentials plays a critical role in the formation of the loosely bound deuteron.

For the $\Sigma_{c}\bar D_1$ state with $I(J^{P})=3/2(1/2^{+})$ and the $\Sigma_{c} \bar D_2^*$ state with $I(J^{P})=3/2(3/2^{+})$, there do not exist bound states with the OPE model, even if the $S$-$D$ wave mixing effects are considered.

Following the procedure discussed above, we present the binding energy, RMS radius, and probabilities for different components for the $\Sigma_{c}^{*}\bar D_{1}(\bar D_{2}^{*})$ systems with the OPE model in Table \ref{jg2p}.

\renewcommand\tabcolsep{0.43cm}
\renewcommand{\arraystretch}{1.15}
\begin{table*}[!htbp]
\caption{Bound state properties for the $\Sigma_{c}^{*}\bar D_{1}(\bar D_{2}^{*})$ systems with the OPE model. Conventions are the same as Table~\ref{jg1p}.}\label{jg2p}
\begin{tabular}{c|c|ccc|cccl}\toprule[1.5pt]
System&$I(J^P)$&$\Lambda$ &$E$  &$r_{\rm RMS}$ &$\Lambda$ &$E$  &$r_{\rm RMS}$ &P(${}^2\mathbb{S}_{\frac{1}{2}}/{}^4\mathbb{D}_{\frac{1}{2}}/{}^6\mathbb{D}_{\frac{1}{2}}$)\\
\multirow{17}{*}{$\Sigma_{c}^{*}\bar D_{1}$}&$\frac{1}{2}(\frac{1}{2}^+)$&0.87&$-0.24$ &4.74                &0.82&$-0.27$ &4.08&\textbf{99.23}/0.47/0.30\\
                                                                         &&0.97&$-3.92$ &1.57                &0.92&$-3.78$ &1.66&\textbf{98.57}/0.89/0.54\\
                                                                        &&1.07&$-12.48$ &0.96                &1.02&$-11.93$ &1.03&\textbf{98.38}/~1.02/0.60\\
&$\frac{3}{2}(\frac{1}{2}^+)$&$\times$&$\times$ &$\times$                 &3.80&$-0.34$ &4.57&\textbf{96.54}/1.62/1.83\\
                            &&$\times$&$\times$ &$\times$                 &3.90&$-1.50$ &2.55&\textbf{93.12}/3.15/3.73\\
                            &&$\times$&$\times$ &$\times$                 &4.00&$-3.75$ &1.68&\textbf{89.55}/4.67/5.78\\
                             \cline{2-9}
&$I(J^P)$&$\Lambda$ &$E$ &$r_{\rm RMS}$ &$\Lambda$ &$E$ &$r_{\rm RMS}$ &P(${}^4\mathbb{S}_{\frac{3}{2}}/{}^2\mathbb{D}_{\frac{3}{2}}/{}^4\mathbb{D}_{\frac{3}{2}}/{}^6\mathbb{D}_{\frac{3}{2}}$) \\
&$\frac{1}{2}(\frac{3}{2}^+)$&1.84&$-0.27$ &4.59                &1.31&$-0.24$ &4.89&\textbf{98.24}/0.44/1.19/0.14 \\
                            &&1.98&$-4.23$ &1.43                &1.46&$-3.78$ &1.68&\textbf{95.84}/1.03/2.82/0.31\\
                            &&2.11&$-12.56$ &0.87                &1.61&$-12.53$ &1.01&\textbf{94.53}/1.35/3.72/0.40\\
                             \cline{2-9}
&$I(J^P)$&$\Lambda$ &$E$ &$r_{\rm RMS}$ &$\Lambda$ &$E$ &$r_{\rm RMS}$ &P(${}^6\mathbb{S}_{\frac{5}{2}}/{}^2\mathbb{D}_{\frac{5}{2}}/{}^4\mathbb{D}_{\frac{5}{2}}/{}^6\mathbb{D}_{\frac{5}{2}}$) \\
&$\frac{1}{2}(\frac{5}{2}^+)$&$\times$&$\times$ &$\times$                 &1.98&$-0.37$ &4.65&\textbf{95.76}/0.14/0.12/3.97 \\
                            &&$\times$&$\times$ &$\times$                 &2.22&$-3.88$ &1.83&\textbf{89.04}/0.26/0.32/10.37\\
                            &&$\times$&$\times$ &$\times$                 &2.45&$-12.50$ &1.15&\textbf{83.55}/0.30/0.48/15.34\\
&$\frac{3}{2}(\frac{5}{2}^+)$&2.38&$-0.24$ &4.73                &1.81&$-0.26$ &4.95&\textbf{98.75}/0.18/0.05/1.02\\
                            &&2.53&$-4.11$ &1.43                &1.98&$-3.96$ &1.57&\textbf{96.72}/0.47/0.13/2.68\\
                            &&2.67&$-12.33$ &0.86                &2.14&$-12.55$ &0.95&\textbf{95.61}/0.63/0.17/3.59\\\midrule[1.0pt]
System&$I(J^P)$&$\Lambda$ &$E$ &$r_{\rm RMS}$&$\Lambda$ &$E$ &$r_{\rm RMS}$&P(${}^2\mathbb{S}_{\frac{1}{2}}/{}^4\mathbb{D}_{\frac{1}{2}}/{}^6\mathbb{D}_{\frac{1}{2}}$)\\
\multirow{27}{*}{$\Sigma_{c}^{*}\bar D_{2}^{*}$}&$\frac{1}{2}(\frac{1}{2}^+)$&0.82&$-0.29$ &4.55                &0.79&$-0.75$ &3.35&\textbf{98.82}/0.60/0.58 \\
                            &&0.91&$-3.64$ &1.63                &0.88&$-4.80$ &1.52&\textbf{98.27}/0.90/0.83 \\
                            &&1.01&$-12.29$ &0.97                &0.96&$-11.86$ &1.05&\textbf{98.11}/0.99/0.80\\
&$\frac{3}{2}(\frac{1}{2}^+)$&$\times$&$\times$ &$\times$                &3.70&$-0.56$ &3.89&\textbf{95.89}/1.81/2.31\\
                            &&$\times$&$\times$ &$\times$                &3.85&$-2.49$ &2.05&\textbf{91.84}/3.52/4.64\\
                            &&$\times$&$\times$ &$\times$                &4.00&$-6.33$ &1.36&\textbf{87.92}/5.13/6.95\\
                             \cline{2-9}
&$I(J^P)$&$\Lambda$ &$E$  &$r_{\rm RMS}$ &$\Lambda$ &$E$  &$r_{\rm RMS}$ &P(${}^4\mathbb{S}_{\frac{3}{2}}/{}^2\mathbb{D}_{\frac{3}{2}}/{}^4\mathbb{D}_{\frac{3}{2}}/{}^6\mathbb{D}_{\frac{3}{2}}/{}^8\mathbb{D}_{\frac{3}{2}}$) \\
&$\frac{1}{2}(\frac{3}{2}^+)$&1.11&$-0.21$ &4.86                &0.97&$-0.28$ &4.68&\textbf{98.68}/0.34/0.62/0.12/0.23\\
                            &&1.23&$-4.16$ &1.48                &1.09&$-3.97$ &1.62&\textbf{97.30}/0.71/1.31/0.25/0.43\\
                            &&1.34&$-12.62$ &0.91                &1.21&$-12.80$ &0.99&\textbf{96.71}/0.88/1.61/0.30/0.48\\
&$\frac{3}{2}(\frac{3}{2}^+)$&$\times$&$\times$ &$\times$                &3.68&$-0.26$ &4.88&\textbf{97.15}/0.72/0.96/0.23/0.95\\
                            &&$\times$&$\times$ &$\times$                &3.89&$-3.11$ &1.81&\textbf{91.51}/2.05/2.69/0.66/3.08\\
                            &&$\times$&$\times$ &$\times$                &4.00&$-6.29$ &1.32&\textbf{88.66}/2.68/3.50/0.87/4.29\\
                             \cline{2-9}
&$I(J^P)$&$\Lambda$ &$E$ &$r_{\rm RMS}$&$\Lambda$ &$E$ &$r_{\rm RMS}$&P(${}^6\mathbb{S}_{\frac{5}{2}}/{}^2\mathbb{D}_{\frac{5}{2}}/{}^4\mathbb{D}_{\frac{5}{2}}/{}^6\mathbb{D}_{\frac{5}{2}}/{}^8\mathbb{D}_{\frac{5}{2}}$) \\
&$\frac{1}{2}(\frac{5}{2}^+)$&$\times$&$\times$ &$\times$                &1.57&$-0.25$ &4.93&\textbf{97.21}/0.29/0.14/2.29/0.07\\
                            &&$\times$&$\times$ &$\times$                &1.75&$-3.61$ &1.79&\textbf{92.85}/0.67/0.37/5.96/0.16\\
                            &&$\times$&$\times$ &$\times$                &1.93&$-12.09$ &1.07&\textbf{89.90}/0.87/0.52/8.49/0.22\\
&$\frac{3}{2}(\frac{5}{2}^+)$&$\times$&$\times$ &$\times$                &3.40&$-0.27$ &4.76&\textbf{97.81}/0.59/0.15/1.39/0.06\\
                            &&$\times$&$\times$ &$\times$                &3.65&$-3.77$ &1.62&\textbf{93.73}/1.75/0.45/3.93/0.15\\
                            &&$\times$&$\times$ &$\times$                &3.90&$-12.49$ &0.96&\textbf{90.50}/2.71/0.69/5.88/0.23\\
                             \cline{2-9}
&$I(J^P)$&$\Lambda$ &$E$  &$r_{\rm RMS}$&$\Lambda$ &$E$  &$r_{\rm RMS}$ &P(${}^8\mathbb{S}_{\frac{7}{2}}/{}^4\mathbb{D}_{\frac{7}{2}}/{}^6\mathbb{D}_{\frac{7}{2}}/{}^8\mathbb{D}_{\frac{7}{2}}$)\\
&$\frac{1}{2}(\frac{7}{2}^+)$&$\times$&$\times$ &$\times$                &1.88&$-0.29$ &4.97&\textbf{95.75}/~0.11/~0.01/~4.09 \\
                            &&$\times$&$\times$ &$\times$                &2.12&$-3.67$ &1.89&\textbf{88.01}/~0.22/~0.14/~11.62\\
                            &&$\times$&$\times$ &$\times$                &2.35&$-12.38$ &1.17&\textbf{82.00}/~0.25/~0.22/~17.53\\
&$\frac{3}{2}(\frac{7}{2}^+)$&2.00&$-0.21$ &4.85                &1.63&$-0.24$ &4.81&\textbf{98.86}/0.14/0.02/0.98\\
                            &&2.14&$-3.82$ &1.48                &1.79&$-4.06$ &1.54&\textbf{97.20}/0.34/0.04/2.42\\
                            &&2.28&$-12.27$ &0.86                &1.94&$-12.70$ &0.93&\textbf{96.35}/0.45/0.05/3.15\\
\bottomrule[1.5pt]
\end{tabular}
\end{table*}

From the numerical results in Table \ref{jg2p}, the investigated $\Sigma_{c}^* \bar D_1$ state with $I(J^{P})=1/2(1/2^{+})$ and the $\Sigma_{c}^* \bar  D_2^*$ states with $I(J^{P})=1/2(1/2^{+}, 3/2^{+})$ can be good candidates of the new types of $P_c$ states since they have small binding energies and suitable RMS radii under reasonable cutoffs, which is because the $\pi$ exchange interactions give strongly attractive forces around 1.0 fm and strongly attractive interactions at long range can be used to generate these reasonable loosely bound states. The attractive interactions from the $\pi$ exchanges contribute at short range for the $\Sigma_{c}^{*}\bar D_1$ state with $I(J^{P})=3/2(5/2^{+})$ and the $\Sigma_{c}^{*} \bar D_2^*$ state with $I(J^{P})=3/2(7/2^{+})$, which makes loosely bound state solutions possible, we find that there exist weakly bound state solutions when the values of the cutoff parameters are taken around 2.4 and 2.0 GeV, respectively. Additionally, the effective potential from the $\pi$ exchange provides a very weak attractive force for the $\Sigma_{c}^{*} \bar D_2^*$ state with $I(J^{P})=1/2(5/2^{+})$ if we only consider the $S$-wave component, and we cannot obtain bound state solutions until we increase the cutoff parameter to be around 4.0 GeV.

In contrast to the above situation, the $\pi$ exchange effective potentials are repulsive for the $\Sigma_{c}^{*}\bar D_1$ states with $I(J^{P})=3/2(1/2^{+},3/2^{+}), 1/2(5/2^{+})$ and the $\Sigma_{c}^{*} \bar D_2^*$ states with $I(J^{P})=3/2(1/2^{+},3/2^{+},5/2^{+}), 1/2(7/2^{+})$, and there do not exist bound state solutions with the OPE model if the contributions of the $D$-wave channels are not considered.  The tensor forces from the $S$-$D$ wave mixing effects are essential in the formation of these loosely bound states. There are no bound solutions for the $\Sigma_{c}^* \bar D_1$ state with $I(J^{P})=3/2(3/2^{+})$ with the OPE model, even if we consider the $S$-$D$ wave mixing effects.

\subsection{$\Sigma_{c}^{(*)}\bar D_{1}(\bar D_{2}^{*})$ systems with the OBE model}

%In 2019, the LHCb Collaboration updated the observations of $P_c(4312)$, $P_c(4440)$, and $P_c(4457)$ \cite{Aaij:2019vzc}. Among them, many investigations indicate that the $P_c(4312)$ and $P_c(4457)$ can be identified as the $\Sigma_{c} \bar  D$ state with $I(J^{P})=1/2(1/2^{+})$ and $\Sigma_{c} \bar D^*$ state with $I(J^{P})=1/2(3/2^{+})$, respectively \cite{Chen:2019asm}, and the vector exchange interactions play a crucial role in the formation of these two $P_c$ states in these theoretical explanations.
Usually, the pion exchange interaction plays a dominant role in forming the loosely bound states at long range due to the molecular state being a loosely bound state; the interactions from other light meson exchanges, especially the vector exchange interaction, may play an important role to modify the bound state properties to a large extent at intermediate and short ranges. In fact, the $\eta$ exchange interaction usually plays a minor role in generating the loosely bound states \cite{Chen:2016ryt}, and the contribution from the $\sigma$ exchange always gives an attractive force \cite{Chen:2017vai}.

The OBE model is a little more complex, and we carry out an analysis similar to that for OPE. For the $\Sigma_{c}\bar D_1$ and $\Sigma_{c} \bar  D_2^*$ systems, the $\pi$, $\sigma$, $\eta$, $\rho,$ and $\omega$ exchanges contribute to the total effective potentials. In Table \ref{jg1b}, we present the corresponding bound properties for the $\Sigma_{c}\bar D_1$ and $\Sigma_{c} \bar  D_2^*$ systems within the OBE model.

\renewcommand\tabcolsep{0.08cm}
\renewcommand{\arraystretch}{1.50}
\begin{table}[!htbp]
\caption{Bound state properties for the $\Sigma_{c}\bar D_{1}(\bar D_{2}^{*})$ systems with the OBE model. Conventions are the same as Table~\ref{jg1p}.}\label{jg1b}
\begin{tabular}{c|ccc|cccl}\toprule[1.5pt]
\multicolumn{8}{c}{$\Sigma_{c}\bar D_{1}$}\\\midrule[1.0pt]
$I(J^P)$&$\Lambda$ &$E$ &$r_{\rm RMS}$  &$\Lambda$ &$E$ &$r_{\rm RMS}$ &P(${}^2\mathbb{S}_{\frac{1}{2}}/{}^4\mathbb{D}_{\frac{1}{2}}$)\\
$\frac{1}{2}(\frac{1}{2}^+)$&0.91&$-0.40$ &4.24        &0.88&$-0.22$ &4.98 &\textbf{99.61}/0.39\\
                            &1.01&$-5.07$&1.49         &0.98&$-4.65$ &1.57 &\textbf{99.20}/0.80\\
                            &1.10&$-12.30$&1.05        &1.07&$-12.28$ &1.07 &\textbf{99.04}/0.96\\
$\frac{3}{2}(\frac{1}{2}^+)$&2.106&$-0.48$&$3.27$       &1.96&$-1.64$ &2.06 &\textbf{89.81}/10.19\\
                            &2.112&$-5.63$&0.79       &1.97&$-5.30$ &1.14 &\textbf{86.95}/13.05\\
                            &2.118&$-12.02$ &0.53             &1.98&$-10.20$ &0.84 &\textbf{85.98}/~14.02\\\hline
$I(J^P)$&$\Lambda$ &$E$&$r_{\rm RMS}$ &$\Lambda$ &$E$&$r_{\rm RMS}$  &P(${}^4\mathbb{S}_{\frac{3}{2}}/{}^2\mathbb{D}_{\frac{3}{2}}/{}^4\mathbb{D}_{\frac{3}{2}}$)\\
$\frac{1}{2}(\frac{3}{2}^+)$&1.59&$-0.26$ &4.99       &1.24&$-0.35$ &4.71&\textbf{97.88}/0.34/1.78 \\
                            &1.78&$-4.11$&1.68        &1.35&$-3.93$ &1.83&\textbf{94.91}/0.81/4.29 \\
                            &1.97&$-12.71$ &1.04      &1.46&$-12.24$ &1.16&\textbf{92.92}/1.11/5.97 \\
$\frac{3}{2}(\frac{3}{2}^+)$&$\times$&$\times$ &$\times$       &2.02&$-0.40$ &4.28&\textbf{94.71}/1.32/3.97\\
                            &$\times$&$\times$ &$\times$       &2.11&$-4.55$ &1.58&\textbf{86.34}/3.50/10.16\\
                            &$\times$&$\times$ &$\times$       &2.19&$-12.28$ &1.07&\textbf{81.06}/4.98/13.96\\\midrule[1.0pt]
\multicolumn{8}{c}{$\Sigma_{c}\bar D_{2}^{*}$}\\\midrule[1.0pt]
$I(J^P)$&$\Lambda$ &$E$&$r_{\rm RMS}$ &$\Lambda$ &$E$&$r_{\rm RMS}$ &P(${}^4\mathbb{S}_{\frac{3}{2}}/{}^4\mathbb{D}_{\frac{3}{2}}/{}^6\mathbb{D}_{\frac{3}{2}}$)\\
$\frac{1}{2}(\frac{3}{2}^+)$&0.94&$-0.38$ &4.32               &0.90&$-0.37$ &4.38&\textbf{99.21}/0.11/0.68\\
                            &1.04&$-5.39$&1.46                &0.99&$-4.39$ &1.62&\textbf{98.53}/0.21/1.26\\
                            &1.14&$-12.36$ &1.05              &1.08&$-11.98$ &1.09&\textbf{98.21}/0.26/1.53\\
$\frac{3}{2}(\frac{3}{2}^+)$&2.202&$-0.66$ &2.65      &1.93&$-1.01$ &2.76&\textbf{90.84}/1.95/7.21\\
                            &2.208&$-5.92$ &0.75            &1.95&$-5.75$ &1.19&\textbf{85.30}/3.19/11.51\\
                            &2.214&$-12.27$ &0.52            &1.96&$-9.21$ &0.96&\textbf{84.13}/3.48/12.40\\\hline
$I(J^P)$&$\Lambda$ &$E$&$r_{\rm RMS}$ &$\Lambda$ &$E$&$r_{\rm RMS}$ &P(${}^6\mathbb{S}_{\frac{5}{2}}/{}^4\mathbb{D}_{\frac{5}{2}}/{}^6\mathbb{D}_{\frac{5}{2}}$) \\
$\frac{1}{2}(\frac{5}{2}^+)$&1.65&$-0.30$ &4.78      &1.24&$-0.29$ &4.93&\textbf{97.72}/0.69/1.58\\
                            &1.84&$-4.21$ &1.66      &1.35&$-3.80$ &1.86&\textbf{94.09}/1.78/4.13\\
                            &2.02&$-12.61$ &1.03     &1.46&$-12.32$ &1.16&\textbf{91.58}/2.52/5.89\\
$\frac{3}{2}(\frac{5}{2}^+)$&$\times$&$\times$&$\times$      &2.08&$-0.22$ &5.02&\textbf{96.24}/1.64/2.12\\
                            &$\times$&$\times$&$\times$      &2.18&$-3.89$ &1.69&\textbf{88.61}/5.11/6.28\\
                            &$\times$&$\times$&$\times$      &2.28&$-12.54$ &1.07&\textbf{83.14}/7.79/9.07\\\hline
\bottomrule[1.5pt]
\end{tabular}
\end{table}

As shown in Table \ref{jg1b}, we can obtain more loosely bound states with reasonable cutoffs after considering the scalar and vector meson exchange effective potentials. For the $\Sigma_{c}\bar D_1$ state with $I(J^{P})=1/2(1/2^{+})$ and the $\Sigma_{c} \bar  D_2^*$ state with $I(J^{P})=1/2(3/2^{+})$, we find loosely bound state solutions within reasonable cutoffs for both the OPE model and the OBE model, and the energies and radii are similar, which means that the total effective potentials mainly come from the $\pi$ exchanges, and the scalar and vector meson exchange effective potentials play a minor role in the formation of these bound states when cutoff parameters $\Lambda$ are around 0.95 GeV. Therefore, we conclude that the $\Sigma_{c}\bar D_1$ state with $I(J^{P})=1/2(1/2^{+})$ and the $\Sigma_{c} \bar  D_2^*$ state with $I(J^{P})=1/2(3/2^{+})$ are favored to be the better candidates of the new types of $P_c$ states. By comparing the numerical results of the OPE model, the cutoffs are relatively more reasonable for the loosely bound states in the $\Sigma_{c}\bar D_1$ system with $I(J^{P})=1/2(3/2^{+})$ and the $\Sigma_{c} \bar  D_2^*$ system with $I(J^{P})=1/2(5/2^{+})$ within OBE model. Here, the $\rho$ and $\omega$ exchange attractive forces play a significant role in the formation of these loosely bound states with cutoffs of about 1.65 GeV.

The binding energies of the $\Sigma_{c}\bar D_1$ state with $I(J^{P})=3/2(1/2^{+})$ and the $\Sigma_{c} \bar  D_2^*$ state with $I(J^{P})=3/2(3/2^{+})$ are very sensitive to the cutoffs, which differs from other states greatly. The main reason is that the $\rho$ and $\omega$ exchange interactions give strongly attractive forces at relatively short range when the cutoff parameter $\Lambda$ is around 2.2 GeV, and there are almost no attractive forces at intermediate and long ranges. Especially, the effective potentials related to the ${\bf{q}}^2/(q^2+m^2)$ term in momentum space have a great influence on these bound states, and the short-range dynamics make the numerical results more sensitive to the values of the cutoffs.

In many  $\Sigma_{c}\bar D_1(\bar  D_2^*)$ systems, the binding energy in the OBE model is larger than that in the OPE model with the same cutoff parameters since the intermediate and short range interactions often play an essential role in the formation of the loosely bound states. Comparing the results with and without the $S$-$D$ wave mixing effects, it is obvious that the $S$-$D$ wave mixing effects are important in the formation of several molecular states, especially for the $\Sigma_{c}\bar D_1$ state with $I(J^{P})=3/2(3/2^{+})$ and the $\Sigma_{c} \bar  D_2^*$ state with $I(J^{P})=3/2(5/2^{+})$.

Similarly to the $\Sigma_{c}\bar D_{1}(\bar D_{2}^{*})$ systems, we present the bound properties for the $\Sigma_{c}^{*}\bar D_{1}(\bar D_{2}^{*})$ systems with the OBE model in Table \ref{jg2b}. From Tables \ref{jg2b} and \ref{jg2p}, it is obvious that the bound state properties will change accordingly after we consider the contributions from the scalar and vector meson exchange effective potentials.
\renewcommand\tabcolsep{0.43cm}
\renewcommand{\arraystretch}{1.15}
\begin{table*}[!htbp]
\caption{Bound state properties for the $\Sigma_{c}^{*}\bar D_{1}(\bar D_{2}^{*})$ systems with the OBE model. Conventions are the same as Table~\ref{jg1p}.}\label{jg2b}
\begin{tabular}{c|c|ccc|cccl}\toprule[1.5pt]
System&$I(J^P)$&$\Lambda$ &$E$  &$r_{\rm RMS}$ &$\Lambda$ &$E$  &$r_{\rm RMS}$ &P(${}^2\mathbb{S}_{\frac{1}{2}}/{}^4\mathbb{D}_{\frac{1}{2}}/{}^6\mathbb{D}_{\frac{1}{2}}$)\\
\multirow{20}{*}{$\Sigma_{c}^{*}\bar D_{1}$}&$\frac{1}{2}(\frac{1}{2}^+)$&0.82&$-0.31$ &4.50       &0.79&$-0.35$ &4.40&\textbf{99.26}/0.45/0.29\\
                            &&0.91&$-4.68$ &1.52       &0.88&$-4.72$&1.56 &\textbf{98.71}/0.80/0.49\\
                            &&1.00&$-12.98$ &1.02      &0.96&$-12.49$ &1.06&\textbf{98.51}/0.93/0.56\\
&$\frac{3}{2}(\frac{1}{2}^+)$&1.917&$-0.29$ &4.42       &1.71&$-0.25$ &4.73&\textbf{93.76}/2.85/3.39\\
                            &&1.923&$-5.51$ &0.82      &1.73&$-4.11$&1.43&\textbf{84.65}/6.96/8.39\\
                            &&1.929&$-12.21$ &0.54          &1.75&$-11.55$ &0.90&\textbf{81.92}/8.25/9.84\\
                            \cline{2-9}
&$I(J^P)$&$\Lambda$ &$E$ &$r_{\rm RMS}$ &$\Lambda$ &$E$ &$r_{\rm RMS}$ &P(${}^4\mathbb{S}_{\frac{3}{2}}/{}^2\mathbb{D}_{\frac{3}{2}}/{}^4\mathbb{D}_{\frac{3}{2}}/{}^6\mathbb{D}_{\frac{3}{2}}$) \\
&$\frac{1}{2}(\frac{3}{2}^+)$&1.09&$-0.36$ &4.44       &1.00&$-0.41$ &4.32&\textbf{98.63}/0.36/0.91/0.11\\
                            &&1.22&$-5.06$ &1.51        &1.10&$-4.69$ &1.62&\textbf{97.15}/0.75/1.89/0.21\\
                            &&1.34&$-12.54$ &1.05       &1.19&$-12.08$ &1.12&\textbf{96.37}/0.95/~2.42/0.26\\
&$\frac{3}{2}(\frac{3}{2}^+)$&2.979&$-0.36$ &3.49                       &2.02&$-0.54$ &3.71&\textbf{92.95}/2.58/3.89/0.58\\
                            &&2.985&$-5.80$ &0.67                      &2.06&$-5.86$ &1.27&\textbf{84.19}/5.91/8.62/1.28\\
                            &&2.991&$-12.26$ &0.45                         &2.09&$-13.10$ &0.90&\textbf{80.98}/7.21/10.31/1.50\\
                            \cline{2-9}
&$I(J^P)$&$\Lambda$ &$E$ &$r_{\rm RMS}$ &$\Lambda$ &$E$ &$r_{\rm RMS}$ &P(${}^6\mathbb{S}_{\frac{5}{2}}/{}^2\mathbb{D}_{\frac{5}{2}}/{}^4\mathbb{D}_{\frac{5}{2}}/{}^6\mathbb{D}_{\frac{5}{2}}$) \\
&$\frac{1}{2}(\frac{5}{2}^+)$&1.74&$-0.29$ &4.86          &1.24&$-0.38$ &4.61&\textbf{97.00}/0.15/0.10/2.75\\
                            &&1.92&$-3.85$ &1.72          &1.35&$-4.18$ &1.80&\textbf{92.61}/0.33/0.23/6.82\\
                            &&2.10&$-12.63$ &1.02        &1.45&$-12.36$ &1.18&\textbf{89.80}/0.42/0.33/9.45\\
&$\frac{3}{2}(\frac{5}{2}^+)$&$\times$&$\times$ &$\times$          &1.99&$-0.38$ &4.36&\textbf{95.62}/0.79/0.19/3.40\\
                            &&$\times$&$\times$ &$\times$         &2.10&$-4.40$ &1.63&\textbf{88.59}/2.18/0.50/8.72\\
                            &&$\times$&$\times$ &$\times$         &2.20&$-12.30$ &1.10&\textbf{83.91}/3.26/0.73/12.10\\\midrule[1.0pt]
System&$I(J^P)$&$\Lambda$ &$E$ &$r_{\rm RMS}$&$\Lambda$ &$E$ &$r_{\rm RMS}$&P(${}^2\mathbb{S}_{\frac{1}{2}}/{}^4\mathbb{D}_{\frac{1}{2}}/{}^6\mathbb{D}_{\frac{1}{2}}$)\\
\multirow{27}{*}{$\Sigma_{c}^{*}\bar D_{2}^{*}$}&$\frac{1}{2}(\frac{1}{2}^+)$&0.79&$-0.40$ &4.16             &0.79&$-1.57$ &2.47&\textbf{98.74}/0.64/0.62\\
                           &&0.88&$-5.12$ &1.46             &0.86&$-6.19$&1.40&\textbf{98.42}/0.82/0.76\\
                           &&0.96&$-12.96$ &1.01            &0.92&$-12.60$ &1.06&\textbf{98.26}/0.91/0.83\\
&$\frac{3}{2}(\frac{1}{2}^+)$&1.860&$-0.67$ &2.69             &1.65&$-0.33$ &4.40&\textbf{93.70}/2.71/3.59\\
                            &&1.866&$-6.33$ &0.77            &1.67&$-4.24$&1.42&\textbf{85.61}/6.17/8.22\\
                            &&1.871&$-12.09$ &0.55                &1.69&$-11.81$ &0.90&\textbf{83.01}/7.32/8.67\\
                            \cline{2-9}
&$I(J^P)$&$\Lambda$ &$E$  &$r_{\rm RMS}$ &$\Lambda$ &$E$  &$r_{\rm RMS}$ &P(${}^4\mathbb{S}_{\frac{3}{2}}/{}^2\mathbb{D}_{\frac{3}{2}}/{}^4\mathbb{D}_{\frac{3}{2}}/{}^6\mathbb{D}_{\frac{3}{2}}/{}^8\mathbb{D}_{\frac{3}{2}}$) \\
&$\frac{1}{2}(\frac{3}{2}^+)$&0.93&$-0.28$ &4.64           &0.87&$-0.25$ &4.85&\textbf{98.96}/0.27/0.49/0.10/0.18\\
                            &&1.03&$-4.52$ &1.55          &0.96&$-4.12$ &1.66&\textbf{97.84}/0.58/1.03/0.20/0.35\\
                            &&1.13&$-12.14$ &1.04         &1.05&$-12.04$ &1.09&\textbf{97.30}/0.73/1.31/0.25/0.41\\
&$\frac{3}{2}(\frac{3}{2}^+)$&2.194&$-0.61$ &2.74           &1.78&$-0.24$ &4.77&\textbf{94.09}/1.40/7.84/0.46/2.21\\
                            &&2.200&$-5.86$ &0.74               &1.81&$-4.38$ &1.42&\textbf{84.22}/3.74/4.83/1.21/6.01\\
                            &&2.206&$-12.22$ &0.50              &1.84&$-12.87$ &0.88&\textbf{80.40}/4.70/6.00/1.49/7.40\\
                            \cline{2-9}
&$I(J^P)$&$\Lambda$ &$E$ &$r_{\rm RMS}$&$\Lambda$ &$E$ &$r_{\rm RMS}$&P(${}^6\mathbb{S}_{\frac{5}{2}}/{}^2\mathbb{D}_{\frac{5}{2}}/{}^4\mathbb{D}_{\frac{5}{2}}/~{}^6\mathbb{D}_{\frac{5}{2}}/{}^8\mathbb{D}_{\frac{5}{2}}$) \\
&$\frac{1}{2}(\frac{5}{2}^+)$&1.23&$-0.24$ &4.98             &1.07&$-0.37$ &4.52&\textbf{98.08}/0.25/0.11/1.52/0.04\\
                            &&1.38&$-4.37$ &1.61             &1.17&$-4.26$ &1.71&\textbf{95.68}/0.53/0.24/3.46/0.09\\
                            &&1.53&$-12.15$ &1.06             &1.27&$-12.41$ &1.12&\textbf{94.19}/0.68/0.32/4.68/0.13\\
&$\frac{3}{2}(\frac{5}{2}^+)$&$\times$&$\times$ &$\times$             &2.07&$-0.27$ &4.73&\textbf{95.08}/1.46/0.36/2.97/0.12\\
                            &&$\times$&$\times$ &$\times$             &2.13&$-4.56$ &1.47&\textbf{85.84}/4.41/1.07/8.35/0.34\\
                            &&$\times$&$\times$ &$\times$             &2.18&$-12.17$ &0.98&\textbf{81.25}/6.03/1.43/0.86/0.43\\
                            \cline{2-9}
&$I(J^P)$&$\Lambda$ &$E$  &$r_{\rm RMS}$ &$\Lambda$ &$E$  &$r_{\rm RMS}$ &P(${}^8\mathbb{S}_{\frac{7}{2}}/{}^4\mathbb{D}_{\frac{7}{2}}/{}^6\mathbb{D}_{\frac{7}{2}}/{}^8\mathbb{D}_{\frac{7}{2}}$)\\
&$\frac{1}{2}(\frac{7}{2}^+)$&1.83&$-0.29$ &4.85               &1.24&$-0.36$ &4.71&\textbf{96.65}/0.13/0.05/3.17\\
                            &&2.00&$-3.74$ &1.72               &1.35&$-4.10$ &1.82&\textbf{91.44}/0.29/0.13/8.15\\
                            &&2.16&$-12.16$ &1.11              &1.45&$-12.47$ &1.17&\textbf{88.03}/0.36/0.18/11.43\\
&$\frac{3}{2}(\frac{7}{2}^+)$&$\times$&$\times$ &$\times$               &1.98&$-0.32$ &4.50&\textbf{96.24}/0.63/0.04/3.09\\
                            &&$\times$&$\times$ &$\times$               &2.10&$-3.98$ &1.70&\textbf{89.94}/1.82/0.12/8.12\\
                            &&$\times$&$\times$ &$\times$              &2.22&$-12.19$ &1.10&\textbf{85.31}/2.86/0.17/11.66\\
\bottomrule[1.5pt]
\end{tabular}
\end{table*}

We can obtain loosely bound states in the $\Sigma_{c}^* \bar D_1$ system with $I(J^{P})=1/2(1/2^{+})$ and the $\Sigma_{c}^* \bar  D_2^*$ system with $I(J^{P})=1/2(1/2^{+}, 3/2^{+})$ with the cutoffs around 1.00 GeV by using both the OPE model and the OBE model. Different from the OPE case, the $\rho$ and $\omega$ exchange attractive forces play an important role in the formation of the $\Sigma_{c}^* \bar D_1$ states with $I(J^{P})=1/2(3/2^{+}, 5/2^{+})$ and the $\Sigma_{c}^* \bar  D_2^*$ state with $I(J^{P})=1/2(5/2^{+}, 7/2^{+})$, and they are loosely bound molecular states with the OBE model, especially when we consider the contributions of the $D$-wave channels. The situation of the $\Sigma_{c}^* \bar D_1$ states with $I(J^{P})=3/2(1/2^{+}, 3/2^{+})$ and the $\Sigma_{c}^* \bar  D_2^*$ states with $I(J^{P})=3/2(1/2^{+}, 3/2^{+})$ is similar to that of the $\Sigma_{c}\bar D_1$ state with $I(J^{P})=3/2(1/2^{+})$ and the $\Sigma_{c} \bar  D_2^*$ state with $I(J^{P})=3/2(3/2^{+})$ qualitatively, and their properties significantly depend sensitively on the cutoff parameter, and thus our results indicate that these bound states cannot be good candidates of hadronic molecular states.

\subsection{Discussion and rating for bound states}

The $D$-wave components with small contributions does not obviously change some bound state properties \cite{Chen:2015add}. However, for the $\Sigma_{c}^{(*)}\bar D_{1}(\bar D_{2}^{*})$ states with $I=3/2$ and others, their bound state properties may change considerably after considering the $S$-$D$ wave mixing effects, which is mainly determined by the non diagonal matrix elements of the effective potentials \cite{Li:2012bt}.

The states with $I=1/2$ are easier to bind as a hadronic molecular candidate than those with $I=3/2$ for the low spin $\Sigma_{c}^{(*)} \bar T$-type systems. This is not surprising since the $\pi$ exchange effective potentials are weakly repulsive with $I=3/2$ and strongly attractive with $I=1/2$ for the low spin hidden-charm pentaquark states.

Empirically, the vector exchange interaction may be important for the formation of hadronic molecular states if the cutoff parameter is greater than 1.5 GeV; this is due to the introduction of the form factor  $\mathcal{F}(q^2,m_E^2) = (\Lambda^2-m_E^2)/(\Lambda^2-q^2)$ into the effective potential in the present work. Of course, when the cutoff parameter is relatively small, it is obvious that the $\pi$ exchange interaction plays an irreplaceable role in generating these loosely bound  molecular states.

As shown in Tables~\ref{jg1p}-\ref{jg2b}, we can predict a series of  possible new types of $P_c$ states with the form $\Sigma_{c}^{(*)} \bar T$. We have considered four different scenarios: (A) OPE without $S$-$D$ mixing effects, (B) OPE with $S$-$D$ mixing effects, (C) OBE without $S$-$D$ mixing effects, and (D) OBE with $S$-$D$ mixing effects. We notice that some bound states exist in some scenarios but disappear in other scenarios. Shall we treat them equally?

Many hadron molecular states have been predicted by different theoretical groups, and we try to categorize them in some reasonable way and give a rough indication which
states deserve experimental resources.

If a bound state can be formed both with and without $D$-wave contributions, it means that the convergence of partial-wave expansion is good and thus the results are more reliable in our scheme. For similar reasons, the states bound within both OPE and OBE models have smaller theoretical uncertainties.

Therefore, we provide a preliminary rating scheme. If a bound state can exist in $n$ scenarios with reasonable cutoff  around 1.0 GeV, we rate it as an $n$-star state. For example, $\Sigma_{c}^{*}\bar D_{1}$ with $I(J^{P})=1/2(3/2^{+})$ can be bound in scenarios (B), (C), and (D), and thus we assign it three stars. In this rating system, $P_c(4440)$ is a four-star state. In Table \ref{Summary1}, we present our rating for new types of $P_c$ states in $\Sigma_{c}^{(*)}\bar D_{1}(\bar D_{2}^{*})$ systems .

In this rating scheme, the deuteron has only three stars but is still in the top two among five ranks. There are many zero-star states which cannot be bound in any scenario and are not listed in Table VII. However, in an ideal rating scheme the deuteron should most probably be in the top one rank. The rating scheme in this paper definitely needs efforts from  particle physicists to be improved in future.

\renewcommand\tabcolsep{0.36cm}
\renewcommand{\arraystretch}{1.50}
\begin{table}[!htbp]
\caption{Rating for new types of $P_c$ states in the $\Sigma_{c}^{(*)}\bar D_{1}(\bar D_{2}^{*})$ systems with $(I,J)$.}\label{Summary1}
\begin{tabular}{c|cccc}\toprule[1.5pt]
Rating                 &$\Sigma_{c}\bar D_{1}$            &$\Sigma_{c}\bar D_{2}^{*}$                &$\Sigma_{c}^{*}\bar D_{1}$                 &$\Sigma_{c}^{*}\bar D_{2}^{*}$\\\midrule[1.0pt]
$\star\star\star\star$ &$(\frac{1}{2},\frac{1}{2})$       &$(\frac{1}{2},\frac{3}{2})$               &$(\frac{1}{2},\frac{1}{2})$             & $(\frac{1}{2},\frac{1}{2}/\frac{3}{2})$\\
$\star\star\star$      &                                  &                                          &$(\frac{1}{2},\frac{3}{2})$ &$(\frac{1}{2},\frac{5}{2})$,$(\frac{3}{2},\frac{7}{2})$\\
$\star\star$           &$(\frac{1}{2},\frac{3}{2})$       &                                          &                                           &                           \\
$\star$                &                                  &  $(\frac{1}{2},\frac{5}{2})$             &$(\frac{1}{2},\frac{5}{2})$ &$(\frac{1}{2},\frac{7}{2})$\\
\bottomrule[1.5pt]
\end{tabular}
\end{table}

\begin{figure}[!htbp]
\centering
\begin{tabular}{c}
\includegraphics[width=0.46\textwidth]{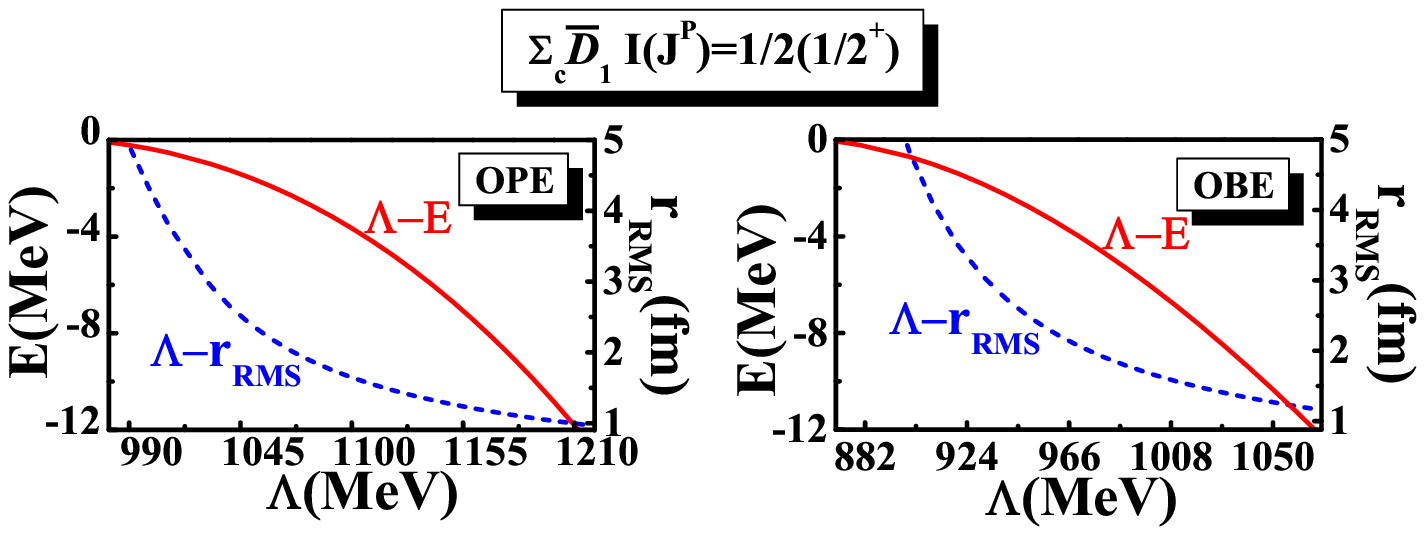}\\
\includegraphics[width=0.46\textwidth]{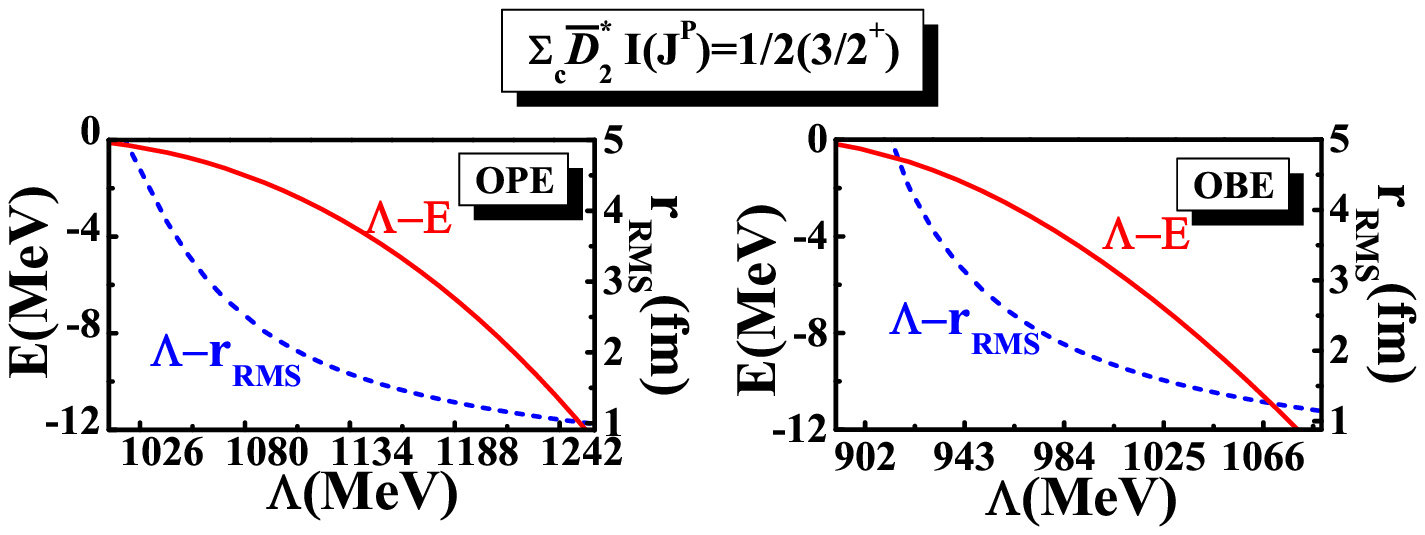}
\end{tabular}
\caption{(color online) The cutoff $\Lambda$ dependence of the binding energy $E$ and root-mean-square radius $r_{RMS}$  for the $\Sigma_c \bar D_1$ state with $I(J^{P})=1/2(1/2^+)$ and $\Sigma_c \bar D_2^*$ state with $I(J^{P})=1/2(3/2^+)$ with the OPE model and OBE model, respectively. }\label{wx}
\end{figure}

To investigate the uncertainties of cutoff for these four-star states within the OPE model and OBE model, in Fig.~\ref{wx} we present the $\Lambda$ dependence of the binding energy and root-mean-square radius for two typical states, $\Sigma_c \bar D_1$ with $I(J^{P})=1/2(1/2^+)$ and $\Sigma_c \bar D_2^*$ with $I(J^{P})=1/2(3/2^+)$. We can see clearly that the binding properties of these four-star states are not significantly dependent on the cutoff $\Lambda$.

In comparison to those large spin states, we find that the low spin states may be bound more tightly for the $\Sigma_{c}^{(*)}\bar D_{1}(\bar D_{2}^{*})$ states with $I=1/2$, which have results similar to the $\Sigma_{c}^{(*)}\bar D^{(*)}$ states with $I=1/2$ \cite{Chen:2019asm}. The observations of three resonance structures, $P_c(4312)$, $P_c(4440)$, and $P_c(4457)$, are from the $\Lambda_b$ baryon decays \cite{Aaij:2019vzc}, and thus one can search for these predicted new types of $P_c$ states through the $\Lambda_b$ baryon weak decays reported by the LHCb Collaboration. By simple partial wave analysis, we suggest first searching for these low spin $\Sigma_{c}^{(*)}\bar D_{1}(\bar D_{2}^{*})$ states with $I=1/2$ in the future. For these new types of $P_c$ states with the structures of $\Sigma_{c}\bar D_1$, $\Sigma_{c} \bar  D_2^*$, $\Sigma_{c}^{*}\bar D_1$, and $\Sigma_{c}^{*} \bar  D_2^*$, the masses were predicted to be around 4.87, 4.91, 4.93, and 4.97 GeV, respectively.

\section{New types of $P_c$ states---$\Lambda_{c}\bar D_{1}(\bar D_{2}^{*})$ systems}\label{sec4}

In this section, we study the $\Lambda_{c}\bar D_{1}(\bar D_{2}^{*})$ systems only within the OBE model since the pion exchange interaction is forbidden because of spin-parity conservation. In addition, there are no tensor force in the effective potentials for the $\Lambda_{c}\bar D_{1}(\bar D_{2}^{*})$ systems (see Appendix \ref{app0202} for more details), and therefore we do not consider the $S$-$D$ wave mixing effects in this investigation.

The interactions of the $\Lambda_{c}\bar D_{1}(\bar D_{2}^{*})$ systems are quite simple, because spin-parity conservation forbids the vertexes $\Lambda_c\Lambda_c\pi/\eta/\rho$. Thus there are only $\omega$ and $\sigma$ exchange interactions. Moreover, the $\omega$ exchange potentials are repulsive while the $\sigma$ exchange potentials are attractive, and thus the total potentials are very shallow. In fact, we cannot find the bound states for the $\Lambda_{c}\bar D_{1}$ systems with $I(J^{P})=1/2(1/2^{+},3/2^{+})$ and the $\Lambda_{c}\bar D_{2}^*$ systems with $I(J^{P})=1/2(3/2^{+},5/2^{+})$ when cutoffs $\Lambda$ are tuned from 0.79 to 4.00 GeV. This is similar to the results of the $\Lambda_{c}\bar D^{(*)}$ systems; that is, $\Lambda_{c}$ does not combine with $\bar D^{(*)}$ to form hidden-charm molecular pentaquarks for the single channel calculations \cite{Yang:2011wz}.

Since the pion exchange interactions are allowed for the processes $\Lambda_{c}\bar T\to \Sigma_{c}^{(*)}\bar T$, we expect that the coupled channel effects play an important role to generate the loosely bound hidden-charm molecular states for the $\Lambda_{c}\bar T$ systems. Thus, we further briefly discuss the $\mathcal{B}_{c}^{(*)}\bar T$ coupled systems. In the coupled channel analysis, the binding energy is provided relative to the lowest-lying threshold \cite{Chen:2017xat}.

Take the $\mathcal{B}_{c}^{(*)}\bar T$ coupled systems with $I(J^{P})=1/2(1/2^{+})$ as an example. The corresponding spatial wave function $|\psi\rangle$, the kinetic term $\mathcal{K}$, and the effective potential $\mathcal{V}$ can be
written as
\begin{eqnarray}
&&|\psi\rangle=\left(|\Lambda_c\bar{D}_1\rangle,\,|\Sigma_c\bar{D}_1\rangle,\,
|\Sigma_c^*\bar{D}_1\rangle,\,|\Sigma_c^*\bar{D}_2^*\rangle\right)^T,\\
&&\mathcal{K}=\mathrm{diag}\left(-\frac{\triangledown^2}{2\mu_1},\, -\frac{\triangledown^2}{2\mu_2}+d_1,\, -\frac{\triangledown^2}{2\mu_3}+d_2,\,-\frac{\triangledown^2}{2\mu_4}+d_3\right),\nonumber\\\\
&&\mathcal{V}\nonumber\\
&&={\left(\begin{array}{cccc}
\mathcal{V}^{\Lambda_c\bar{D}_1\to\Lambda_c\bar{D}_1}
&\mathcal{V}^{\Lambda_c\bar{D}_1\to\Sigma_c\bar{D}_1}
              &\mathcal{V}^{\Lambda_c\bar{D}_1\to\Sigma_c^*\bar{D}_1} &\mathcal{V}^{\Lambda_c\bar{D}_1\to\Sigma_c^*\bar{D}_2^*}\\
\mathcal{V}^{\Sigma_c\bar{D}_1\to\Lambda_c\bar{D}_1}
&\mathcal{V}^{\Sigma_c\bar{D}_1\to\Sigma_c\bar{D}_1}
              &\mathcal{V}^{\Sigma_c\bar{D}_1\to\Sigma_c^*\bar{D}_1}
              &\mathcal{V}^{\Sigma_c\bar{D}_1\to\Sigma_c^*\bar{D}_2^*}\\
\mathcal{V}^{\Sigma_c^*\bar{D}_1\to\Lambda_c\bar{D}_1}
&\mathcal{V}^{\Sigma_c^*\bar{D}_1\to\Sigma_c\bar{D}_1}
              &\mathcal{V}^{\Sigma_c^*\bar{D}_1\to\Sigma_c^*\bar{D}_1}
              &\mathcal{V}^{\Sigma_c^*\bar{D}_1\to\Sigma_c^*\bar{D}_2^*}\\
\mathcal{V}^{\Sigma_c^*\bar{D}_2^*\to\Lambda_c\bar{D}_1}
&\mathcal{V}^{\Sigma_c^*\bar{D}_2^*\to\Sigma_c\bar{D}_1}
              &\mathcal{V}^{\Sigma_c^*\bar{D}_2^*\to\Sigma_c^*\bar{D}_1}
              &\mathcal{V}^{\Sigma_c^*\bar{D}_2^*\to\Sigma_c^*\bar{D}_2^*}\\\end{array}\right)},\label{VV}\nonumber\\
\end{eqnarray}
respectively. The specific effective potentials for each component are presented in Appendixes \ref{app0201}-\ref{app0202}. $\mu_1$, $\mu_2$, $\mu_3$, and $\mu_4$ are the reduced masses of the $\Lambda_{c}\bar D_1$, $\Sigma_{c}\bar D_1$, $\Sigma_{c}^*\bar D_1$, and $\Sigma_{c}^*\bar D_2^*$ systems, respectively. In addition,  $d_i$ are defined as $d_1=m_{\Sigma_{c}}-m_{\Lambda_{c}}$, $d_2=m_{\Sigma_{c}^*}-m_{\Lambda_{c}}$, and $d_3=m_{\Sigma_{c}^*}+m_{{D}_2^*}-m_{\Lambda_{c}}-m_{{D}_1}$. Then, we solve the coupled channel Schr\"odinger equation and try to find the corresponding bound state solutions.

\begin{figure}[!htbp]
\centering
\begin{tabular}{c}
\includegraphics[width=0.46\textwidth]{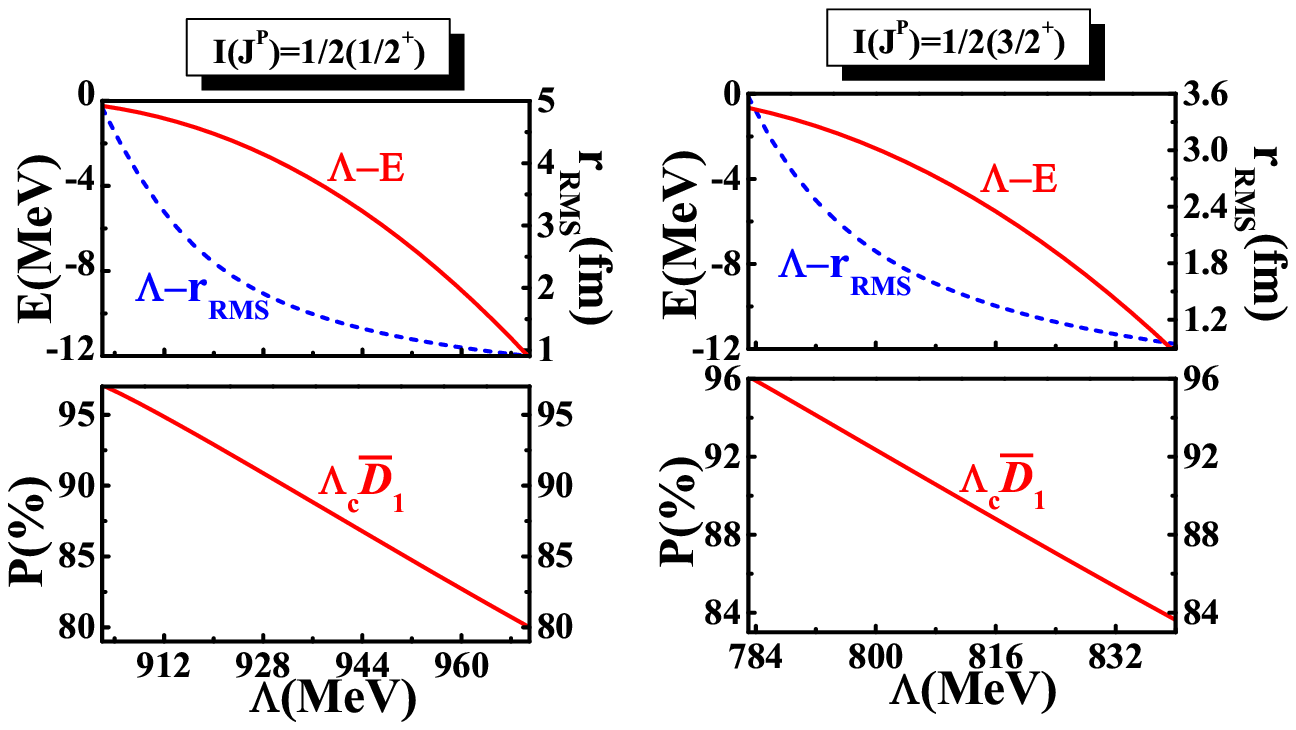}\\
\includegraphics[width=0.46\textwidth]{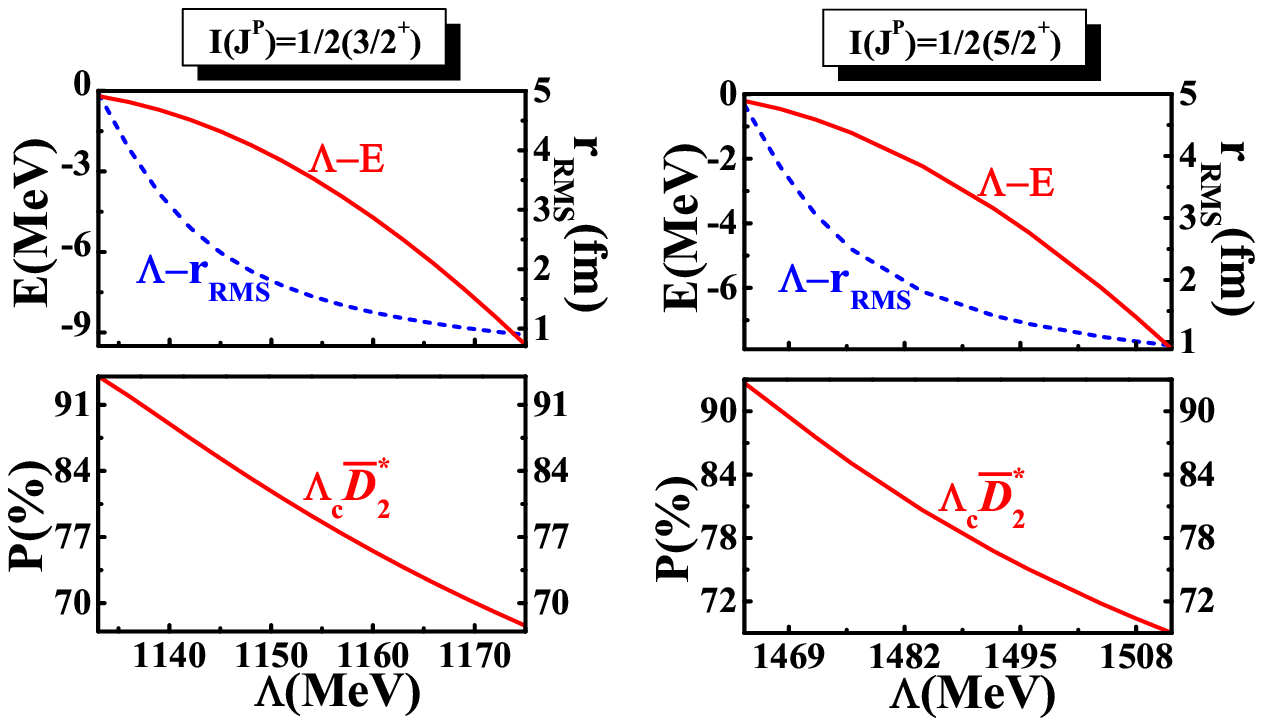}
\end{tabular}
\caption{(color online) The cutoff $\Lambda$ dependence of the binding energy $E$, root-mean-square radius $r_{RMS}$, and the largest probability $P(\%)$ for the $\mathcal{B}_{c}^{(*)}\bar T$ coupled systems with $I(J^{P})=1/2(1/2^{+},3/2^{+})$ and $I(J^{P})=1/2(3/2^{+},5/2^{+})$; the masses of bound states are defined relative to the $\Lambda_c\bar{D}_1$ and $\Lambda_c\bar{D}_2^*$ thresholds, respectively. }\label{cr}
\end{figure}

In Fig.~\ref{cr}, we present the cutoff $\Lambda$ dependence of the binding energy $E$, root-mean-square radius $r_{RMS}$, and the largest probability $P(\%)$ for the $\mathcal{B}_{c}^{(*)}\bar T$ coupled systems with $I(J^{P})=1/2(1/2^{+},3/2^{+})$ and $I(J^{P})=1/2(3/2^{+},5/2^{+})$. The masses of bound states are defined relative to the $\Lambda_c\bar{D}_1$ and $\Lambda_c\bar{D}_2^*$ thresholds, respectively.

After including the coupled channel effects, the bound solutions for the $\mathcal{B}_{c}^{(*)}\bar T$ coupled systems with $I(J^{P})= 1/2(1/2^+, 3/2^+)$ and $I(J^{P})= 1/2(3/2^+, 5/2^+)$ are obtained when the cutoffs $\Lambda$ vary in a reasonable range, and they are mainly composed of the $\Lambda_{c}\bar D_{1}$ channel with probability over $80\%$ and the $\Lambda_{c}\bar D_{2}^*$ channel with almost $70\%$ probability, respectively. Their RMS radii are around or larger than 1.0 fm, and the binding energies are a few MeV. Therefore, the $\Lambda_{c}\bar D_{1}$ states with $I(J^{P})= 1/2(1/2^+, 3/2^+)$ and $\Lambda_{c}\bar D_{2}^*$ states with $I(J^{P})= 1/2(3/2^+, 5/2^+)$ are possible candidates of the new types of $P_c$ states with the coupled channel effects.

\section{Doubly charmed molecular pentaquarks}\label{sec5}

%In Ref. \cite{Shimizu:2017xrg}, Yuki Shimizu and Masayasu Harada once predicted some possible double-charm molecular pentaquarks between a charmed baryon $\mathcal{B}_c^{(*)}=\Lambda_c/\Sigma_c/\Sigma_c^*$ and a charmed meson $(D,D^*)$.

In this work, we extend the obtained OBE effective potentials to search for possible doubly charmed molecular pentaquarks composed of a charmed baryon $\mathcal{B}_c^{(*)}=\Lambda_c/\Sigma_c/\Sigma_c^*$ and an excited charmed meson in a $T$ doublet $({D}_1,{D}_2^*)$, which is an interesting research topic full of opportunities and challenges.

The effective potentials for the doubly charmed $\mathcal{B}_{c}^{(*)} T$ systems are related to those for hidden-charmed $\mathcal{B}_{c}^{(*)} \bar T$ systems via the $G$-parity transformation. The potentials induced by the $\sigma$, $\eta$, and $\rho$ exchanges are same, while those resulting from the $\pi$ and $\omega$ exchanges have opposite signs between the corresponding doubly charmed and hidden-charmed systems with the same isospin and spin-parity quantum numbers \cite{Klempt:2002ap}.

\subsection{$\Sigma_{c}^{(*)}D_{1}(D_{2}^{*})$ systems}

We show the numerical results for the $\Sigma_{c} D_{1}(D_{2}^{*})$ and $\Sigma_{c}^{*}D_{1}(D_{2}^{*})$ systems with the OPE model in Tables \ref{jg1pu} and \ref{jg2pu}, respectively.

\renewcommand\tabcolsep{0.08cm}
\renewcommand{\arraystretch}{1.50}
\begin{table}[!tbp]
\caption{Bound state properties for the $\Sigma_{c}^{*}D_{1}(D_{2}^{*})$ systems with the OPE model. Conventions are the same as Table~\ref{jg1p}.}\label{jg1pu}
\begin{tabular}{c|ccc|cccl}\toprule[1.5pt]
\multicolumn{8}{c}{$\Sigma_{c} D_{1}$}\\\midrule[1.0pt]
$I(J^P)$&$\Lambda$ &$E$ &$r_{\rm RMS}$    &$\Lambda$ &$E$ &$r_{\rm RMS}$ &P(${}^2\mathbb{S}_{\frac{1}{2}}/{}^4\mathbb{D}_{\frac{1}{2}}$)\\
$\frac{1}{2}(\frac{1}{2}^+)$&$\times$&$\times$ &$\times$                 &2.62&$-0.34$ &4.68&\textbf{95.15}/4.85\\
                            &$\times$&$\times$ &$\times$                 &2.78&$-3.59$ &1.80&\textbf{85.39}/14.61\\
                            &$\times$&$\times$ &$\times$                &2.94&$-12.22$ &1.08&\textbf{75.90}/24.10\\
$\frac{3}{2}(\frac{1}{2}^+)$&1.86&$-0.26$ &4.64        &1.76&$-0.28$ &4.62&\textbf{99.65}/0.35\\
                            &2.00&$-4.22$ &1.44        &1.90&$-3.95$ &1.52&\textbf{99.23}/0.77\\
                            &2.13&$-12.56$ &0.87        &2.04&$-12.46$ &0.90&\textbf{99.05}/0.95\\\hline
$I(J^P)$&$\Lambda$ &$E$&$r_{\rm RMS}$&$\Lambda$ &$E$&$r_{\rm RMS}$  &P(${}^4\mathbb{S}_{\frac{3}{2}}/{}^2\mathbb{D}_{\frac{3}{2}}/{}^4\mathbb{D}_{\frac{3}{2}}$)\\
$\frac{1}{2}(\frac{3}{2}^+)$&1.86&$-0.26$ &4.64        &1.22&$-0.24$ &4.96&\textbf{97.51}/0.59/1.90 \\
                            &2.00&$-4.22$ &1.44        &1.37&$-3.91$&1.70&\textbf{93.83}/1.45/4.72\\
                            &2.13&$-12.56$ &0.87         &1.51&$-12.25$ &1.07&\textbf{91.85}/1.89/6.26\\\midrule[1.0pt]
\multicolumn{8}{c}{$\Sigma_{c} D_{2}^{*}$}\\\midrule[1.0pt]
$I(J^P)$&$\Lambda$ &$E$&$r_{\rm RMS}$ &$\Lambda$ &$E$&$r_{\rm RMS}$ &P(${}^4\mathbb{S}_{\frac{3}{2}}/{}^4\mathbb{D}_{\frac{3}{2}}/{}^6\mathbb{D}_{\frac{3}{2}}$)\\
$\frac{1}{2}(\frac{3}{2}^+)$&$\times$&$\times$ &$\times$                 &2.30&$-0.31$ &4.81&\textbf{95.69}/0.94/3.37\\
                            &$\times$&$\times$ &$\times$                 &2.50&$-3.68$&1.83 &\textbf{87.29}/2.82/9.89\\
                            &$\times$&$\times$ &$\times$                &2.70&$-12.83$ &1.09&\textbf{79.66}/4.52/15.82\\
$\frac{3}{2}(\frac{3}{2}^+)$&2.03&$-0.29$ &4.51       &1.81&$-0.20$ &4.99&\textbf{99.42}/0.08/0.50\\
                           &2.17&$-4.14$ &1.44        &1.96&$-3.66$ &1.58&\textbf{98.59}/0.19/1.22\\
                           &2.30&$-12.10$ &0.88        &2.11&$-12.04$ &0.93&\textbf{98.18}/0.25/1.56\\\hline
$I(J^P)$&$\Lambda$ &$E$&$r_{\rm RMS}$ &$\Lambda$ &$E$&$r_{\rm RMS}$ &P(${}^6\mathbb{S}_{\frac{5}{2}}/{}^4\mathbb{D}_{\frac{5}{2}}/{}^6\mathbb{D}_{\frac{5}{2}}$) \\
$\frac{1}{2}(\frac{5}{2}^+)$&1.57&$-0.20$ &4.92       &1.16&$-0.32$ &4.58&\textbf{97.68}/0.95/1.37\\
                            &1.70&$-3.83$ &1.52       &1.30&$-4.05$ &1.66&\textbf{94.95}/2.06/2.99\\
                            &1.83&$-12.39$ &0.89        &1.43&$-12.12$ &1.05&\textbf{93.60}/2.57/3.82\\\hline
\bottomrule[1.5pt]
\end{tabular}
\end{table}

\renewcommand\tabcolsep{0.45cm}
\renewcommand{\arraystretch}{1.15}
\begin{table*}[!htbp]
\caption{Bound state properties for the $\Sigma_{c}^{*}D_{1}(D_{2}^{*})$ systems with the OPE model. Conventions are the same as Table~\ref{jg1p}.}\label{jg2pu}
\begin{tabular}{c|c|ccc|cccl}\toprule[1.5pt]
System&$I(J^P)$&$\Lambda$ &$E$  &$r_{\rm RMS}$ &$\Lambda$ &$E$  &$r_{\rm RMS}$ &P(${}^2\mathbb{S}_{\frac{1}{2}}/{}^4\mathbb{D}_{\frac{1}{2}}/{}^6\mathbb{D}_{\frac{1}{2}}$)\\
\multirow{20}{*}{$\Sigma_{c}^{*} D_{1}$}&$\frac{1}{2}(\frac{1}{2}^+)$&$\times$&$\times$ &$\times$                            &1.95&$-0.34$ &4.79&\textbf{94.52}/2.57/2.91\\
                            &&$\times$&$\times$ &$\times$                            &2.13&$-3.80$ &1.85&\textbf{83.60}/7.10/9.30\\
                            &&$\times$&$\times$ &$\times$                          &2.30&$-12.72$ &1.14&\textbf{73.93}/10.42/15.65\\
&$\frac{3}{2}(\frac{1}{2}^+)$&1.51&$-0.20$ &4.93                  &1.41&$-0.29$ &4.56&\textbf{99.44}/0.34/0.22\\
                            &&1.64&$-3.88$ &1.51                  &1.54&$-3.88$ &1.55&\textbf{98.83}/0.72/0.45\\
                            &&1.77&$-12.62$ &0.89                  &1.67&$-12.18$ &0.94&\textbf{98.57}/0.89/0.54\\
                            \cline{2-9}
&$I(J^P)$&$\Lambda$ &$E$ &$r_{\rm RMS}$ &$\Lambda$ &$E$ &$r_{\rm RMS}$ &P(${}^4\mathbb{S}_{\frac{3}{2}}/{}^2\mathbb{D}_{\frac{3}{2}}/{}^4\mathbb{D}_{\frac{3}{2}}/{}^6\mathbb{D}_{\frac{3}{2}}$) \\
&$\frac{1}{2}(\frac{3}{2}^+)$&$\times$&$\times$ &$\times$                          &2.10&$-0.35$ &4.62&\textbf{95.68}/1.57/2.42/0.33 \\
                            &&$\times$&$\times$ &$\times$                          &2.32&$-3.84$ &1.79&\textbf{88.74}/4.15/6.20/0.91\\
                            &&$\times$&$\times$ &$\times$                          &2.54&$-12.66$ &1.11&\textbf{83.27}/6.21/9.08/1.44\\
&$\frac{3}{2}(\frac{3}{2}^+)$&3.47&$-0.21$ &4.84                 &2.40&$-0.30$ &4.56&\textbf{98.78}/0.31/0.81/0.10\\
                            &&3.63&$-3.98$ &1.51                 &2.58&$-3.88$ &1.56&\textbf{97.01}/0.75/2.01/0.23\\
                            &&3.79&$-12.55$ &0.83                  &2.76&$-12.33$ &0.93&\textbf{95.87}/1.03/2.79/0.31\\
                            \cline{2-9}
&$I(J^P)$&$\Lambda$ &$E$ &$r_{\rm RMS}$ &$\Lambda$ &$E$ &$r_{\rm RMS}$ &P(${}^6\mathbb{S}_{\frac{5}{2}}/{}^2\mathbb{D}_{\frac{5}{2}}/{}^4\mathbb{D}_{\frac{5}{2}}/{}^6\mathbb{D}_{\frac{5}{2}}$) \\
&$\frac{1}{2}(\frac{5}{2}^+)$&1.30&$-0.26$ &4.63                 &1.02&$-0.22$ &5.00&\textbf{98.10}/0.28/0.08/1.55 \\
                            &&1.42&$-3.98$ &1.51                 &1.15&$-3.77$ &1.71&\textbf{95.62}/0.65/0.17/3.56\\
                            &&1.54&$-12.57$ &0.91                 &1.28&$-12.53$ &1.04&\textbf{94.42}/0.81/0.22/4.55\\
&$\frac{3}{2}(\frac{5}{2}^+)$&$\times$&$\times$ &$\times$                          &3.80&$-0.29$ &4.78&\textbf{97.60}/0.10/0.07/2.24\\
                            &&$\times$&$\times$ &$\times$                          &3.90&$-0.99$ &3.06&\textbf{95.95}/0.14/0.12/3.79\\
                            &&$\times$&$\times$ &$\times$                          &4.00&$-2.20$ &2.14&\textbf{94.30}/0.19/0.17/5.34\\\midrule[1.0pt]
System&$I(J^P)$&$\Lambda$ &$E$ &$r_{\rm RMS}$&$\Lambda$ &$E$ &$r_{\rm RMS}$&P(${}^2\mathbb{S}_{\frac{1}{2}}/{}^4\mathbb{D}_{\frac{1}{2}}/{}^6\mathbb{D}_{\frac{1}{2}}$)\\
\multirow{27}{*}{$\Sigma_{c}^{*}D_{2}^{*}$}&$\frac{1}{2}(\frac{1}{2}^+)$&$\times$&$\times$ &$\times$                          &1.86&$-0.35$ &4.80&\textbf{94.65}/2.37/2.98\\
                            &&$\times$&$\times$ &$\times$                          &2.06&$-3.53$ &1.95&\textbf{85.06}/6.38/8.56\\
                            &&$\times$&$\times$ &$\times$                        &2.26&$-12.10$ &1.20&\textbf{76.58}/9.66/13.76\\
&$\frac{3}{2}(\frac{1}{2}^+)$&1.41&$-0.27$ &4.59                &1.30&$-0.25$ &4.75&\textbf{99.38}/0.32/0.31\\
                            &&1.54&$-4.29$ &1.44                &1.43&$-3.85$ &1.57&\textbf{98.64}/0.70/0.66\\
                            &&1.66&$-12.75$ &0.89                &1.56&$-12.40$ &0.94&\textbf{98.33}/0.87/0.80\\
                            \cline{2-9}
&$I(J^P)$&$\Lambda$ &$E$  &$r_{\rm RMS}$ &$\Lambda$ &$E$  &$r_{\rm RMS}$ &P(${}^4\mathbb{S}_{\frac{3}{2}}/{}^2\mathbb{D}_{\frac{3}{2}}/{}^4\mathbb{D}_{\frac{3}{2}}/{}^6\mathbb{D}_{\frac{3}{2}}/{}^8\mathbb{D}_{\frac{3}{2}}$) \\
&$\frac{1}{2}(\frac{3}{2}^+)$&$\times$&$\times$ &$\times$                         &1.90&$-0.27$ &5.00&\textbf{95.31}/1.19/1.56/0.37/1.58\\
                            &&$\times$&$\times$ &$\times$                         &2.10&$-3.64$ &1.87&\textbf{85.84}/3.33/4.30/1.07/5.46\\
                            &&$\times$&$\times$ &$\times$                         &2.29&$-12.50$ &1.14&\textbf{77.75}/4.80/6.17/1.61/9.67\\
&$\frac{3}{2}(\frac{3}{2}^+)$&2.00&$-0.21$ &4.85                &1.70&$-0.21$ &4.95&\textbf{99.17}/0.22/0.39/0.08/0.14\\
                            &&2.14&$-3.82$ &1.48                &1.86&$-3.99$ &1.53&\textbf{97.92}/0.55/1.00/0.20/0.34\\
                            &&2.28&$-12.27$ &0.86                &2.01&$-12.60$ &0.91&\textbf{97.30}/0.72/1.30/0.25/0.42\\
                            \cline{2-9}
&$I(J^P)$&$\Lambda$ &$E$ &$r_{\rm RMS}$&$\Lambda$ &$E$ &$r_{\rm RMS}$&P(${}^6\mathbb{S}_{\frac{5}{2}}/{}^2\mathbb{D}_{\frac{5}{2}}/{}^4\mathbb{D}_{\frac{5}{2}}/{}^6\mathbb{D}_{\frac{5}{2}}/{}^8\mathbb{D}_{\frac{5}{2}}$) \\
&$\frac{1}{2}(\frac{5}{2}^+)$&$\times$&$\times$ &$\times$                &1.80&$-0.30$ &4.76&\textbf{96.27}/1.04/0.27/2.33/0.08\\
                            &&$\times$&$\times$ &$\times$                &2.01&$-3.83$ &1.76&\textbf{90.33}/2.80/0.71/5.94/0.23\\
                            &&$\times$&$\times$ &$\times$                &2.21&$-12.20$ &1.10&\textbf{86.25}/4.06/1.01/8.35/0.34\\
&$\frac{3}{2}(\frac{5}{2}^+)$&$\times$&$\times$ &$\times$                &2.94&$-0.37$ &4.33&\textbf{98.01}/0.22/0.10/1.62/4.68\\
                            &&$\times$&$\times$ &$\times$                &3.15&$-3.89$ &1.58&\textbf{95.07}/0.50/0.26/4.06/0.11\\
                            &&$\times$&$\times$ &$\times$                &3.36&$-12.10$ &0.95&\textbf{92.88}/0.68/0.37/5.91/0.16\\
                            \cline{2-9}
&$I(J^P)$&$\Lambda$ &$E$  &$r_{\rm RMS}$&$\Lambda$ &$E$  &$r_{\rm RMS}$ &P(${}^8\mathbb{S}_{\frac{7}{2}}/{}^4\mathbb{D}_{\frac{7}{2}}/{}^6\mathbb{D}_{\frac{7}{2}}/{}^8\mathbb{D}_{\frac{7}{2}}$)\\
&$\frac{1}{2}(\frac{7}{2}^+)$&1.10&$-0.21$ &4.86                &0.93&$-0.23$ &4.93&\textbf{98.35}/~0.20/~0.02/~1.42 \\
                            &&1.22&$-3.64$ &1.57                &1.05&$-3.76$ &1.69&\textbf{96.39}/~0.45/~0.05/~3.11\\
                            &&1.34&$-12.62$ &0.91                &1.17&$-12.40$ &1.03&\textbf{95.52}/~0.56/~0.06/~3.87\\
&$\frac{3}{2}(\frac{7}{2}^+)$&$\times$&$\times$ &$\times$                         &3.65&$-0.37$ &4.45&\textbf{97.06}/0.09/0.04/2.81\\
                            &&$\times$&$\times$ &$\times$                         &3.88&$-3.04$ &1.85&\textbf{92.74}/0.18/0.10/6.98\\
                            &&$\times$&$\times$ &$\times$                         &4.00&$-5.77$ &1.39&\textbf{90.66}/0.22/0.13/9.00\\
\bottomrule[1.5pt]
\end{tabular}
\end{table*}

In Table \ref{jg1pu}, the binding energies and RMS radii are the same for the bound states in the $\Sigma_{c} D_1$ system with $I(J^{P})=3/2(1/2^{+})$ and the $\Sigma_{c} D_1$ system with $I(J^{P})= 1/2(3/2^{+})$ when we do not consider the contributions of the $D$-wave channels, owing to the same product of the matrix elements for the operators $\mathcal{A}_{2}$ and the isospin factors $\mathcal{H}(I)$ giving rise to identical and attractive $\pi$ exchange interactions. Even if we add the contributions of the $S$-$D$ wave mixing effects, the bound state solutions are still absent for the $\Sigma_{c} D_1$ state with $I(J^{P})=3/2(3/2^{+})$ and the $\Sigma_{c} D_2^*$ state with $I(J^{P})=3/2(5/2^{+})$ with the OPE model.
From Table \ref{jg2pu}, we also notice that although the contribution of the $\pi$ exchange effective potential gives a slightly attractive force for the $\Sigma_{c}^* D_2^*$ state with $I(J^{P})= 3/2(5/2^{+})$ when we only consider the contributions of the $S$-wave channel, and we cannot obtain bound state solutions until we increase the cutoffs to be around 4.00 GeV.

We have shown the OPE results above. To compare the properties of the bound states with OPE and OBE models, we still need to present the numerical results with the OBE model. We list the bound state properties of the $\Sigma_{c}D_{1}(D_{2}^{*})$ systems and the $\Sigma_{c}^{*} D_{1}(D_{2}^{*})$ systems with the OBE model in Tables \ref{jg1bu} and \ref{jg2bu}. The exchanges of $\pi$, $\sigma$, $\eta$, $\rho$, and $\omega$ are allowed.

\renewcommand\tabcolsep{0.08cm}
\renewcommand{\arraystretch}{1.50}
\begin{table}[!tbp]
\caption{Bound state properties for the $\Sigma_{c}D_{1}(D_{2}^{*})$ systems with the OBE model. Conventions are the same as Table~\ref{jg1p}.}\label{jg1bu}
\begin{tabular}{c|ccc|cccl}\toprule[1.5pt]
\multicolumn{8}{c}{$\Sigma_{c}D_{1}$}\\\midrule[1.0pt]
$I(J^P)$&$\Lambda$ &$E$ &$r_{\rm RMS}$  &$\Lambda$ &$E$ &$r_{\rm RMS}$ &P(${}^2\mathbb{S}_{\frac{1}{2}}/{}^4\mathbb{D}_{\frac{1}{2}}$)\\
$\frac{1}{2}(\frac{1}{2}^+)$&1.31&$-0.36$ &4.81         &1.26&$-0.32$ &4.97 &\textbf{99.20}/0.80\\
                            &1.48&$-4.53$&1.86         &1.45&$-4.80$ &1.84 &\textbf{98.47}/1.53\\
                            &1.65&$-12.59$&1.29        &1.63&$-12.95$ &1.29 &\textbf{98.69}/1.31\\
$\frac{3}{2}(\frac{1}{2}^+)$&1.38&$-0.28$&4.70      &1.34&$-0.24$ &4.90 &\textbf{99.82}/0.18\\
                            &1.59&$-4.24$&1.56         &1.55&$-4.25$ &1.57 &\textbf{99.58}/0.42\\
                            &1.80&$-12.69$ &0.97        &1.75&$-12.48$ &0.99 &\textbf{99.47}/~0.53\\\hline
$I(J^P)$&$\Lambda$ &$E$&$r_{\rm RMS}$ &$\Lambda$ &$E$&$r_{\rm RMS}$  &P(${}^4\mathbb{S}_{\frac{3}{2}}/{}^2\mathbb{D}_{\frac{3}{2}}/{}^4\mathbb{D}_{\frac{3}{2}}$)\\
$\frac{1}{2}(\frac{3}{2}^+)$&0.94&$-0.39$ &4.31       &0.91&$-0.55$ &3.92&\textbf{98.19}/0.41/1.40 \\
                            &0.98&$-4.72$&1.55        &0.96&$-5.30$ &1.54&\textbf{97.32}/0.59/2.09 \\
                            &1.01&$-11.20$ &1.09      &1.00&$-13.35$ &1.06&\textbf{97.49}/0.54/1.97 \\
$\frac{3}{2}(\frac{3}{2}^+)$&$\times$&$\times$ &$\times$       &3.48&$-0.28$ &4.87&\textbf{98.96}/0.17/0.87\\
                            &$\times$&$\times$ &$\times$       &3.79&$-1.28$ &2.78&\textbf{98.02}/0.31/1.66\\
                            &$\times$&$\times$ &$\times$       &4.00&$-2.48$ &2.07&\textbf{97.39}/0.41/2.20\\\midrule[1.0pt]
\multicolumn{8}{c}{$\Sigma_{c}D_{2}^{*}$}\\\midrule[1.0pt]
$I(J^P)$&$\Lambda$ &$E$&$r_{\rm RMS}$ &$\Lambda$ &$E$&$r_{\rm RMS}$ &P(${}^4\mathbb{S}_{\frac{3}{2}}/{}^4\mathbb{D}_{\frac{3}{2}}/{}^6\mathbb{D}_{\frac{3}{2}}$)\\
$\frac{1}{2}(\frac{3}{2}^+)$&1.30&$-0.40$ &4.62                &1.22&$-0.32$ &4.96&\textbf{98.80}/0.24/0.96\\
                            &1.47&$-4.81$&1.80                &1.40&$-4.47$ &1.88&\textbf{97.65}/0.48/1.88\\
                            &1.63&$-12.48$ &1.28              &1.58&$-12.22$ &1.31&\textbf{97.87}/0.44/1.69\\
$\frac{3}{2}(\frac{3}{2}^+)$&1.44&$-0.23$ &4.94              &1.38 &$-0.29$ &4.67&\textbf{99.66}/0.05/0.29\\
                            &1.67&$-4.03$ &1.59              &1.59&$-4.06$ &1.61&\textbf{99.25}/0.11/0.64\\
                            &1.90&$-12.34$ &0.97              &1.80&$-12.06$ &1.00&\textbf{99.03}/0.14/0.83\\\hline
$I(J^P)$&$\Lambda$ &$E$&$r_{\rm RMS}$ &$\Lambda$ &$E$&$r_{\rm RMS}$ &P(${}^6\mathbb{S}_{\frac{5}{2}}/{}^4\mathbb{D}_{\frac{5}{2}}/{}^6\mathbb{D}_{\frac{5}{2}}$) \\
$\frac{1}{2}(\frac{5}{2}^+)$&0.92&$-0.45$ &4.11             &0.89&$-0.48$ &4.10&\textbf{98.39}/0.64/0.97\\
                            &0.96&$-5.19$ &1.47             &0.94&$-5.34$ &1.52&\textbf{97.58}/0.94/1.48\\
                            &0.99&$-12.33$ &1.04            &0.98&$-13.98$ &1.03&\textbf{97.77}/0.84/1.39\\
$\frac{3}{2}(\frac{5}{2}^+)$&$\times$&$\times$&$\times$      &3.50&$-0.29$ &4.80&\textbf{98.81}/0.36/0.83\\
                            &$\times$&$\times$&$\times$      &3.80&$-1.35$ &2.72&\textbf{97.72}/0.69/1.59\\
                            &$\times$&$\times$&$\times$      &4.00&$-2.59$ &2.03&\textbf{96.98}/0.90/2.12\\\hline
\bottomrule[1.5pt]
\end{tabular}
\end{table}

\renewcommand\tabcolsep{0.45cm}
\renewcommand{\arraystretch}{1.15}
\begin{table*}[!htbp]
\caption{Bound state properties for the $\Sigma_{c}^{*} D_{1}(D_{2}^{*})$ systems with the OBE model. Conventions are the same as Table~\ref{jg1p}.}\label{jg2bu}
\begin{tabular}{c|c|ccc|cccl}\toprule[1.5pt]
System&$I(J^P)$&$\Lambda$ &$E$  &$r_{\rm RMS}$ &$\Lambda$ &$E$  &$r_{\rm RMS}$ &P(${}^2\mathbb{S}_{\frac{1}{2}}/{}^4\mathbb{D}_{\frac{1}{2}}/{}^6\mathbb{D}_{\frac{1}{2}}$)\\
\multirow{20}{*}{$\Sigma_{c}^{*} D_{1}$}&$\frac{1}{2}(\frac{1}{2}^+)$&1.32&$-0.36$ &4.83       &1.22&$-0.33$ &4.97&\textbf{98.27}/0.95/0.78\\
                            &&1.49&$-4.50$ &1.89       &1.41&$-4.49$&1.92 &\textbf{96.54}/1.91/1.55\\
                            &&1.65&$-12.21$ &1.33     &1.59&$-12.15$ &1.36&\textbf{96.81}/1.80/1.39\\
&$\frac{3}{2}(\frac{1}{2}^+)$&1.21&$-0.23$ &4.88      &1.17&$-0.28$ &4.71&\textbf{99.66}/0.20/0.14\\
                            &&1.38&$-4.07$ &1.58      &1.34&$-4.36$&1.56&\textbf{99.28}/0.44/0.28\\
                            &&1.55&$-12.31$ &0.98      &1.50&$-12.47$ &1.00&\textbf{99.13}/0.54/0.33\\
                            \cline{2-9}
&$I(J^P)$&$\Lambda$ &$E$ &$r_{\rm RMS}$ &$\Lambda$ &$E$ &$r_{\rm RMS}$ &P(${}^4\mathbb{S}_{\frac{3}{2}}/{}^2\mathbb{D}_{\frac{3}{2}}/{}^4\mathbb{D}_{\frac{3}{2}}/{}^6\mathbb{D}_{\frac{3}{2}}$) \\
&$\frac{1}{2}(\frac{3}{2}^+)$&1.21&$-0.33$ &4.78       &1.14&$-0.42$ &4.51&\textbf{98.36}/0.56/0.96/0.12\\
                            &&1.35&$-4.83$ &1.72       &1.28&$-4.72$ &1.77&\textbf{97.06}/1.01/1.72/0.22\\
                            &&1.49&$-12.84$ &1.20      &1.42&$-12.06$ &1.26&\textbf{97.17}/0.97/1.65/0.21\\
&$\frac{3}{2}(\frac{3}{2}^+)$&1.94&$-0.25$ &4.85       &1.68&$-0.25$ &4.90&\textbf{99.47}/0.15/0.35/0.04\\
                            &&2.35&$-4.03$ &1.58       &1.99&$-4.00$ &1.63&\textbf{98.58}/0.39/0.93/0.11\\
                            &&2.76&$-12.23$ &0.97       &2.29&$-12.07$ &1.01&\textbf{97.97}/0.55/1.33/0.15\\
                            \cline{2-9}
&$I(J^P)$&$\Lambda$ &$E$ &$r_{\rm RMS}$ &$\Lambda$ &$E$ &$r_{\rm RMS}$ &P(${}^6\mathbb{S}_{\frac{5}{2}}/{}^2\mathbb{D}_{\frac{5}{2}}/{}^4\mathbb{D}_{\frac{5}{2}}/{}^6\mathbb{D}_{\frac{5}{2}}$) \\
&$\frac{1}{2}(\frac{5}{2}^+)$&0.89&$-0.26$ &4.37          &0.86&$-0.53$ &3.93&\textbf{98.23}/0.25/0.07/1.45\\
                            &&0.93&$-5.10$ &1.47          &0.90&$-4.00$ &1.70&\textbf{97.43}/0.35/0.09/2.12\\
                            &&0.96&$-12.62$ &1.02         &0.94&$-11.81$ &1.10&\textbf{97.52}/0.32/0.09/2.07\\
&$\frac{3}{2}(\frac{5}{2}^+)$&$\times$&$\times$ &$\times$         &3.40&$-0.29$ &4.83&\textbf{98.60}/0.08/0.04/1.28\\
                            &&$\times$&$\times$ &$\times$         &3.70&$-1.45$ &2.63&\textbf{97.20}/0.15/0.09/2.57\\
                            &&$\times$&$\times$ &$\times$         &4.00&$-3.81$ &1.72&\textbf{95.77}/0.20/0.13/3.89\\\midrule[1.0pt]
System&$I(J^P)$&$\Lambda$ &$E$ &$r_{\rm RMS}$&$\Lambda$ &$E$ &$r_{\rm RMS}$&P(${}^2\mathbb{S}_{\frac{1}{2}}/{}^4\mathbb{D}_{\frac{1}{2}}/{}^6\mathbb{D}_{\frac{1}{2}}$)\\
\multirow{27}{*}{$\Sigma_{c}^{*} D_{2}^{*}$}&$\frac{1}{2}(\frac{1}{2}^+)$&1.32&$-0.37$ &4.80             &1.20&$-0.34$ &4.96&\textbf{97.88}/1.00/1.11\\
                            &&1.49&$-4.57$ &1.89             &1.39&$-4.46$&1.94&\textbf{95.67}/2.08/2.25\\
                            &&1.65&$-12.45$ &1.33            &1.57&$-12.10$ &1.37&\textbf{95.85}/2.03/2.11\\
&$\frac{3}{2}(\frac{1}{2}^+)$&1.16&$-0.29$ &4.60            &1.10&$-0.26$ &4.78&\textbf{99.59}/0.21/0.20\\
                            &&1.32&$-4.35$ &1.53            &1.27&$-4.36$&1.56&\textbf{99.13}/0.44/0.43\\
                            &&1.47&$-12.20$ &0.98            &1.42&$-12.51$ &1.00&\textbf{98.95}/0.54/0.51\\
                            \cline{2-9}
&$I(J^P)$&$\Lambda$ &$E$  &$r_{\rm RMS}$ &$\Lambda$ &$E$  &$r_{\rm RMS}$ &P(${}^4\mathbb{S}_{\frac{3}{2}}/{}^2\mathbb{D}_{\frac{3}{2}}/{}^4\mathbb{D}_{\frac{3}{2}}/{}^6\mathbb{D}_{\frac{3}{2}}/{}^8\mathbb{D}_{\frac{3}{2}}$) \\
&$\frac{1}{2}(\frac{3}{2}^+)$&1.29&$-0.34$ &4.82           &1.18&$-0.33$ &4.90&\textbf{98.09}/0.54/0.74/0.16/0.48\\
                            &&1.46&$-4.65$ &1.81          &1.37&$-4.78$ &1.83&\textbf{96.14}/1.10/1.48/0.31/0.97\\
                            &&1.63&$-12.88$ &1.26        &1.55&$-12.47$ &1.30&\textbf{96.37}/1.07/1.42/0.29/0.86\\
&$\frac{3}{2}(\frac{3}{2}^+)$&1.43&$-0.27$ &4.70          &1.33&$-0.26$ &4.78&\textbf{99.51}/0.13/0.22/0.05/0.09\\
                            &&1.66&$-4.28$ &1.53          &1.54&$-4.18$ &1.58&\textbf{98.86}/0.31/0.53/0.11/0.20\\
                            &&1.88&$-12.39$ &0.96          &1.74&$-12.20$ &0.99&\textbf{98.53}/0.40/0.69/0.14/0.25\\
                            \cline{2-9}
&$I(J^P)$&$\Lambda$ &$E$ &$r_{\rm RMS}$&$\Lambda$ &$E$ &$r_{\rm RMS}$&P(${}^6\mathbb{S}_{\frac{5}{2}}/{}^2\mathbb{D}_{\frac{5}{2}}/{}^4\mathbb{D}_{\frac{5}{2}}/~{}^6\mathbb{D}_{\frac{5}{2}}/{}^8\mathbb{D}_{\frac{5}{2}}$) \\
&$\frac{1}{2}(\frac{5}{2}^+)$&1.10&$-0.38$ &4.49             &1.04&$-0.36$ &4.60&\textbf{98.33}/0.41/0.11/1.11/0.03\\
                            &&1.18&$-4.77$ &1.63             &1.13&$-4.65$ &1.70&\textbf{97.00}/0.73/0.20/1.99/0.07\\
                            &&1.25&$-11.96$ &1.13            &1.21&$-12.08$ &1.17&\textbf{97.07}/0.70/0.19/1.96/0.07\\
&$\frac{3}{2}(\frac{5}{2}^+)$&3.50&$-0.31$ &4.61             &2.07&$-0.26$ &4.85&\textbf{99.18}/0.12/0.05/0.64/0.02\\
                            &&3.80&$-0.71$ &3.43             &2.49&$-3.83$ &1.66&\textbf{97.66}/0.32/0.13/1.84/0.05\\
                            &&4.00&$-1.05$ &2.90            &2.91&$-11.45$ &1.03&\textbf{96.48}/0.46/0.19/2.79/0.08\\
                            \cline{2-9}
&$I(J^P)$&$\Lambda$ &$E$  &$r_{\rm RMS}$ &$\Lambda$ &$E$  &$r_{\rm RMS}$ &P(${}^8\mathbb{S}_{\frac{7}{2}}/{}^4\mathbb{D}_{\frac{7}{2}}/{}^6\mathbb{D}_{\frac{7}{2}}/{}^8\mathbb{D}_{\frac{7}{2}}$)\\
&$\frac{1}{2}(\frac{7}{2}^+)$&0.86&$-0.27$ &4.69               &0.83&$-0.47$ &4.06&\textbf{98.36}/0.19/0.02/1.42\\
                            &&0.90&$-4.68$ &1.50               &0.88&$-5.04$ &1.53&\textbf{97.52}/0.29/0.03/2.16\\
                            &&0.93&$-12.09$ &1.02              &0.92&$-13.93$ &1.01&\textbf{97.67}/0.26/0.03/2.04\\
&$\frac{3}{2}(\frac{7}{2}^+)$&$\times$&$\times$ &$\times$             &3.40&$-0.34$ &4.62&\textbf{98.32}/0.07/0.02/1.59\\
                            &&$\times$&$\times$ &$\times$             &3.70&$-1.66$ &2.48&\textbf{96.62}/0.13/0.05/3.20\\
                            &&$\times$&$\times$ &$\times$             &4.00&$-4.36$ &1.62&\textbf{94.89}/0.18/0.07/4.85\\
\bottomrule[1.5pt]
\end{tabular}
\end{table*}

For the $\Sigma_{c} D_1$ state with $I(J^{P})=1/2(3/2^{+})$ and the $\Sigma_{c} D_2^*$ state with $I(J^{P})=1/2(5/2^{+})$, the effective potentials of the $\sigma$, $\pi$, $\eta$, $\rho$, and $\omega$ exchanges all supply attractive forces, and therefore we can obtain loosely bound state solutions with reasonable cutoff values according to the experience of the deuteron \cite{Tornqvist:1993ng,Tornqvist:1993vu}, and thus they can be regarded as good candidates of double-charm molecular pentaquarks. For the $\Sigma_{c} D_1$ state with $I(J^{P})=1/2(1/2^{+})$ and the $\Sigma_{c} D_2^*$ state with $I(J^{P})=1/2(3/2^{+})$, the $\pi$ exchange potentials provide repulsive forces in the whole range, but the total effective potentials are strongly repulsive in the range from 0.1 fm to 0.4 fm and slightly attractive in the range $0.4 < r < 1.7~\rm {fm}$. Our results suggest that there exist bound state solutions with cutoffs around 1.3 GeV as indicated in Table \ref{jg1bu}, and the attractive force that binds these states mainly comes from the $\rho$ and $\omega$ exchanges. For the $\Sigma_{c} D_1$ state with $I(J^{P})=3/2(3/2^{+})$ and the $\Sigma_{c} D_2^*$ state with $I(J^{P})=3/2(5/2^{+})$, the $\pi$ and $\rho$ exchange potentials are repulsive while the $\sigma$, $\eta$, and $\omega$ exchanges provide the attractive force; nevertheless the contributions of the $\rho$ and $\omega$ exchanges cancel each other almost exactly, and thus the total effective potentials are repulsive in the whole range and dominated by the $\pi$ exchange potentials. In most cases of the $\Sigma_{c}D_{1}(D_{2}^{*})$ systems and the $\Sigma_{c}^{*} D_{1}(D_{2}^{*})$ systems, if we take the same cutoff parameters, it is clear that the numerical results within the OBE model are similar to those within the OPE model except that the bindings become deeper.

From Tables \ref{jg1bu} and \ref{jg2bu}, we notice there are some similarities between the $\Sigma_{c} D_{1}(D_{2}^{*})$. For instance, in the $\Sigma_{c}^* D_1$ system with $I(J^{P})=1/2(1/2^{+},3/2^{+})$ and the $\Sigma_{c}^* D_2^*$ system with $I(J^{P})=1/2(1/2^{+},3/2^{+},5/2^{+})$, the bound states can also be obtained with cutoff parameters of about 1.3 GeV.

Compared with the results with the OPE model in Table \ref{jg2pu}, the $\Sigma_{c}^* D_{1}(D_{2}^{*})$ states with the OBE model are more likely to be bound in many of the cases from Table \ref{jg2bu}. The main reason is that the $\sigma$ exchange interaction always provides an attractive force, and the vector meson exchange effective potential usually plays an essential role in forming the loosely bound molecular states for many of the $\Sigma_{c}^* D_{1}(D_{2}^{*})$ states.

To summarize, we can predict a series of possible candidates of double-charm molecular pentaquarks with the structures of $\Sigma_{c}^{(*)} T$ as shown in Tables~\ref{jg2pu}-\ref{jg2bu}. If we use the same criteria as the $\Sigma_{c}^{(*)} \bar T$-type hidden-charm molecular pentaquarks, we can also present the ratings for these double-charm molecular pentaquarks in Table \ref{Summary2}, which provides crucial information for further experimental search. According to our calculations, we find that there exist ideal candidates of the double-charm molecular pentaquarks, such as the $\Sigma_{c}  D_2^*$ state with $I(J^{P})=1/2(5/2^{+})$, the $\Sigma_c^* {D}_1$ states with $I(J^P)=1/2(5/2^+), 3/2(1/2^+)$, and the $\Sigma_c^*\bar{D}_2^*$ states with $I(J^P)=1/2(7/2^+), 3/2(1/2^+)$.

\renewcommand\tabcolsep{0.16cm}
\renewcommand{\arraystretch}{1.50}
\begin{table}[!htbp]
\caption{Rating for predicted double-charm molecular pentaquarks with $\Sigma_{c}^{(*)} D_{1}(D_{2}^{*})$ systems $(I,J)$.}\label{Summary2}
\begin{tabular}{c|cccc}\toprule[1.5pt]
Rating                 &$\Sigma_{c} D_{1}$            &$\Sigma_{c} D_{2}^{*}$                &$\Sigma_{c}^{*} D_{1}$                 &$\Sigma_{c}^{*} D_{2}^{*}$\\\midrule[1.0pt]
$\star\star\star\star$
&     &$(\frac{1}{2},\frac{5}{2})$&$(\frac{1}{2},\frac{5}{2})$,$(\frac{3}{2},\frac{1}{2})$& $(\frac{1}{2},\frac{7}{2})$,$(\frac{3}{2},\frac{1}{2})$\\
$\star\star\star$
& $(\frac{1}{2},\frac{3}{2})$  &  &  & \\
$\star\star$
&$(\frac{1}{2}/\frac{3}{2},\frac{1}{2})$&$(\frac{1}{2}/\frac{3}{2},\frac{3}{2})$& $(\frac{1}{2},\frac{1}{2}/\frac{3}{2})$& $(\frac{1}{2},\frac{1}{2}/\frac{3}{2}/\frac{5}{2})$,$(\frac{3}{2},\frac{3}{2})$ \\
$\star$
&          &      &                                &  \\
\bottomrule[1.5pt]
\end{tabular}
\end{table}

\subsection{$\Lambda_{c} D_{1}(D_{2}^{*})$ systems}

We also extend our work to investigate the bound properties of the $\Lambda_{c} D_{1}(D_{2}^{*})$ systems. Similarly to the $\Lambda_{c} \bar D_{1}(\bar D_{2}^{*})$ systems, we also mainly focus on the results of the coupled channel effects in forming these bound states within the OBE model. For the $\Lambda_{c} D_{1}(D_{2}^{*})$ systems, only the $\sigma$ and  $\omega$ exchanges are allowed and both provide attractive potentials. They enhance each other and lead to a strong interaction. By solving the coupled channel Schr\"odinger equation, we can obtain bound solutions for the $\Lambda_{c} D_{1}$ states with $I(J^{P})=1/2(1/2^{+},3/2^{+})$ and for the $\Lambda_{c} D_{2}^*$ states with $I(J^{P})=1/2(3/2^{+},5/2^{+})$, the corresponding numerical results are presented in Table \ref{jg3u}. Thus, we conclude that these four states are possible candidates of double-charm molecular pentaquarks with coupled channel effects.

\renewcommand\tabcolsep{0.23cm}
\renewcommand{\arraystretch}{1.40}
\begin{table}[!tbp]
\caption{Bound state properties for the $\Lambda_{c} D_{1}( D_{2}^{*})$ systems. Conventions are the same as Table~\ref{jg1p}.}\label{jg3u}
\begin{tabular}{c|ccc|ccc}\toprule[1.0pt]
Systems&$\Lambda$ &$E$  &$r_{\rm RMS}$&$\Lambda$ &$E$  &$r_{\rm RMS}$\\\midrule[1.0pt]
\multirow{4}{*}{$\Lambda_{c} D_{1}$}&\multicolumn{3}{c|}{$(J=1/2,I=1/2)$}&\multicolumn{3}{c}{$(J=3/2,I=1/2)$}\\
&0.96&$-0.45$ &4.15       &0.82&$-0.30$ &4.63\\
&1.00&$-4.30$&1.56        &0.86&$-4.40$&1.52\\
&1.04&$-12.25$ &0.98      &0.90&$-12.96$ &0.94\\\hline
\multirow{4}{*}{$\Lambda_{c} D_{2}^{*}$}&\multicolumn{3}{c|}{$(J=3/2,I=1/2)$}&\multicolumn{3}{c}{$(J=5/2,I=1/2)$}\\
&1.07&$-0.59$ &3.73       &1.09&$-0.24$ &4.94\\
&1.09&$-5.03$&1.31        &1.11&$-4.24$&1.42\\
&1.10&$-10.44$ &0.86      &1.12&$-12.00$ &0.73\\
\bottomrule[1.0pt]
\end{tabular}
\end{table}

\section{Summary}\label{sec6}

Very recently, the LHCb Collaboration updated the observations of $P_c$ states by analyzing $\Lambda_b^0\to J/\psi pK^-$. They not only reported a new structure, $P_c(4312)$, but also found that the $P_c(4450)$ previously reported consist of two narrow overlapping peaks, $P_c(4440)$ and $P_c(4457)$ \cite{Aaij:2019vzc}. These $P_c$ states can be naturally interpreted as hidden-charm molecular pentaquarks composed of an $S$-wave charmed baryon and an $S$-wave anticharmed meson since the masses of the three $P_c$ states are just below the $\Sigma_c \bar D^{(*)}$ thresholds \cite{Chen:2019asm}.

In this work, we try to search for possible new types of $P_c$ states composed of an $S$-wave charmed baryon and an anticharmed meson in a $T$-doublet with the structures of $\mathcal{B}_{c}^{(*)}\bar T$. In the present study, both the $S$-$D$ wave mixing effects and the coupled channel effects are taken into consideration.  As for the unknown coupling constants, we estimate them within the quark model. Our results indicate that there exist ideal candidates of new types of $P_c$ states, the $\Sigma_c\bar{D}_1$ state with $I(J^P)=1/2(1/2^+)$, the $\Sigma_c\bar{D}_2^*$ state with $I(J^P)=1/2(3/2^+)$, the $\Sigma_c^*\bar{D}_1$ state with $I(J^P)=1/2(1/2^+)$, and the $\Sigma_c^*\bar{D}_2^*$ states with $I(J^P)=1/2(1/2^+, 3/2^+)$. Moreover, the low spin $\Sigma_{c}^{(*)}\bar T$ states with $I=1/2$ configuration can be more stable than the $I=3/2$ states with the same spin-parity configuration, and we also find that the coupled channel effects play an essential role for the $\Lambda_{c} \bar T$ systems; especially the $\Lambda_{c}\bar D_{1}$ states with $I(J^P)=1/2(1/2^+, 3/2^+)$ and $\Lambda_{c}\bar D_{2}^{*}$ states with $I(J^P)=1/2(3/2^+, 5/2^+)$ disappear without considering the coupled channel effects. Besides the above predictions, we extend our study to the interactions between an $S$-wave ground charmed baryon and a charmed meson in a $T$ doublet, and can predict the existence of double-charm molecular pentaquarks with typical quark configuration $ccqq\bar q$.

Experimental searches for these predicted new types of $P_c$ states are an interesting research topic. Obviously, we suggest that the $\Lambda_b$ baryon decay is an ideal process to produce these predicted new types of $P_c$ states, which can decay into a ground charmed baryon with a ground anticharmed meson, and a charmonium with a light baryon if kinetically allowed, like $\Lambda_c \bar D^{(*)}$, $\Sigma_c^{(*)} \bar D^{(*)}$, $\eta_c(nS)N/\Delta~(n\leq2)$, $\psi(nS)N/\Delta~(n\leq2)$, $\chi_{cJ}(1P)N/\Delta~(J\leq2)$. Experimental searches for these possible new types of $P_c$ states can be also a crucial test of hadronic molecular state assignments to the $P_c$ states. The $P_c$ states are observed in the $\Lambda_b^0 \to J/\psi p \pi^{-}$ process by the LHCb Collaboration \cite{Aaij:2019vzc}, and thus it would provide a sign for the success of our work if these predicted states are detected in the process $\Lambda_b^0 \to \psi(2S) p \pi^{-}$ since these and $P_c$ states have similar structures in the molecule picture. Actually, the decay $\Lambda_b^0 \to \psi(2S) p \pi^{-}$ has been reported in the search for hidden-charm pentaquark states with higher mass by the LHCb experiment \cite{Aaij:2018jlf}. Unfortunately, no significant exotic pentaquarks are observed due to the bin size of experimental events being a little large. As commonly believed, a resonance could not be identified in such rough data, especially a narrow resonance. Thus, we hope further experiments can provide a more precise measurement of the process $\Lambda_b^0 \to \psi(2S) p \pi^{-}$, and the existence of more $P_c$ states with larger mass should be revealed in the $\psi(2S) p$ invariant mass spectrum, especially for the low spin $\Sigma_{c}^{(*)}\bar D_{1}(\bar D_{2}^{*})$ with $I=1/2$ states.

The possible doubly charmed molecular pentaquarks with the structures of $\mathcal{B}_{c}^{(*)} T$ can be searched for in their possible two-body strong decay channels
\begin{eqnarray}\label{decay}
\Lambda_c D^{(*)},\, \Sigma_c^{(*)} D^{(*)},\, \Xi_{cc} \pi,\, \Xi_{cc} \rho,\, \Xi_{cc} \omega,\, \Xi_{cc} \eta^{(\prime)},\nonumber
\end{eqnarray}
which helps to further explore them in experiment. These doubly charmed molecular pentaquarks may be produced via proton-proton collisions, and we hope experimental colleagues will bring us more surprises after the LHCb observations of three $P_c$ states \cite{Aaij:2019vzc}. Nevertheless, we notice that no open-charm multi quark states have been reported up to now. Experimental searched for these possible doubly-charmed molecular pentaquarks are  still difficult compared with the hidden-charm exotic hadronic states.

Since 2003, the observations of abundant charmoniumlike $XYZ$ states have stimulated extensive discussion of the interactions between charmed mesons and anticharmed mesons \cite{Chen:2016qju}. With the observation of three $P_c$ states by LHCb, searching for heavy flavor pentaquarks will become a hot topic of hadron physics; more opportunities and challenges are waiting for us. Different from $P_c$ states composed of an $S$-wave charmed baryon and an $S$-wave anticharmed meson \cite{Liu:2019tjn}, we find that new types of $P_c$ states with the structures of $\mathcal{B}_{c}^{(*)}\bar T$ are more abundant due to different quantum number configurations, and we have reason to believe that these predicted new types of $P_c$ states can be accessible in future experiment in the next decades.

\appendix

\section{The detailed Lagrangians}\label{app01}

The detailed effective Lagrangians for anticharmed mesons in a $\bar T$-doublet and light mesons are expressed as
\begin{eqnarray}
\mathcal{L}_{\bar{T}\bar{T}\sigma} &=&-2g_\sigma^{\prime\prime}\bar{D}_{1a\mu}\bar{D}^{\mu\dagger}_{1a}\sigma
+2g_\sigma^{\prime\prime}\bar{D}^{*\dagger}_{2a\mu\nu}\bar{D}^{*\mu\nu}_{2a}\sigma,\\
%%%
\mathcal {L}_{\bar{T} \bar{T}\mathbb{P}} &=&-\frac{5ik}{3f_\pi}\varepsilon^{\mu\nu\rho\tau}v_\nu\bar{D}^{\dag}_{1a\rho}
\bar{D}_{1b\tau}\partial_\mu\mathbb{P}_{ba}\nonumber\\
    &&+\frac{2ik}{f_\pi}\varepsilon^{\mu\nu\rho\tau}v_\nu\bar{D}^{*\alpha\dag}_{2a\rho}
    \bar{D}^{*}_{2b\alpha\tau}\partial_\mu\mathbb{P}_{ba}\nonumber\\
    &&+\sqrt{\frac{2}{3}}\frac{k}{f_\pi}\left(\bar{D}^{\dagger}_{1a\mu}
    \bar{D}^{*\mu\lambda}_{2b}+\bar{D}_{1b\mu}
    \bar{D}^{*\mu\lambda\dagger}_{2a}\right)\partial_\lambda\mathbb{P}_{ba},\\
%%%
\mathcal {L}_{\bar{T}\bar{T}\mathbb{V}} &=&\sqrt{2}\beta^{\prime\prime}g_{V}\left(v\cdot\mathbb{V}_{ba}\right)\bar{D}_{1b\mu}
\bar{D}^{\mu\dagger}_{1a}\nonumber\\
    &&+\frac{5\sqrt{2}i\lambda^{\prime\prime}g_{V}}{3}\left(\bar{D}^{\nu}_{1b}
    \bar{D}^{\mu\dagger}_{1a}
    -\bar{D}^{\nu\dagger}_{1a}\bar{D}^{\mu}_{1b}\right)\partial_\mu\mathbb{V}_{ba\nu}\nonumber\\
    &&-\sqrt{2}\beta^{\prime \prime}g_{V}\left(v\cdot\mathbb{V}_{ba}\right) \bar{D}_{2b}^{*\lambda\nu} \bar{D}^{*\dagger}_{2a{\lambda\nu}}\nonumber\\
    &&+2\sqrt{2}i\lambda^{\prime\prime} g_{V}\left(\bar{D}^{*\lambda\nu\dagger}_{2a}
    \bar{D}^{*\mu}_{2b\lambda}-\bar{D}^{*\lambda\nu}_{2b} \bar{D}^{*\mu\dagger}_{2a\lambda}\right)\partial_\mu \mathbb{V}_{ba\nu}\nonumber\\
    &&+\frac{i\beta^{\prime\prime}g_{V}}{\sqrt{3}}\varepsilon^{\lambda\alpha\rho\tau}v_{\rho}
    \left(v\cdot\mathbb{V}_{ba}\right)\left(\bar{D}^{\dagger}_{1a\alpha} \bar{D}^{*}_{2b\lambda\tau}-\bar{D}_{1b\alpha}
    \bar{D}^{\dagger*}_{2a\lambda\tau}\right)\nonumber\\
    &&+\frac{2\lambda^{\prime\prime}g_{V}}{\sqrt{3}}
    \left[3\varepsilon^{\mu\lambda\nu\tau}v_\lambda\left(\bar{D}^{\alpha\dagger}_{1a}
    \bar{D}^{*}_{2b\alpha\tau}+\bar{D}^{\alpha}_{1b}
    \bar{D}^{*\dagger}_{2a\alpha\tau}\right)\partial_\mu\mathbb{V}_{ba\nu}\right.\nonumber\\
    &&\left.+2\varepsilon^{\lambda\alpha\rho\nu}v_\rho
    \left(\bar{D}^{\dagger}_{1a\alpha}\bar{D}^{*\mu}_{2b\lambda}
    +\bar{D}_{1b\alpha}\bar{D}^{\dagger\mu*}_{2a\lambda}\right)\right.\nonumber\\
    &&\left.\times\left(\partial_\mu \mathbb{V}_{ba\nu}-\partial_\nu \mathbb{V}_{ba\mu}\right)\right],
\end{eqnarray}
and those for $S$-wave charmed baryons and light mesons are
\begin{eqnarray}
\mathcal{L}_{\mathcal{B}_{\bar{3}}\mathcal{B}_{\bar{3}}\sigma} &=& l_B\langle \bar{\mathcal{B}}_{\bar{3}}\sigma\mathcal{B}_{\bar{3}}\rangle,\\
%%%
\mathcal{L}_{\mathcal{B}_{\bar{3}}\mathcal{B}_{\bar{3}}\mathbb{V}}&=&
\frac{1}{\sqrt{2}}\beta_Bg_V\langle\bar{\mathcal{B}}_{\bar{3}}v\cdot\mathbb{V}
\mathcal{B}_{\bar{3}}\rangle,\\
%%%
\mathcal{L}_{\mathcal{B}_{6}^{(*)}\mathcal{B}_{6}^{(*)}\sigma} &=&-l_S\langle\bar{\mathcal{B}}_6\sigma\mathcal{B}_6\rangle+l_S\langle
\bar{\mathcal{B}}_{6\mu}^{*}
\sigma\mathcal{B}_6^{*\mu}\rangle\nonumber\\
    &&-\frac{l_S}{\sqrt{3}}\langle\bar{\mathcal{B}}_{6\mu}^{*}\sigma
    \left(\gamma^{\mu}+v^{\mu}\right)\gamma^5\mathcal{B}_6\rangle+h.c.,\\
%%%
\mathcal{L}_{\mathcal{B}_6^{(*)}\mathcal{B}_6^{(*)}\mathbb{P}} &=&i\frac{g_1}{2f_{\pi}}\varepsilon^{\mu\nu\lambda\kappa}v_{\kappa}\langle\bar{\mathcal{B}}_6
\gamma_{\mu}\gamma_{\lambda}\partial_{\nu}\mathbb{P}\mathcal{B}_6\rangle\nonumber\\
    &&+i\frac{\sqrt{3}g_1}{2f_{\pi}}v_{\kappa}\varepsilon^{\mu\nu\lambda\kappa}
    \langle\bar{\mathcal{B}}_{6\mu}^*\partial_{\nu}\mathbb{P}{\gamma_{\lambda}\gamma^5}
      \mathcal{B}_6\rangle+h.c.\nonumber\\
    &&-i\frac{3g_1}{2f_{\pi}}\varepsilon^{\mu\nu\lambda\kappa}v_{\kappa}
    \langle\bar{\mathcal{B}}_{6\mu}^{*}\partial_{\nu}\mathbb{P}\mathcal{B}_{6\lambda}^*\rangle,\\
%%%
\mathcal{L}_{\mathcal{B}_6^{(*)}\mathcal{B}_6^{(*)}\mathbb{V}}&=&
-\frac{\beta_Sg_V}{\sqrt{2}}\langle\bar{\mathcal{B}}_6v\cdot\mathbb{V}
\mathcal{B}_6\rangle+\frac{\beta_Sg_V}{\sqrt{2}}\langle\bar{\mathcal{B}}_{6\mu}^*v\cdot {V}\mathcal{B}_6^{*\mu}\rangle\nonumber\\
    &&-i\frac{\lambda_S g_V}{3\sqrt{2}}\langle\bar{\mathcal{B}}_6\gamma_{\mu}\gamma_{\nu}
    \left(\partial^{\mu}\mathbb{V}^{\nu}-\partial^{\nu}\mathbb{V}^{\mu}\right)
    \mathcal{B}_6\rangle\nonumber\\
    &&+i\frac{\lambda_Sg_V}{\sqrt{2}}\langle\bar{\mathcal{B}}_{6\mu}^*
    \left(\partial^{\mu}\mathbb{V}^{\nu}-\partial^{\nu}\mathbb{V}^{\mu}\right)
    \mathcal{B}_{6\nu}^*\rangle+h.c.\nonumber\\
    &&-i\frac{\lambda_Sg_V}{\sqrt{6}}\langle\bar{\mathcal{B}}_{6\mu}^*
    \left(\partial^{\mu}\mathbb{V}^{\nu}-\partial^{\nu}\mathbb{V}^{\mu}\right)
    \left(\gamma_{\nu}+v_{\nu}\right)\gamma^5\mathcal{B}_6\rangle\nonumber\\
    &&-\frac{\beta_Sg_V}{\sqrt{6}}\langle\bar{\mathcal{B}}_{6\mu}^*v\cdot \mathbb{V}\left(\gamma^{\mu}+v^{\mu}\right)\gamma^5\mathcal{B}_6\rangle,\\
%%%
\mathcal{L}_{\mathcal{B}_{\bar{3}}\mathcal{B}_6^{(*)}\mathbb{P}} &=& -\sqrt{\frac{1}{3}}\frac{g_4}{f_{\pi}}\langle\bar{\mathcal{B}}_6\gamma^5\left(\gamma^{\mu}
+v^{\mu}\right)\partial_{\mu}\mathbb{P}\mathcal{B}_{\bar{3}}\rangle\nonumber\\
    &&-\frac{g_4}{f_{\pi}}\langle\bar{\mathcal{B}}_{6\mu}^*\partial^{\mu} \mathbb{P}\mathcal{B}_{\bar{3}}\rangle+h.c.,\\
%%%
\mathcal{L}_{\mathcal{B}_{\bar{3}}\mathcal{B}_6^{(*)}\mathbb{V}} &=&
-\frac{\lambda_Ig_V}{\sqrt{2}}\varepsilon^{\mu\nu\lambda\kappa}v_{\mu}\langle \bar{\mathcal{B}}_{6\nu}^*\left(\partial_{\lambda}\mathbb{V}_{\kappa}
-\partial_{\kappa}\mathbb{V}_{\lambda}\right)
          \mathcal{B}_{\bar{3}}\rangle+h.c.\nonumber\\
    &&-\frac{\lambda_Ig_V}{\sqrt{6}}\varepsilon^{\mu\nu\lambda\kappa}v_{\mu}\langle \bar{\mathcal{B}}_6\gamma^5\gamma_{\nu}
        \left(\partial_{\lambda}\mathbb{V}_{\kappa}-\partial_{\kappa} \mathbb{V}_{\lambda}\right)\mathcal{B}_{\bar{3}}\rangle.\nonumber\\
\end{eqnarray}

\section{Relevant subpotentials}\label{app02}

For convenience, we define the following functions
\begin{eqnarray}
&&\mathcal{H}(I=1/2)Y(\Lambda,m_P,r)=-Y(\Lambda,m_\pi,r)
+\frac{1}{6}Y(\Lambda,m_\eta,r),\nonumber\\
&&\mathcal{G}(I=1/2)Y(\Lambda,m_V,r)=-Y(\Lambda,m_\rho,r)
+\frac{1}{2}Y(\Lambda,m_\omega,r),\nonumber\\
&&\mathcal{H}(I=3/2)Y(\Lambda,m_P,r)=\frac{1}{2}Y(\Lambda,m_\pi,r)
+\frac{1}{6}Y(\Lambda,m_\eta,r),\nonumber\\
&&\mathcal{G}(I=3/2)Y(\Lambda,m_V,r)=\frac{1}{2}Y(\Lambda,m_\rho,r)
+\frac{1}{2}Y(\Lambda,m_\omega,r),\nonumber\\
&&Y_E\equiv Y(\Lambda,m_E,r)=\frac{1}{4\pi r}(e^{-m_Er}-e^{-\Lambda r})-\frac{\Lambda^2-m_E^2}{8 \pi\Lambda}e^{-\Lambda r},\nonumber\\
\end{eqnarray}
respectively. Here, $\mathcal{H}(I)$ and $\mathcal{G}(I)$ are the isospin factors for the $\Sigma_{c}^{(*)}\bar D_{1}(\bar D_{2}^{*})$ systems, and $I$ denotes the isospin. Additionally, we also define several variables:  $A=l_{B}g_{\sigma}^{\prime\prime}$, $B=\beta_B \beta^{\prime\prime} g_{V}^2$, $C=l_{S}g_{\sigma}^{\prime\prime}$, $D=g_1 k/f_\pi^2$, $E=\beta_S \beta^{\prime\prime} g_{V}^2$, $F=\lambda_S \lambda^{\prime\prime}g_V^2$, $G=g_4 k/f_\pi^2$, and $H=\lambda_I \lambda^{\prime\prime}g_V^2$.

\subsection{The effective potentials involved in $\Sigma_{c}^{(*)}\bar D_{1}(\bar D_{2}^{*})$ systems}\label{app0201}

With the above preparation, the OBE effective potentials in the analysis of the $\Sigma_{c}^{(*)}\bar D_{1}(\bar D_{2}^{*})$ systems are given by
\begin{eqnarray}\label{eq1}
\mathcal{V}^{\Sigma_{c}\bar D_{1}\rightarrow\Sigma_{c}\bar D_{1}}({\bf{r}})&=&CY_\sigma\mathcal{A}_1+\frac{1}{2}E\mathcal{G}(I)Y_V\mathcal{A}_1\nonumber\\
&&+\frac{5}{18}D\mathcal{H}(I)\left[\mathcal{A}_2\nabla^2+\mathcal{A}_3r\frac{\partial}{\partial r}\frac{1}{r}\frac{\partial}{\partial r}\right]Y_P\nonumber\\
&&-\frac{5}{27}F\mathcal{G}(I)\left[2\mathcal{A}_2\nabla^2-\mathcal{A}_3 r\frac{\partial}{\partial r}\frac{1}{r}\frac{\partial}{\partial r}\right]Y_V,\nonumber\\\\
%%%%%%%%
\mathcal{V}^{\Sigma_{c}\bar D_{2}^{*}\rightarrow\Sigma_{c}\bar D_{2}^{*}}({\bf{r}})&=&CY_\sigma\mathcal{A}_4+\frac{1}{2}E\mathcal{G}(I)Y_V\mathcal{A}_4
\nonumber\\
&&+\frac{1}{3}D\mathcal{H}(I)\left[\mathcal{A}_5\nabla^2+\mathcal{A}_6r\frac{\partial}{\partial r}\frac{1}{r}\frac{\partial}{\partial r}\right]Y_P\nonumber\\
&&-\frac{2}{9}F\mathcal{G}(I)\left[2\mathcal{A}_5\nabla^2-\mathcal{A}_6 r\frac{\partial}{\partial r}\frac{1}{r}\frac{\partial}{\partial r}\right]Y_V,\nonumber\\\\
%%%%%%%
\mathcal{V}^{\Sigma_{c}^{*}\bar D_{1}\rightarrow\Sigma_{c}^{*}\bar D_{1}}({\bf{r}})&=&CY_\sigma\mathcal{A}_7+\frac{1}{2}E\mathcal{G}(I)Y_V\mathcal{A}_7\nonumber\\
&&+\frac{5}{12}D\mathcal{H}(I)\left[\mathcal{A}_8\nabla^2+\mathcal{A}_9r\frac{\partial}{\partial r}\frac{1}{r}\frac{\partial}{\partial r}\right]Y_P\nonumber\\
&&-\frac{5}{18}F\mathcal{G}(I)\left[2\mathcal{A}_8\nabla^2-\mathcal{A}_9 r\frac{\partial}{\partial r}\frac{1}{r}\frac{\partial}{\partial r}\right]Y_V,\nonumber\\\\
%%%%%%%%
\mathcal{V}^{\Sigma_{c}^{*}\bar D_{2}^{*}\rightarrow\Sigma_{c}^{*}\bar D_{2}^{*}}({\bf{r}})&=&CY_\sigma\mathcal{A}_{10}+\frac{1}{2}E\mathcal{G}(I)Y_V
\mathcal{A}_{10}\nonumber\\
&&+\frac{1}{2}D\mathcal{H}(I)\left[\mathcal{A}_{11}\nabla^2+\mathcal{A}_{12}
r\frac{\partial}{\partial r}\frac{1}{r}\frac{\partial}{\partial r}\right]Y_P\nonumber\\
&&-\frac{1}{3}F\mathcal{G}(I)\left[2\mathcal{A}_{11}\nabla^2-\mathcal{A}_{12}
r\frac{\partial}{\partial r}\frac{1}{r}\frac{\partial}{\partial r}\right]Y_V.\nonumber\\\label{eq2}
\end{eqnarray}

In Eqs. (\ref{eq1})-(\ref{eq2}), we define several operators, i.e.,
\begin{eqnarray*}\label{op}
\mathcal{A}_{1}&=&{\bm\epsilon^{\dagger}_4}\cdot{\bm\epsilon_2}\chi^{\dagger}_3\chi_1,\nonumber\\
\mathcal{A}_{2}&=&\chi^{\dagger}_3 \left[{\bm\sigma}\cdot\left(i{\bm\epsilon^{\dagger}_4}\times {\bm\epsilon_2}\right)\right]\chi_1,\nonumber\\
\mathcal{A}_{3}&=&\chi^{\dagger}_3S({\bm\sigma},i{\bm\epsilon^{\dagger}_4}
\times{\bm\epsilon_2},\hat{\bm{r}})\chi_1,\nonumber\\
\mathcal{A}_{4}&=&\mathcal{X}\mathcal{Y}\left({\bm\epsilon^{\dagger}_{4m}}
\cdot{\bm\epsilon_{2a}}\right)\left({\bm\epsilon^{\dagger}_{4n}}\cdot {\bm\epsilon_{2b}}\right)\chi^{\dagger}_3\chi_1,\nonumber\\
\mathcal{A}_{5}&=&\mathcal{X}\mathcal{Y}\left({\bm\epsilon^{\dagger}_{4m}}
\cdot{\bm\epsilon_{2a}}\right)\chi^{\dagger}_3[{\bm\sigma}\cdot(i{\bm\epsilon^{\dagger}_{4n}}
\times {\bm\epsilon_{2b}})]\chi_1,\nonumber\\
\mathcal{A}_{6}&=&\mathcal{X}\mathcal{Y}\left({\bm\epsilon^{\dagger}_{4m}}
\cdot{\bm\epsilon_{2a}}\right)\chi^{\dagger}_3S({\bm\sigma},i{\bm \epsilon^{\dagger}_{4n}}\times{\bm\epsilon_{2b}},\hat{\bm{r}})\chi_1,\nonumber\\
\mathcal{A}_{7}&=&\mathcal{Z}\mathcal{T}\mathcal{Q}\left({\bm\epsilon^{\dagger m_2^\prime}_{3}}\cdot{\bm\epsilon_{1}}^{m_2}\right)\left({\bm\epsilon^{\dagger}_{4}}\cdot {\bm\epsilon_{2}}\right),\nonumber\\
\mathcal{A}_{8}&=&\mathcal{Z}\mathcal{T}\mathcal{Q}\left[\left({\bm\epsilon^{\dagger m_2^\prime}_{3}}\times{\bm\epsilon_{1}}^{m_2}\right)\cdot\left({\bm\epsilon^{\dagger}_{4}}\times {\bm\epsilon_{2}}\right)\right],\nonumber\\
\mathcal{A}_{9}&=&\mathcal{Z}\mathcal{T}\mathcal{Q}S({\bm\epsilon^{\dagger m_2^\prime}_{3}}\times{\bm\epsilon_{1}}^{m_2},{\bm\epsilon^{\dagger}_{4}}\times {\bm\epsilon_{2}},\hat{\bm{r}}),\nonumber\\
\mathcal{A}_{10}&=&\mathcal{X}\mathcal{Y}\mathcal{Z}\mathcal{T}\mathcal{Q}
\left({\bm\epsilon^{\dagger m_2^\prime}_{3}}\cdot{\bm\epsilon_{1}}^{m_2}\right)\left({\bm\epsilon^{\dagger}_{4m}}
\cdot{\bm\epsilon_{2a}}\right)\left({\bm\epsilon^{\dagger}_{4n}}\cdot {\bm\epsilon_{2b}}\right),\nonumber\\
\mathcal{A}_{11}&=&\mathcal{X}\mathcal{Y}\mathcal{Z}\mathcal{T}\mathcal{Q}
\left({\bm\epsilon^{\dagger}_{4m}}
\cdot{\bm\epsilon_{2a}}\right)\left[\left({\bm\epsilon^{\dagger m_2^\prime}_{3}}\times{\bm\epsilon_{1}}^{m_2}\right)\cdot\left({\bm\epsilon^{\dagger}_{4n}}\times {\bm\epsilon_{2b}}\right)\right],\nonumber\\
\mathcal{A}_{12}&=&\mathcal{X}\mathcal{Y}\mathcal{Z}\mathcal{T}\mathcal{Q}
\left({\bm\epsilon^{\dagger}_{4m}}\cdot{\bm\epsilon_{2a}}\right)S({\bm\epsilon^{\dagger m_2^\prime}_{3}}\times{\bm\epsilon_{1}}^{m_2},{\bm\epsilon^{\dagger}_{4n}}\times {\bm\epsilon_{2b}},\hat{\bm{r}}).\nonumber\\
\end{eqnarray*}
Here, $\mathcal{X}=\sum_{m,n}C^{2,m+n}_{1m,1n}$, $\mathcal{Y}=\sum_{a,b}C^{2,a+b}_{1a,1b}$, $\mathcal{Z}=\sum_{m_1,m_2}C^{\frac{3}{2},m_1+m_2}_{\frac{1}{2}m_1,1m_2}$, $\mathcal{T}=\sum_{m_1^\prime,m_2^\prime}C^{\frac{3}{2},m_1^\prime+m_2^\prime}_{\frac{1}{2}m_1^\prime,1m_2^\prime}$, $\mathcal{Q}=\chi^{\dagger m_1^\prime}_3\chi_1^{m_1}$, and $S({\bm x},{\bm y},\hat{\bm r})= 3\left(\hat{\bm r} \cdot {\bm x}\right)\left(\hat{\bm r} \cdot {\bm y}\right)-{\bm x} \cdot {\bm y}$. We present the corresponding matrix elements $\langle f|\mathcal{A}_k|i\rangle$ in Table~\ref{matrix1}, which are obtained by sandwiching these operators between the relevant spin-orbit wave functions.
\renewcommand\tabcolsep{0.01cm}
\renewcommand{\arraystretch}{1.80}
\begin{table*}[!htbp]
  \caption{Matrix elements $\langle f|\mathcal{A}_k|i\rangle$ in various channels for operators $\mathcal{A}_k$ in the effective potentials.  \label{matrix1}}
  \begin{tabular}{c|cccc}\toprule[1pt]
   {{{Spin}}}
                    & $J=1/2$     & $J=3/2$     & $J=5/2$  & $J=7/2$ \\\midrule[1pt]
 $\langle\mathcal{A}_1\rangle$
            &diag(1,1) &diag(1,1,1) &$/$&$/$\\
 $\langle\mathcal{A}_2\rangle$
            &diag(2,$-1$) &diag($-1$,2,$-1$) &$/$&$/$\\
 $\langle\mathcal{A}_3\rangle$
           &$\left(\begin{array}{cc} 0 & \sqrt{2} \\ \sqrt{2} & 2\end{array}\right)$&$\left(\begin{array}{ccc} 0 & -1& -2 \\ -1 & 0& 1 \\ -2 & 1& 0 \end{array}\right)$&$/$&$/$\\
 $\langle\mathcal{A}_4\rangle$
            &$/$ &diag(1,1,1) &diag(1,1,1)&$/$\\
 $\langle\mathcal{A}_5\rangle$
            &$/$ &diag($\frac{3}{2}$,$\frac{3}{2}$,$-1$) &diag($-1$,$\frac{3}{2}$,$-1$)&$/$\\
 $\langle\mathcal{A}_6\rangle$
           &$/$
           &$\left(\begin{array}{ccc} 0 & \frac{3}{5}& \frac{3\sqrt{21}}{10} \\ \frac{3}{5} & 0& \frac{3\sqrt{21}}{14} \\ \frac{3\sqrt{21}}{10} & \frac{3\sqrt{21}}{14}& \frac{4}{7} \end{array}\right)$
           &$\left(\begin{array}{ccc} 0 & -\frac{3\sqrt{14}}{10}& -\frac{2\sqrt{14}}{5} \\ -\frac{3\sqrt{14}}{10} & \frac{3}{7}& \frac{3}{7} \\ -\frac{2\sqrt{14}}{5} & \frac{3}{7}& -\frac{4}{7} \end{array}\right)$&$/$\\
 $\langle\mathcal{A}_7\rangle$
            &diag(1,1,1) &diag(1,1,1,1) &diag(1,1,1,1)&$/$\\
 $\langle\mathcal{A}_8\rangle$
            &diag($\frac{5}{3}$,$\frac{2}{3}$,$-1$) &diag($\frac{2}{3}$,$\frac{5}{3}$,$\frac{2}{3}$,$-1$) &diag($-1$,$\frac{5}{3}$,$\frac{2}{3}$,$-1$)&$/$  \\
 $\langle\mathcal{A}_9\rangle$
           &$\left(\begin{array}{ccc} 0 & -\frac{7}{3\sqrt{5}}& \frac{2}{\sqrt{5}} \\ -\frac{7}{3\sqrt{5}} & \frac{16}{15}& -\frac{1}{5} \\ \frac{2}{\sqrt{5}} &-\frac{1}{5}& \frac{8}{5} \end{array}\right)$
           &$\left(\begin{array}{cccc} 0 & \frac{7}{3\sqrt{10}}& -\frac{16}{15}& -\frac{\sqrt{7}}{5\sqrt{2}}\\ \frac{7}{3\sqrt{10}} & 0& -\frac{7}{3\sqrt{10}} & -\frac{2}{\sqrt{35}} \\ -\frac{16}{15} & -\frac{7}{3\sqrt{10}}& 0& -\frac{1}{\sqrt{14}} \\-\frac{\sqrt{7}}{5\sqrt{2}}&-\frac{2}{\sqrt{35}} &-\frac{1}{\sqrt{14}}&\frac{4}{7}\end{array}\right)$
           &$\left(\begin{array}{cccc} 0 & \frac{2}{\sqrt{15}}& \frac{\sqrt{7}}{5\sqrt{3}}& -\frac{2\sqrt{14}}{5}\\ \frac{2}{\sqrt{15}} & 0& \frac{\sqrt{7}}{3\sqrt{5}} & -\frac{4\sqrt{2}}{\sqrt{105}} \\ \frac{\sqrt{7}}{5\sqrt{3}} & \frac{\sqrt{7}}{3\sqrt{5}}& -\frac{16}{21}& -\frac{\sqrt{2}}{7\sqrt{3}} \\-\frac{2\sqrt{14}}{5}&-\frac{4\sqrt{2}}{\sqrt{105}} &-\frac{\sqrt{2}}{7\sqrt{3}}&-\frac{4}{7}\end{array}\right)$&$/$\\
 $\langle\mathcal{A}_{10}\rangle$
            &diag(1,1,1)  &diag(1,1,1,1,1)  &diag(1,1,1,1,1) &diag(1,1,1,1)  \\
 $\langle\mathcal{A}_{11}\rangle$
            &diag($\frac{3}{2}$,1,$\frac{1}{6}$)&diag(1,$\frac{3}{2}$,1,$\frac{1}{6}$,$-1$)
            &diag($\frac{1}{6}$,$\frac{3}{2}$,1,$\frac{1}{6}$,$-1$) &diag($-1$,1,$\frac{1}{6}$,$-1$)  \\
  $\langle\mathcal{A}_{12}\rangle$
   &$\left(\begin{array}{ccc} 0 & \frac{9}{10}& \frac{\sqrt{21}}{5} \\ \frac{9}{10} & \frac{4}{5}& \frac{3\sqrt{21}}{70} \\ \frac{\sqrt{21}}{5} & \frac{3\sqrt{21}}{70}& \frac{116}{105} \end{array}\right)$
           &$\left(\begin{array}{ccccc}0 & -\frac{9\sqrt{2}}{20} & -\frac{4}{5}& \frac{3\sqrt{6}}{20} & \frac{2\sqrt{2}}{5}\\ -\frac{9\sqrt{2}}{20} & 0& \frac{9\sqrt{2}}{20} & -\frac{\sqrt{3}}{5}& 0\\ -\frac{4}{5} & \frac{9\sqrt{2}}{20}& 0& \frac{3\sqrt{6}}{28}& -\frac{4\sqrt{2}}{35}\\ \frac{3\sqrt{6}}{20} & -\frac{\sqrt{3}}{5}&\frac{3\sqrt{6}}{28}& \frac{58}{147} & \frac{18\sqrt{3}}{245}\\ \frac{2\sqrt{2}}{5} & 0& -\frac{4\sqrt{2}}{35}& \frac{18\sqrt{3}}{245}& \frac{60}{49}\end{array}\right)$
           &$\left(\begin{array}{ccccc}0 & \frac{\sqrt{7}}{5} & -\frac{3}{10}& -\frac{29\sqrt{2}}{15\sqrt{7}} & \frac{\sqrt{6}}{5\sqrt{7}}\\ \frac{\sqrt{7}}{5} & 0& \frac{-9}{10\sqrt{7}} & -\frac{2\sqrt{2}}{5}& 0\\ -\frac{3}{10} & -\frac{9}{10\sqrt{7}}& -\frac{4}{7}& \frac{3}{7\sqrt{14}}& -\frac{12\sqrt{6}}{35\sqrt{7}}\\ -\frac{29\sqrt{2}}{15\sqrt{7}} & -\frac{2\sqrt{2}}{5}&\frac{3}{7\sqrt{14}}& -\frac{58}{147} & \frac{17\sqrt{3}}{245}\\ \frac{\sqrt{6}}{5\sqrt{7}} & 0& -\frac{12\sqrt{6}}{35\sqrt{7}}& \frac{17\sqrt{3}}{245}& \frac{10}{49}\end{array}\right)$
           &$\left(\begin{array}{cccc} 0 & \frac{2}{5}& -\frac{3}{5\sqrt{14}}& -\frac{2\sqrt{21}}{7}\\ \frac{2}{5} & \frac{8}{35}& -\frac{\sqrt{27}}{35\sqrt{14}} & -\frac{4\sqrt{21}}{49} \\ -\frac{3}{5\sqrt{14}} & -\frac{27}{35\sqrt{14}}& -\frac{493}{735}& \frac{\sqrt{3}}{49\sqrt{2}} \\-\frac{2\sqrt{21}}{7}&-\frac{4\sqrt{21}}{49} &\frac{\sqrt{3}}{49\sqrt{2}}&-\frac{32}{49}\end{array}\right)$\\
           \bottomrule[1pt]
  \end{tabular}
\end{table*}

\subsection{The effective potentials involved in $\Lambda_{c}\bar D_{1}(\bar D_{2}^{*})$ systems}\label{app0202}

The OBE effective potentials in the analysis of the $\Lambda_{c}\bar D_{1}(\bar D_{2}^{*})$ systems can be written as
\begin{eqnarray}\label{eq3}
\mathcal{V}^{\Lambda_{c}\bar D_{1}\rightarrow\Lambda_{c}\bar D_{1}}({\bf{r}})&=&-2AY_\sigma\mathcal{A}_1-\frac{1}{2}BY_\omega\mathcal{A}_1,\\
%%%%%%%%
\mathcal{V}^{\Lambda_{c}\bar D_{2}^{*}\rightarrow\Lambda_{c}\bar D_{2}^{*}}({\bf{r}})&=&-2AY_\sigma\mathcal{A}_4-\frac{1}{2}BY_\omega\mathcal{A}_4,\\
%%%%%%%%
\mathcal{V}^{\Sigma_{c}\bar D_{1}\rightarrow\Sigma_{c}\bar D_{2}^{*}}({\bf{r}})&=&-\frac{D\mathcal{H}(I)}{3\sqrt{6}}\left[\mathcal{D}_1\nabla^2
+\mathcal{D}_2r\frac{\partial}{\partial r}\frac{1}{r}\frac{\partial}{\partial r}\right]Y_{P0}\nonumber\\
&&-\frac{2 F\mathcal{G}(I)}{9\sqrt{6}}\left[2\mathcal{D}_1\nabla^2-\mathcal{D}_2r\frac{\partial}{\partial r}\frac{1}{r}\frac{\partial}{\partial r}\right]Y_{V0},\label{eq4}\nonumber\\\\
%%%%%%%%
\mathcal{V}^{\Sigma_{c}\bar D_{1}\rightarrow\Sigma_{c}^{*}\bar D_{1}}({\bf{r}})&=&-\frac{1}{\sqrt{3}}CY_{\sigma1}\mathcal{D}_3-\frac{1}{2\sqrt{3}}
E\mathcal{G}(I)Y_{V1}\mathcal{D}_3\nonumber\\
&&-\frac{5D\mathcal{H}(I)}{12\sqrt{3}}\left[\mathcal{D}_4\nabla^2+\mathcal{D}_5
r\frac{\partial}{\partial r}\frac{1}{r}\frac{\partial}{\partial r}\right]Y_{P1}\nonumber\\
&&+\frac{5F\mathcal{G}(I)}{18\sqrt{3}}\left[2\mathcal{D}_4\nabla^2-\mathcal{D}_5
r\frac{\partial}{\partial r}\frac{1}{r}\frac{\partial}{\partial r}\right]Y_{V1},\nonumber\\\\
%%%%%%%%
\mathcal{V}^{\Sigma_{c}\bar D_{1}\rightarrow\Sigma_{c}^{*}\bar D_{2}^{*}}({\bf{r}})&=&-\frac{D\mathcal{H}(I)}{6\sqrt{2}}
\left[\mathcal{D}_6\nabla^2-\mathcal{D}_7r\frac{\partial}{\partial r}\frac{1}{r}\frac{\partial}{\partial r}\right]Y_{P2}\nonumber\\
&&+\frac{F\mathcal{G}(I)}{9\sqrt{2}}\left[2\mathcal{D}_6\nabla^2
-\mathcal{D}_7r\frac{\partial}{\partial r}\frac{1}{r}\frac{\partial}{\partial r}\right]Y_{V2},\nonumber\\\\
%%%%%%%%
\mathcal{V}^{\Sigma_{c}\bar D_{2}^{*}\rightarrow\Sigma_{c}^{*}\bar D_{1}}({\bf{r}})&=&-\frac{D\mathcal{H}(I)}{6\sqrt{2}}\left[\mathcal{D}_8\nabla^2
+\mathcal{D}_9r\frac{\partial}{\partial r}\frac{1}{r}\frac{\partial}{\partial r}\right]Y_{P3}\nonumber\\
&&+\frac{F\mathcal{G}(I)}{9\sqrt{2}}\left[2\mathcal{D}_8\nabla^2
-\mathcal{D}_9r\frac{\partial}{\partial r}\frac{1}{r}\frac{\partial}{\partial r}\right]Y_{V3},\nonumber\\\\
%%%%%%%%
\mathcal{V}^{\Sigma_{c}\bar D_{2}^{*}\rightarrow\Sigma_{c}^{*}\bar D_{2}^{*}}({\bf{r}})&=&-\frac{1}{\sqrt{3}}CY_{\sigma4}\mathcal{D}_{10}
-\frac{1}{2\sqrt{3}}E\mathcal{G}(I)Y_{V4}\mathcal{D}_{10}\nonumber\\
&&-\frac{D\mathcal{H}(I)}{2\sqrt{3}}\left[\mathcal{D}_{11}\nabla^2
+\mathcal{D}_{12}r\frac{\partial}{\partial r}\frac{1}{r}\frac{\partial}{\partial r}\right]Y_{P4}\nonumber\\
&&+\frac{E\mathcal{G}(I)}{3\sqrt{3}}\left[2\mathcal{D}_{11}\nabla^2
-\mathcal{D}_{12}r\frac{\partial}{\partial r}\frac{1}{r}\frac{\partial}{\partial r}\right]Y_{V4},\nonumber\\\\
%%%%%%%%
\mathcal{V}^{\Sigma_{c}^{*}\bar D_{1}\rightarrow\Sigma_{c}^{*}\bar D_{2}^{*}}({\bf{r}})&=&\frac{D\mathcal{H}(I)}{2\sqrt{6}}\left[\mathcal{D}_{13}\nabla^2
+\mathcal{D}_{14}r\frac{\partial}{\partial r}\frac{1}{r}\frac{\partial}{\partial r}\right]Y_{P5}\nonumber\\
&&-\frac{ F\mathcal{G}(I)}{3\sqrt{6}}\left[2\mathcal{D}_{13}\nabla^2-\mathcal{D}_{14}
r\frac{\partial}{\partial r}\frac{1}{r}\frac{\partial}{\partial r}\right]Y_{V5},\nonumber\\\\
%%%%%%%%
\mathcal{V}^{\Lambda_{c}\bar D_{1}\rightarrow\Sigma_{c}\bar D_{1}}({\bf{r}})&=&-\frac{5G}{18\sqrt{2}}\left[\mathcal{A}_{2}\nabla^2
+\mathcal{A}_{3}r\frac{\partial}{\partial r}\frac{1}{r}\frac{\partial}{\partial r}\right]Y_{\pi6}\nonumber\\
&&-\frac{5H}{9\sqrt{2}}\left[2\mathcal{A}_{2}\nabla^2-\mathcal{A}_{3}r\frac{\partial}{\partial r}\frac{1}{r}\frac{\partial}{\partial r}\right]Y_{\rho6},\\
%%%%%%%%
\mathcal{V}^{\Lambda_{c}\bar D_{1}\rightarrow\Sigma_{c}\bar D_{2}^{*}}({\bf{r}})&=&\frac{\sqrt{3}G}{18}\left[\mathcal{D}_{1}\nabla^2+\mathcal{D}_{2}
r\frac{\partial}{\partial r}\frac{1}{r}\frac{\partial}{\partial r}\right]Y_{\pi7}\nonumber\\
&&+\frac{\sqrt{3}H}{9}\left[2\mathcal{D}_{1}\nabla^2-\mathcal{D}_{2}r\frac{\partial}{\partial r}\frac{1}{r}\frac{\partial}{\partial r}\right]Y_{\rho7},\nonumber\\\\
%%%%%%%%
\mathcal{V}^{\Lambda_{c}\bar D_{1}\rightarrow\Sigma_{c}^{*}\bar D_{1}}({\bf{r}})&=&\frac{5\sqrt{6}}{12}G\left[\mathcal{D}_{15}\nabla^2+\mathcal{D}_{16}
r\frac{\partial}{\partial r}\frac{1}{r}\frac{\partial}{\partial r}\right]Y_{\pi8}\nonumber\\
&&-\frac{5\sqrt{6}}{18}H\left[2\mathcal{D}_{15}\nabla^2-\mathcal{D}_{16}
r\frac{\partial}{\partial r}\frac{1}{r}\frac{\partial}{\partial r}\right]Y_{\rho8},\nonumber\\\\
%%%%%%%%
\mathcal{V}^{\Lambda_{c}\bar D_{1}\rightarrow\Sigma_{c}^{*}\bar D_{2}^{*}}({\bf{r}})&=&-\frac{G}{6}\left[\mathcal{D}_{17}\nabla^2
+\mathcal{D}_{18}r\frac{\partial}{\partial r}\frac{1}{r}\frac{\partial}{\partial r}\right]Y_{\pi9}\nonumber\\
&&-\frac{H}{3}\left[2\mathcal{D}_{17}\nabla^2-\mathcal{D}_{18}r\frac{\partial}{\partial r}\frac{1}{r}\frac{\partial}{\partial r}\right]Y_{\rho9},\\
%%%%%%%%
\mathcal{V}^{\Lambda_{c}\bar D_{2}^{*}\rightarrow\Sigma_{c}\bar D_{1}}({\bf{r}})&=&\frac{\sqrt{3}}{18}G\left[\mathcal{D}_{19}\nabla^2
+\mathcal{D}_{20}r\frac{\partial}{\partial r}\frac{1}{r}\frac{\partial}{\partial r}\right]Y_{\pi10}\nonumber\\
&&+\frac{1}{3\sqrt{3}}H\left[2\mathcal{D}_{19}\nabla^2-\mathcal{D}_{20}
r\frac{\partial}{\partial r}\frac{1}{r}\frac{\partial}{\partial r}\right]Y_{\rho10},\nonumber\\\\
%%%%%%%%
\mathcal{V}^{\Lambda_{c}\bar D_{2}^{*}\rightarrow\Sigma_{c}\bar D_{2}^{*}}({\bf{r}})&=&-\frac{\sqrt{2}G}{6}\left[\mathcal{A}_{5}\nabla^2
+\mathcal{A}_{6}r\frac{\partial}{\partial r}\frac{1}{r}\frac{\partial}{\partial r}\right]Y_{\pi11}\nonumber\\
&&+\frac{\sqrt{2}H}{3}\left[2\mathcal{A}_{5}\nabla^2-\mathcal{A}_{6}r\frac{\partial}{\partial r}\frac{1}{r}\frac{\partial}{\partial r}\right]Y_{\rho11},\nonumber\\\\
%%%%%%%%
\mathcal{V}^{\Lambda_{c}\bar D_{2}^{*}\rightarrow\Sigma_{c}^{*}\bar D_{1}}({\bf{r}})&=&-\frac{G}{6}\left[\mathcal{D}_{21}\nabla^2+\mathcal{D}_{22}
r\frac{\partial}{\partial r}\frac{1}{r}\frac{\partial}{\partial r}\right]Y_{\pi12}\nonumber\\
&&-\frac{H}{3}\left[2\mathcal{D}_{21}\nabla^2-\mathcal{D}_{22}r\frac{\partial}{\partial r}\frac{1}{r}\frac{\partial}{\partial r}\right]Y_{\rho12},\\
%%%%%%%%
\mathcal{V}^{\Lambda_{c}\bar D_{2}^{*}\rightarrow\Sigma_{c}^{*}\bar D_{2}^{*}}({\bf{r}})&=&\frac{G}{\sqrt{6}}\left[\mathcal{D}_{23}\nabla^2
+\mathcal{D}_{24}r\frac{\partial}{\partial r}\frac{1}{r}\frac{\partial}{\partial r}\right]Y_{\pi13}\nonumber\\
&&+\frac{\sqrt{6}H}{3}\left[2\mathcal{D}_{23}\nabla^2-\mathcal{D}_{24}r\frac{\partial}{\partial r}\frac{1}{r}\frac{\partial}{\partial r}\right]Y_{\rho13}.\nonumber\\\label{eq5}
\end{eqnarray}
Here, the effective potentials in Eqs.(\ref{eq4})-(\ref{eq5}) are only concerned with the coupled channel analysis. Of course, we need to emphasis that the relevant expressions of Appendix \ref{app0201} will also be involved in the coupled channel analysis. Variables $q_i\,(i=0,\,1,...,\,13)$ in the above OBE effective potentials are defined as
\begin{eqnarray}
\left.\begin{array}{ll}
q_0 = \frac{m_{D_{2}^{*}}^2-m_{D_{1}}^2}{2(m_{\Sigma_{c}}+m_{D_{2}^{*}})},\quad\quad\quad\quad\quad
&q_1= \frac{m_{\Sigma_c^*}^2-m_{\Sigma_c}^2}{2(m_{\Sigma_c^*}+m_{D_{1}})},\\
q_2= \frac{m_{\Sigma_c^*}^2+m_{D_{1}}^2-m_{\Sigma_c}^2-m_{D_{2}^{*}}^2}{2(m_{\Sigma_c^*}+m_{D_{2}^{*}})},\quad
&q_3=\frac{m_{\Sigma_c^*}^2+m_{D_{2}^{*}}^2-m_{\Sigma_c}^2-m_{D_{1}}^2}{2(m_{\Sigma_c^*}+m_{D_{1}})},\\
q_4= \frac{m_{\Sigma_c^*}^2-m_{\Sigma_c}^2}{2(m_{\Sigma_c^*}+m_{D_{1}})},\quad
&q_5= \frac{m_{D_{2}^{*}}^2-m_{D_{1}}^2}{2(m_{\Sigma_c^*}+m_{D_{2}^{*}})},\\
q_6= \frac{m_{\Sigma_c}^2-m_{\Lambda_c}^2}{2(m_{\Lambda_c}+m_{D_{1}})},\quad
&q_7 = \frac{m_{\Sigma_c}^2+m_{D_{1}}^2-m_{\Lambda_c}^2-m_{D_{2}^{*}}^2}{2(m_{\Sigma_c}
+m_{D_{2}^{*}})},\\
q_8= \frac{m_{\Sigma_c^{*}}^2-m_{\Lambda_c}^2}{2(m_{\Sigma_c^{*}}+m_{D_{1}})},\quad
&q_9 = \frac{m_{\Sigma_c^{*}}^2+m_{D_{1}}^2-m_{\Lambda_c}^2-m_{D_{2}^{*}}^2}{2(m_{\Sigma_c^{*}}
+m_{D_{2}^{*}})},\\
q_{10} = \frac{m_{\Sigma_c}^2+m_{D_{2}^{*}}^2-m_{\Lambda_c}^2-m_{D_{1}}^2}{2(m_{\Sigma_c}
+m_{D_{1}})},\quad
&q_{11}= \frac{m_{\Sigma_c}^2-m_{\Lambda_c}^2}{2(m_{\Sigma_c}+m_{D_{2}^{*}})},\\
q_{12} = \frac{m_{\Sigma_c^{*}}^2+m_{D_{2}^{*}}^2-m_{\Lambda_c}^2-m_{D_{1}}^2}{2(m_{\Sigma_c^{*}}
+m_{D_{1}})},\quad
&q_{13}= \frac{m_{\Sigma_c^{*}}^2-m_{\Lambda_c}^2}{2(m_{\Sigma_c^{*}}+m_{D_{1}})},\\
\Lambda_i^2= \Lambda^2-q_i^2,\quad
&m_{Ei}^2 =m_E^2-q_i^2.\\
\end{array}\right.
\end{eqnarray}

In Eqs. (\ref{eq4})-(\ref{eq5}), we also define several operators, i.e.,
\begin{eqnarray*}\label{op}
\mathcal{D}_{1}&=&\mathcal{X}\left({\bm\epsilon^{\dagger}_{4m}}\cdot{\bm\epsilon_{2}}\right)
\chi^{\dagger}_3\left({\bm\sigma}\cdot{\bm\epsilon^{\dagger}_{4n}}\right)\chi_1,\nonumber\\
\mathcal{D}_{2}&=&\mathcal{X}\left({\bm\epsilon^{\dagger}_{4m}}\cdot{\bm\epsilon_{2}}\right)
\chi^{\dagger}_3S({\bm\sigma},{\bm \epsilon^{\dagger}_{4n}},\hat{\bm{r}})\chi_1,\nonumber\\
\mathcal{D}_{3}&=&\mathcal{Z}\chi^{\dagger m_1}_3\left({\bm\epsilon^{\dagger m_2}_{3}}\cdot{\bm\sigma}\right)\left({\bm\epsilon^{\dagger}_{4}}\cdot {\bm\epsilon_{2}}\right)\chi_1,\nonumber\\
\mathcal{D}_{4}&=&\mathcal{Z}\chi^{\dagger m_1}_3\left[\left({\bm\epsilon^{\dagger m_2}_{3}}\times{\bm\sigma}\right)\cdot\left({\bm\epsilon^{\dagger}_{4}}\times {\bm\epsilon_{2}}\right)\right]\chi_1,\nonumber\\
\mathcal{D}_{5}&=&\mathcal{Z}\chi^{\dagger m_1}_3S({\bm\epsilon^{\dagger m_2}_{3}}\times{\bm\sigma},{\bm\epsilon^{\dagger}_{4}}\times {\bm\epsilon_{2}},\hat{\bm{r}})\chi_1,\nonumber\\
\mathcal{D}_{6}&=&\mathcal{X}\mathcal{Z}\left({\bm\epsilon^{\dagger}_{4m}}\cdot
{\bm\epsilon_{2}}\right)\chi^{\dagger m_1}_3\left[\left(i{\bm\epsilon^{\dagger m_2}_{3}}\times {\bm\sigma}\right)\cdot{\bm \epsilon^{\dagger}_{4n}}\right]\chi_1,\nonumber\\
\mathcal{D}_{7}&=&\mathcal{X}\mathcal{Z}\left({\bm\epsilon^{\dagger}_{4m}}\cdot
{\bm\epsilon_{2}}\right)\chi^{\dagger m_1}_3S(i{\bm\epsilon^{\dagger m_2}_{3}}\times {\bm\sigma},{\bm \epsilon^{\dagger}_{4n}},\hat{\bm{r}})\chi_1,\nonumber\\
\mathcal{D}_{8}&=&\mathcal{X}\mathcal{Z}\left({\bm\epsilon^{\dagger}_{4}}\cdot
{\bm\epsilon_{2m}}\right)\chi^{\dagger m_1}_3\left[\left(i{\bm\epsilon^{\dagger}_{3}}\times {\bm\sigma}\right)\cdot{\bm \epsilon_{2n}}\right]\chi_1,\nonumber\\
\mathcal{D}_{9}&=&\mathcal{X}\mathcal{Z}\left({\bm\epsilon^{\dagger}_{4}}\cdot
{\bm\epsilon_{2m}}\right)\chi^{\dagger m_1}_3S(i{\bm\epsilon^{\dagger}_{3}}\times {\bm\sigma},{\bm \epsilon_{2n}},\hat{\bm{r}})\chi_1,\nonumber\\
\mathcal{D}_{10}&=&\mathcal{X}\mathcal{Y}\mathcal{Z}\chi^{\dagger m_1}_3\left({\bm\epsilon^{\dagger m_2}_{3}}\cdot{\bm\sigma}\right)\left({\bm\epsilon^{\dagger}_{4m}}\cdot
{\bm\epsilon_{2a}}\right)\left({\bm\epsilon^{\dagger}_{4n}}\cdot {\bm\epsilon_{2b}}\right)\chi_1,\nonumber\\
\mathcal{D}_{11}&=&\mathcal{X}\mathcal{Y}\mathcal{Z}\left({\bm\epsilon^{\dagger}_{4m}}
\cdot{\bm\epsilon_{2a}}\right)\chi^{\dagger m_1}_3\left[\left({\bm\epsilon^{\dagger m_2}_{3}}\times{\bm\sigma}\right)\cdot\left({\bm\epsilon^{\dagger}_{4n}}\times {\bm\epsilon_{2b}}\right)\right]\chi_1,\nonumber\\
\mathcal{D}_{12}&=&\mathcal{X}\mathcal{Y}\mathcal{Z}\left({\bm\epsilon^{\dagger}_{4m}}
\cdot{\bm\epsilon_{2a}}\right)\chi^{\dagger m_1}_3S({\bm\epsilon^{\dagger m_2}_{3}}\times{\bm\sigma},{\bm\epsilon^{\dagger}_{4n}}\times {\bm\epsilon_{2b}},\hat{\bm{r}})\chi_1,\nonumber\\
\mathcal{D}_{13}&=&\mathcal{X}\mathcal{Z}\mathcal{T}\mathcal{Q}
\left({\bm\epsilon^{\dagger}_{4m}}
\cdot{\bm\epsilon_{2}}\right)\left[i\left({\bm\epsilon^{\dagger m_2^\prime}_{3}}\times{\bm\epsilon_{1}}^{m_2}\right)\cdot{\bm \epsilon^{\dagger}_{4n}}\right],\nonumber\\
\mathcal{D}_{14}&=&\mathcal{X}\mathcal{Z}\mathcal{T}\mathcal{Q}
\left({\bm\epsilon^{\dagger}_{4m}}
\cdot{\bm\epsilon_{2}}\right)S(i{\bm\epsilon^{\dagger m_2^\prime}_{3}}\times{\bm\epsilon_{1}}^{m_2},{\bm \epsilon^{\dagger}_{4n}},\hat{\bm{r}}),\nonumber\\
\mathcal{D}_{15}&=&\mathcal{Z}\chi^{\dagger m_1}_3\left[i{\bm\epsilon^{\dagger m_2}_{3}}\cdot\left({\bm\epsilon^{\dagger}_{4}}\times {\bm\epsilon_{2}}\right)\right]\chi_1,\nonumber\\
\mathcal{D}_{16}&=&\mathcal{Z}\chi^{\dagger m_1}_3S(i{\bm\epsilon^{\dagger m_2}_{3}},{\bm\epsilon^{\dagger}_{4}}\times {\bm\epsilon_{2}},\hat{\bm{r}})\chi_1,\nonumber\\
\mathcal{D}_{17}&=&\mathcal{X}\mathcal{Z}\left({\bm\epsilon^{\dagger}_{4m}}
\cdot{\bm\epsilon_{2}}\right)
\chi^{\dagger m_1}_3\left({\bm\epsilon^{\dagger m_2}_{3}}\cdot{\bm \epsilon^{\dagger}_{4n}}\right)\chi_1,\nonumber\\
\mathcal{D}_{18}&=&\mathcal{X}\mathcal{Z}\left({\bm\epsilon^{\dagger}_{4m}}
\cdot{\bm\epsilon_{2}}\right)
\chi^{\dagger m_1}_3S(\bm\epsilon^{\dagger m_2}_{3},{\bm \epsilon^{\dagger}_{4n}},\hat{\bm{r}})\chi_1,\nonumber\\
\mathcal{D}_{19}&=&\mathcal{X}\left({\bm\epsilon^{\dagger}_{4}}\cdot{\bm\epsilon_{2m}}\right)
\chi^{\dagger}_3\left({\bm\sigma}\cdot{\bm\epsilon_{2n}}\right)\chi_1,\nonumber\\
\mathcal{D}_{20}&=&\mathcal{X}\left({\bm\epsilon^{\dagger}_{4}}\cdot{\bm\epsilon_{2m}}\right)
\chi^{\dagger}_3S({\bm\sigma},{\bm \epsilon_{2n}},\hat{\bm{r}})\chi_1,\nonumber\\
\mathcal{D}_{21}&=&\mathcal{X}\mathcal{Z}\left({\bm\epsilon^{\dagger}_{4}}
\cdot{\bm\epsilon_{2m}}\right)\chi^{\dagger m_1}_3\left({\bm\epsilon^{\dagger m_2}_{3}}\cdot{\bm \epsilon_{2n}}\right)\chi_1,\nonumber\\
\mathcal{D}_{22}&=&\mathcal{X}\mathcal{Z}\left({\bm\epsilon^{\dagger}_{4}}
\cdot{\bm\epsilon_{2m}}\right)\chi^{\dagger m_1}_3S(\bm\epsilon^{\dagger m_2}_{3},{\bm \epsilon_{2n}},\hat{\bm{r}})\chi_1,\nonumber\\
\mathcal{D}_{23}&=&\mathcal{X}\mathcal{Y}\mathcal{Z}\left({\bm\epsilon^{\dagger}_{4m}}
\cdot{\bm\epsilon_{2a}}\right)\chi^{\dagger m_1}_3\left[i{\bm\epsilon^{\dagger m_2}_{3}}\cdot\left({\bm\epsilon^{\dagger}_{4n}}\times {\bm\epsilon_{2b}}\right)\right]\chi_1,\nonumber\\
\mathcal{D}_{24}&=&\mathcal{X}\mathcal{Y}\mathcal{Z}\left({\bm\epsilon^{\dagger}_{4m}}
\cdot{\bm\epsilon_{2a}}\right)\chi^{\dagger m_1}_3S({\bm\epsilon^{\dagger m_2}_{3}},i{\bm\epsilon^{\dagger}_{4n}}\times {\bm\epsilon_{2b}},\hat{\bm{r}})\chi_1.\nonumber\\
\end{eqnarray*}
The operators $\mathcal{D}_k\,(k=1,\,2,...,\,24)$ are only concerned with the coupled channel analysis. In Table~\ref{matrix2}, we collect the relevant numerical matrices $\langle f|\mathcal{D}_k|i\rangle$ for these operators $\mathcal{D}_k$.
\renewcommand\tabcolsep{0.16cm}
\renewcommand{\arraystretch}{1.80}
\begin{table*}[!htbp]
  \caption{Matrix elements $\langle f|\mathcal{D}_k|i\rangle$ in various channels for operators $\mathcal{D}_k$ in the effective potentials. \label{matrix2}}
  \begin{tabular}{c|cccccccccccc}\toprule[1pt]
   {{{Spin}}}
                    &$\langle\mathcal{D}_1\rangle$    &$\langle\mathcal{D}_2\rangle$    &$\langle\mathcal{D}_3\rangle$    &$\langle\mathcal{D}_4\rangle$   &$\langle\mathcal{D}_5\rangle$    &$\langle\mathcal{D}_6\rangle$& $\langle\mathcal{D}_7\rangle$ &$\langle\mathcal{D}_8\rangle$&$\langle\mathcal{D}_9\rangle$    &$\langle\mathcal{D}_{10}\rangle$    &$\langle\mathcal{D}_{11}\rangle$   &$\langle\mathcal{D}_{12}\rangle$ \\
 $J=1/2$
                      &$/$
                      &$/$
                      &$\left(\begin{array}{c} 0 \end{array}\right)$
                      &$\left(\begin{array}{c} -\sqrt{\frac{2}{3}}  \end{array}\right)$
                      &$\left(\begin{array}{c} 0 \end{array}\right)$
                       &$\left(\begin{array}{c} -\sqrt{\frac{5}{3}} \end{array}\right)$
                      &$\left(\begin{array}{c} 0  \end{array}\right)$
                      &$/$
                      &$/$
                      &$/$
                      &$/$
                      &$/$\\
$J=3/2$
                      &$\left(\begin{array}{c} \sqrt{\frac{5}{2}} \end{array}\right)$
                      &$\left(\begin{array}{c} 0\end{array}\right)$
                      &$\left(\begin{array}{c} 0\end{array}\right)$
                      &$\left(\begin{array}{c} -\sqrt{\frac{5}{3}} \end{array}\right)$
                      &$\left(\begin{array}{c} 0 \end{array}\right)$
                       &$\left(\begin{array}{c} -\sqrt{\frac{5}{6}} \end{array}\right)$
                      &$\left(\begin{array}{c} 0  \end{array}\right)$
                      &$\left(\begin{array}{c} \frac{1}{\sqrt{6}}\end{array}\right)$
                      &$\left(\begin{array}{c} 0\end{array}\right)$
                      &$\left(\begin{array}{c} \frac{2}{5\sqrt{3}}\end{array}\right)$
                      &$\left(\begin{array}{c} -\frac{\sqrt{3}}{2} \end{array}\right)$
                      &$\left(\begin{array}{c} 0 \end{array}\right)$\\
$J=5/2$
                      &$/$
                      &$/$
                      &$/$
                      &$/$
                      &$/$
                      &$/$
                      &$/$
                      &$\left(\begin{array}{c} 1\end{array}\right)$
                      &$\left(\begin{array}{c} 0\end{array}\right)$
                      &$\left(\begin{array}{c} -\sqrt{\frac{8}{21}}\end{array}\right)$
                      &$\left(\begin{array}{c} -\sqrt{\frac{7}{6}}\end{array}\right)$
                      &$\left(\begin{array}{c} 0 \end{array}\right)$\\\midrule[1pt]
 {{{Spin}}}
                       &$\langle\mathcal{D}_{13}\rangle$ &$\langle\mathcal{D}_{14}\rangle$ &$\langle\mathcal{D}_{15}\rangle$    &$\langle\mathcal{D}_{16}\rangle$ &$\langle\mathcal{D}_{17}\rangle$    &$\langle\mathcal{D}_{18}\rangle$   &$\langle\mathcal{D}_{19}\rangle$    &$\langle\mathcal{D}_{20}\rangle$ &$\langle\mathcal{D}_{21}\rangle$ &$\langle\mathcal{D}_{22}\rangle$    &$\langle\mathcal{D}_{23}\rangle$    &$\langle\mathcal{D}_{24}\rangle$\\
 $J=1/2$
                       &$\left(\begin{array}{c} -\sqrt{\frac{5}{18}} \end{array}\right)$
                      &$\left(\begin{array}{c} 0  \end{array}\right)$
                      &$\left(\begin{array}{c} \sqrt{\frac{2}{3}}\end{array}\right)$
                      &$\left(\begin{array}{c} 0\end{array}\right)$
                                            &$\left(\begin{array}{c} -\sqrt{\frac{5}{3}}\end{array}\right)$
                      &$\left(\begin{array}{c} 0\end{array}\right)$
                      &$/$
                       &$/$
                      &$/$
                      &$/$
                      &$/$
                      &$/$\\
 $J=3/2$

                       &$\left(\begin{array}{c} -\frac{2\sqrt{2}}{3} \end{array}\right)$
                      &$\left(\begin{array}{c} 0  \end{array}\right)$
                      &$\left(\begin{array}{c} \sqrt{\frac{5}{3}}\end{array}\right)$
                      &$\left(\begin{array}{c} 0\end{array}\right)$
                      &$\left(\begin{array}{c} -\sqrt{\frac{5}{6}}\end{array}\right)$
                      &$\left(\begin{array}{c} 0\end{array}\right)$
                      &$\left(\begin{array}{c} \sqrt{\frac{5}{2}}\end{array}\right)$
                       &$\left(\begin{array}{c} 0\end{array}\right)$
                      &$\left(\begin{array}{c} \frac{1}{\sqrt{6}}  \end{array}\right)$
                      &$\left(\begin{array}{c}0\end{array}\right)$
                      &$\left(\begin{array}{c} \frac{\sqrt{3}}{2}\end{array}\right)$
                      &$\left(\begin{array}{c} 0\end{array}\right)$\\
 $J=5/2$
                      &$\left(\begin{array}{c} -\sqrt{\frac{7}{6}}\end{array}\right)$
                      &$\left(\begin{array}{c} 0  \end{array}\right)$
                      &$\left(\begin{array}{c} 0\end{array}\right)$
                      &$\left(\begin{array}{c} \sqrt{\frac{2}{5}}\end{array}\right)$
                      &$\left(\begin{array}{c} 0\end{array}\right)$
                      &$\left(\begin{array}{c} \sqrt{\frac{7}{15}}\end{array}\right)$
                      &$/$
                      &$/$
                      &$\left(\begin{array}{c} 1  \end{array}\right)$
                      &$\left(\begin{array}{c}0\end{array}\right)$
                      &$\left(\begin{array}{c} \frac{\sqrt{3}}{2}\end{array}\right)$
                      &$\left(\begin{array}{c} 0\end{array}\right)$\\
   \bottomrule[1pt]
  \end{tabular}
\end{table*}

\section*{ACKNOWLEDGMENTS}

This project is supported by the China National Funds for Distinguished Young Scientists under Grant No. 11825503 and by the National Program for Support of Top-notch Young Professionals and the National Natural Science Foundation of China under Grants No. 11705072 and No. 11965016. R. C. is also supported by the National Postdoctoral Program for Innovative Talent.

\end{document}